\documentclass[12pt,a4paper]{article}

\usepackage{times}
\usepackage[a4paper]{geometry}

\usepackage{amsmath}
\usepackage{amssymb}

\usepackage{multirow}
\usepackage{wasysym}
\usepackage{graphicx}

\usepackage{cite}

\usepackage{color} 

\usepackage[labelfont=bf,labelsep=period,justification=raggedright]{caption}

\makeatletter
\renewcommand{\@biblabel}[1]{\quad#1.}
\makeatother

\makeatletter
\renewcommand{\maketitle}{%
   \begin{flushleft}%
      \sffamily
      {\Large\bfseries\@title\par}%
      \medskip
      {\large\@author\par}%
      \medskip
      {\itshape\@date\par}%
      \bigskip\hrule\vspace*{2pc}%
   \end{flushleft}%
}
\makeatother

\date{}

\begin{document}

\begin{flushleft}
\title{\Large\textsf{An excess of gene expression divergence on the X chromosome in \emph{Drosophila} embryos; implications for the faster-X hypothesis}
}
\maketitle

Melek A. Kayserili$^{1, \dagger}$,
Dave T. Gerrard$^{2, \dagger}$, 
Pavel Tomancak$^{1, \ast}$, 
Alex T. Kalinka$^{1,\ast}$

\vspace{0.5cm}
\small{1} Max Planck Institute for Molecular Cell Biology and Genetics, Pfotenhauerstr. 108, 01307 Dresden, Germany.
\\
\small{2} Faculty of Life Sciences, The University of Manchester, Michael Smith Building, Oxford Road, Manchester M13 9PT, UK.
\\

$\dagger$ These authors contributed equally
$\ast$ E-mail: Corresponding kalinka@mpi-cbg.de, tomancak@mpi-cbg.de

\end{flushleft}

\section*{Abstract}

The X chromosome is present as a single copy in the heterogametic sex, and this hemizygosity is expected to drive unusual patterns of evolution on the X relative to the autosomes. For example, the hemizgosity of the X may lead to a lower chromosomal effective population size compared to the autosomes suggesting that the X might be more strongly affected by genetic drift. However, the X may also experience stronger positive selection than the autosomes because recessive beneficial mutations will be more visible to selection on the X where they will spend less time being masked by the dominant, less beneficial allele - a proposal known as the faster-X hypothesis. Thus, empirical studies demonstrating increased genetic divergence on the X chromosome could be indicative of either adaptive or non-adaptive evolution. We measured gene expression in \emph{Drosophila} species and in \emph{D. melanogaster} inbred strains for both embryos and adults. In the embryos we found that expression divergence is on average more 
than 20\% higher for genes on the X chromosome relative to the autosomes, but in contrast, in the inbred strains gene expression variation is significantly lower on the X chromosome. Furthermore, expression divergence of genes on Muller's D element is significantly greater along the branch leading to the obscura sub-group, in which this element segregates as a neo-X chromosome. In the adults, divergence is greatest on the X chromosome for males, but not for females, yet in both sexes inbred strains harbour the lowest level of gene expression variation on the X chromosome. We consider different explanations for our results and conclude that they are most consistent within the framework of the faster-X hypothesis.

\clearpage

\section*{Author Summary}

There is a single copy of the X chromosome in males, yet two copies in females. This unique inheritance pattern has long been predicted to influence how the X chromosome evolves. In particular, theory suggests that the single copy of the X in males could facilitate faster evolution of the X, although this faster evolution could be either adaptive or non-adaptive. We measured gene expression across the chromosomes in several different \emph{Drosophila} species, and also in several inbred strains of \emph{D. melanogaster} for both embryos and adults. We found that gene expression is evolving significantly faster between species in the embryos, yet harbours significantly less variation within inbred strains. In adults, evolution between species appears to be much slower than in the embryos, yet they also harbour significantly lower levels of gene expression variation on the X chromosome in inbred strains. Overall, our results are consistent with there being an excess of adaptive evolution on the X chromosome in 
\emph{Drosophila} embryos. Finally, we underscore the importance of biological context for understanding how chromosomes evolve in different species.

\section*{Introduction}

It has long been suspected that the distinct properties of the X chromosome might in turn produce distinct patterns of evolution on the X relative to the autosomes \cite{haldane1924, muller1940}. In particular, the hemizygoisty of the X could be responsible for increased adaptive or non-adaptive evolution on this chromosome. Assuming an equal sex ratio and an equal variance in reproductive success in the two sexes, there will be three copies of the X in each mating pair versus four copies of each autosome thereby exposing the X to elevated levels of genetic drift \cite{avery1984}. If, however, we consider adaptive evolution, then the hemizygosity of the X is expected to facilitate the spread of recessive beneficial mutations, the selective benefit of which would otherwise be masked when in a heterozygous state on the autosomes \cite{haldane1924, hartl1971, avery1984, charlesworthetal1987}. Beneficial mutations with additive effects in heterozygotes are selectively equivalent on the X chromosome and on the 
autosomes, and would therefore be expected to evolve at similar rates across the chromosomes, whereas beneficial mutations that are dominant are expected to evolve faster on the autosomes \cite{charlesworthetal1987}. A faster X may also be expected if mutations have sexually antagonistic effects, in which the sign of the selection coefficient is opposite in males and females \cite{rice1984}. In both adaptive and non-adaptive scenarios, it is the hemizygous context of the X chromosome in the heterogametic sex that is expected to drive more rapid evolution relative to the autosomes \cite{vicoso&charlesworth2006}.

Determining the relative importance of different evolutionary forces in shaping the X chromosome is crucial for understanding several phenomena related to the X. For example, Haldane's rule, which is a classic generalization stating that in the hybrids of inter-species crosses the heterogametic sex is most often the inviable or sterile sex \cite{haldane1921}, could be explained by the fixation of recessive species-specific substitutions on the X chromosome which interact epistatically with autosomal loci \cite{charlesworthetal1987}.  Understanding how the X evolves could also help explain unusual distributions of genes across chromosomes \cite{gurbich&bachtrog2008}, such as a disproportionate number of genes involved in cognitive function residing on the X in mammals \cite{skuse2005} or an excess of sexually antagonistic genes on the X in \emph{Drosophila} \cite{innocenti&morrow2010}. A fuller understanding of how selection acts differentially across autosomes and sex chromosomes could also shed light on the 
role of the X chromosome in the evolution of sexually-selected traits \cite{zechneretal2001}.

Empirical studies have sought to quantify the importance of adaptative processes in driving the evolution of the X. While many studies have found that the differences between species can often be attributed to X-linked loci of large effect \cite{dobzhansky1936, templeton1977, coyne&charlesworth1986}, much of the recent work has found inconsistent evidence for an excess of positive selection of X-linked proteins. For example, studies of chimpanzee and human orthologs shows that X-linked loci have higher rates of adaptive protein evolution than autosomal loci \cite{lu&wu2005, chimpanzeeconsortium2005, hvilsometal2012}, whereas in \emph{Drosophila} species, whole-genome comparisons do not reveal any bias towards higher rates of protein evolution on the X chromosome \cite{betancourtetal2002, thorntonetal2006, connallon2007}. Other \emph{Drosophila} studies, which may use biased samples of genes \cite{vicoso&charlesworth2006}, recover the faster-X effect found in mammals \cite{thornton&long2002, 
countermanetal2004, thornton&long2005, huetal2012} including a study that demonstrated accelerated evolution of X-linked genes on the newly-formed X chromosome of \emph{D. miranda} \cite{bachtrogetal2009}.  A recent study in aphids, an X0 sex determination system, found evidence for adaptive evolution of X-linked genes \cite{jaquieryetal2012}, and, interestingly, the same finding was reported for the Z chromosome (the equivalent of the X chromosome in the ZW sex determination system) in a comparison of chicken and zebra finch orthologs \cite{manketal2007}.

While the evidence for adaptive evolution of the X remains somewhat patchy, such discrepancies suggest that differences in the biology of different groups of species could strongly influence their chromosomal evolution. An important parameter in the faster-X theory is the presence or absence of dosage compensation in the heterogametic sex; that is, whether the presence of a single copy of a gene in the heterogametic sex is compensated, in terms of gene expression, to an extent that it is selectively equivalent to the two copies in the homogametic sex. Theory shows that beneficial mutations will evolve faster on the X compared to the autosomes, only if mutations are at least partially recessive \cite{charlesworthetal1987}. Thus, to observe a global fast-X effect, most beneficial mutations must be at least partially recessive. In the absence of dosage compensation, however, theory suggests that beneficial mutations must be more recessive for the X to evolve faster provided that the 
weaker expression in males results in a correspondingly weaker beneficial selection coefficient \cite{charlesworthetal1987} -- this is because dosage compensation equalises the expression of genes expressed on the X in males and females, and is therefore assumed to also equalise their selection coefficients. Thus, fundamental differences in both the extent and mechanism of dosage compensation between different groups of species could have a dramatic effect on the rate of evolution of the X chromosome \cite{charlesworthetal1987}. However, it is also possible that adaptive evolution of protein sequences accounts for a larger fraction of the evolutionary divergence between some groups of species relative to others. Therefore, while we may not see significantly higher adaptive protein evolution on the X in \emph{Drosophila}, it is conceivable that adaptive differences in this group of species are most often seen in \emph{cis}-regulatory, and therefore non-coding, regions of the genome \cite{thorntonetal2006, 
andolfatto2005}.

We aimed to address evolution on the \emph{Drosophila} X chromosome relative to the autosomes at the level of gene expression divergence. By focusing on gene expression, we relax the implicit assumption of previous studies that a majority of adaptive evolution occurs via changes in amino acid sequences. Additionally, by measuring divergence in terms of gene expression rather than coding sequences, we could compare expression divergence in embryos relative to adults and therefore ask whether gene expression is free to evolve independently in different stages of the animal's life-cycle. Our results show that mean gene expression divergence is higher for the X chromosome relative to autosomes and, more surprisingly, this effect is much stronger in the \emph{Drosophila} embryos relative to the adults.

\clearpage

\section*{Results}

\subsection*{Higher mean expression divergence on the X chromosome in \emph{Drosophila} embryos}

Evidence for accelerated evolution of the X in \emph{Drosophila} has been sought in the adaptive evolution of protein sequences, but has so far produced mixed results \cite{thornton&long2002, countermanetal2004, thornton&long2005, thorntonetal2006, connallon2007}. We chose to focus on the evolution of gene expression with the advantage that we could detect the effects of divergence of non-coding regulatory sequences, and in addition we could directly compare evolution in different stages of the animal's life-cycle. To explore gene expression divergence across \emph{Drosophila} chromosomes we used gene expression data from two distinct stages of the life-cycle -- the embryo \cite{kalinkaetal2010} and the adult \cite{zhangetal2007a}. In addition, we extracted RNA from the embryos of 17 inbred strains of \emph{D. melanogaster} and hybridised the samples to whole-genome microarrays to provide insight into the maintenance of gene expression variation across chromosomes but within a single species. Similarly, for 
adult stages we used 
whole-genome microarray data from 40 adult inbred strains of \emph{D. melanogaster} separated into males and females \cite{ayrolesetal2009, mackayetal2012}. Table S\ref{chromsummary} summarises the chromosomal distributions of genes in each dataset.

\begin{figure}[!ht]
\begin{center}
\includegraphics[width=4.75in]{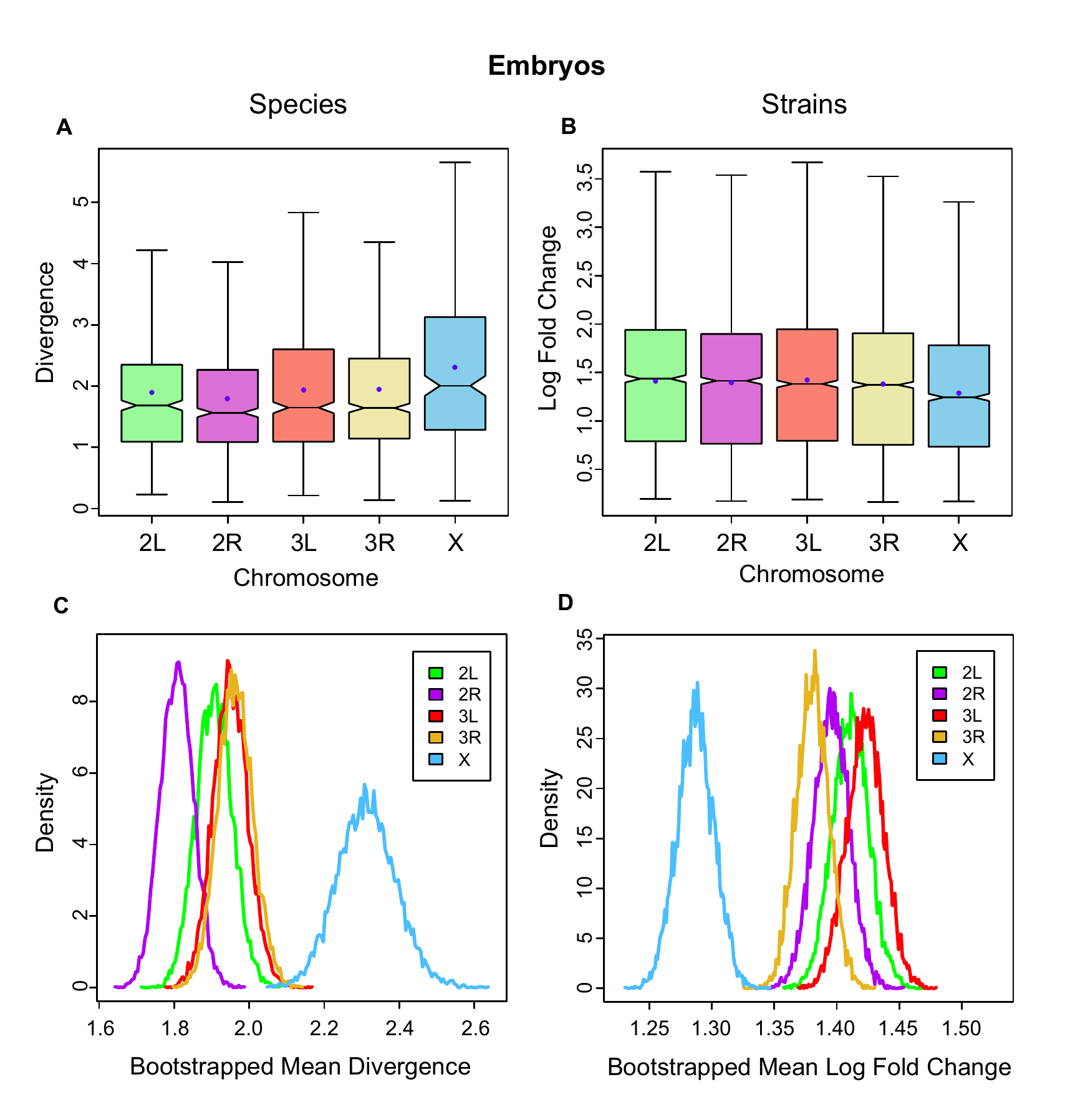}
\end{center}
\caption{
{\bf Gene expression divergence is higher on the X chromosome in \emph{Drosophila} embryos and lower in \emph{D. melanogaster} strains.}\\
The distributions of per gene expression divergence between Drosophila species separated onto each chromosome for {\bf A}, embryos, and {\bf B}, inbred strains of \emph{D. melanogaster}. Divergence is measured per gene as the summed branch lengths for each gene tree for between-species data, and as mean log fold change for inbred strains as described in the Methods. Boxes show the upper and lower quartiles together with the median, error bars encompass data within 1.5 times the inter-quartile range, and blue circles indicate the means. Panels {\bf C} and {\bf D} show, for embryos and strains respectively, the distribution of 10,000 bootstrapped mean divergences for each chromosome using frequency polygons. 
}
\label{embryodiv}
\end{figure}

In the between-species data for embryos, the X chromosome has the highest mean expression divergence ($P = 2.19 \times 10^{-7}$; Figure \ref{embryodiv}A) an effect that ranges from 18\% up to 27\% higher and in all cases is significant (see Table S\ref{contr1} for all chromosomal contrasts). In contrast, the X chromosome shows the lowest level of gene expression variation between the embryos of inbred \emph{D. melanogaster} strains ($P = 1.16 \times 10^{-9}$; Figure \ref{embryodiv}B), ranging from 7\% up to 10\% lower (Table S\ref{contr2}). Bootstrap resampling of the mean divergence across chromosomes confirms that it is significantly higher on the X between species (Figure \ref{embryodiv}C) and significantly lower on the X between strains (Figure \ref{embryodiv}D). In the between-species data, several specific branches in the phylogeny have significantly longer mean lengths judged by bootstrapping individual branches (Figure S\ref{phylogbranches}).

\begin{figure}[!ht]
\begin{center}
\includegraphics[width=3in]{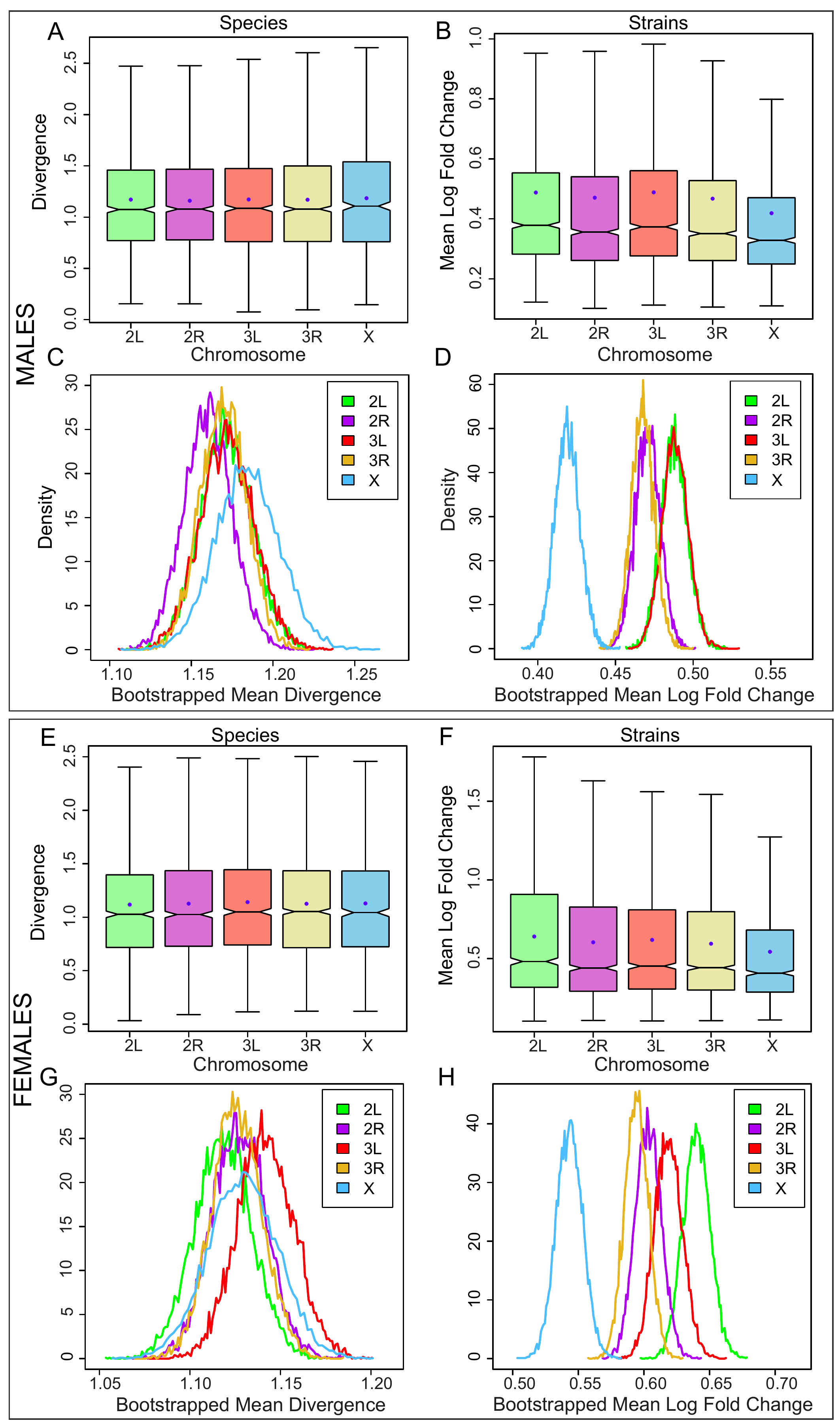}
\end{center}
\caption{
{\bf Gene expression divergence is not higher on the X chromosome in \emph{Drosophila} adults but is lower in \emph{D. melanogaster} adult strains.}\\
The distributions of per gene expression divergence between Drosophila species separated onto each chromosome for {\bf A}, adult males, {\bf B}, inbred adult male strains of \emph{D. melanogaster}, {\bf E}, adult females, and {\bf F}, inbred adult male strains of \emph{D. melanogaster}. Divergence is measured per gene as the summed branch lengths for each gene tree for between-species data, and as mean log fold change for inbred strains as described in the Methods. Boxes show the upper and lower quartiles together with the median, error bars encompass data within 1.5 times the inter-quartile range, and blue circles indicate the means. Panels {\bf C}, {\bf D}, {\bf G}, and {\bf H} show, for adult males, inbred adult strains, adult females, and inbred adult female strains respectively, the distribution of 10,000 bootstrapped mean divergences for each chromosome using frequency polygons.
}
\label{adultdiv}
\end{figure}

In the adults, mean divergence on the X is not higher than the autosomes in females ($P = 0.99$; Figure \ref{adultdiv}A; Table S\ref{contr3}) yet gene expression variation is significantly lower on the X relative to the autosomes in female inbred strains ($P = 7.28 \times 10^{-6}$; Figure \ref{adultdiv}B; Table S\ref{contr4}). In adult males, mean divergence is highest on the X, although it is not significant ($P = 0.35$; Figure \ref{adultdiv}E; Table S\ref{contr5}), but once again mean variation is significantly lower on the X in inbred strains ($P = 9.89 \times 10^{-11}$; Figure \ref{adultdiv}F; Table S\ref{contr6}). Bootstrap resamples confirm that differences between the chromosomes are significant only in the strains (Figures \ref{adultdiv}C,D,G,H). When we reduce genes and species to a common set belonging to both the embryonic and adult between-species data, we find that the X remains more significantly divergent in the embryonic data (Tables S\ref{contrcommon1},\ref{contrcommon2}). In addition, we 
find 
that genes with sex-biased expression patterns also do not display an X effect in either sex confirming that the absence of any effect in adults is not caused by combining genes with different properties in the two sexes (see Methods; Figure S\ref{allsexbias}).

\begin{figure}[!ht]
\begin{center}
\includegraphics[width=6in]{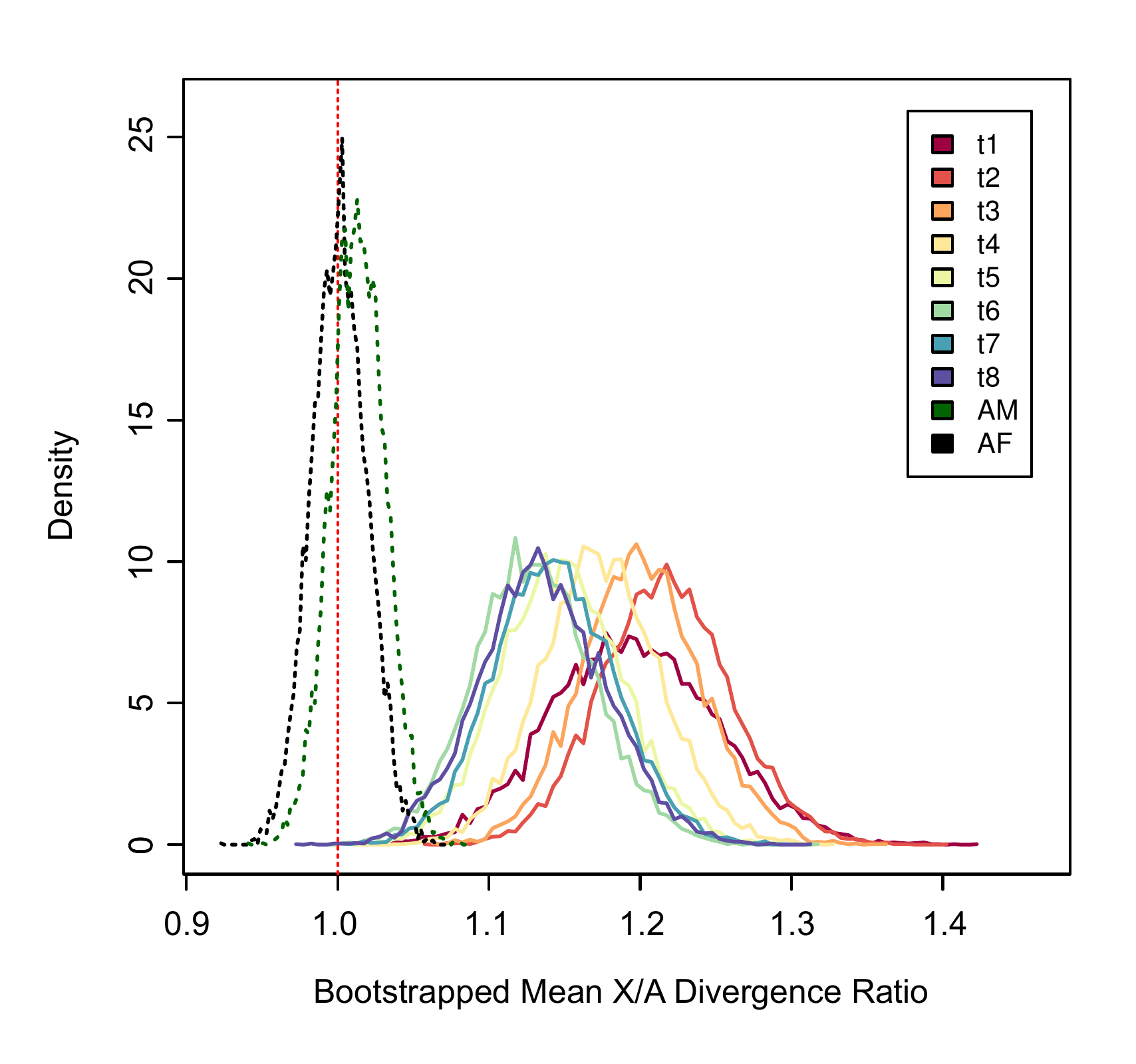}
\end{center}
\caption{
{\bf The X chromosome exhibits an excess of divergence throughout exmbryogenesis.}\\
Bootstrapped mean X/A divergence ratios for each time point throughout embryogenesis. Genes were resampled 10,000 times on each chromosome and the X/A ratio was scored for each time point separately. Bootstrapped distributions are shown as frequency polygons. Dashed green and black lines represent adult males (AM) and adult females (AF) respectively, and the vertical dashed red line marks an X/A ratio of 1.
}
\label{timediv}
\end{figure}

We find that divergence on the X in embryos is not driven by a small subset of time points (Figure \ref{timediv}), nor can it be explained by artifacts caused by extreme expression levels (Figure S\ref{explevelboot}) or by skews in the sex ratio (Figure S\ref{expskewall}; see Methods). Overall, these results indicate that there is a strong and significant excess of gene expression divergence on the X chromosome in \emph{Drosophila} embryos together with a significant reduction of gene expression variation on the X within inbred strains of \emph{D. melanogaster}. Divergence between species coupled with conservation within species is often viewed as a signature of adaptive evolution, and, at the least, is firm evidence against the observed divergence being driven by a relaxation of selective constraints.

\subsection*{Higher divergence on the ancestral branch of the neo-X in \emph{Drosophila} embryos}

\begin{figure}[!ht]
\begin{center}
\includegraphics[width=6in]{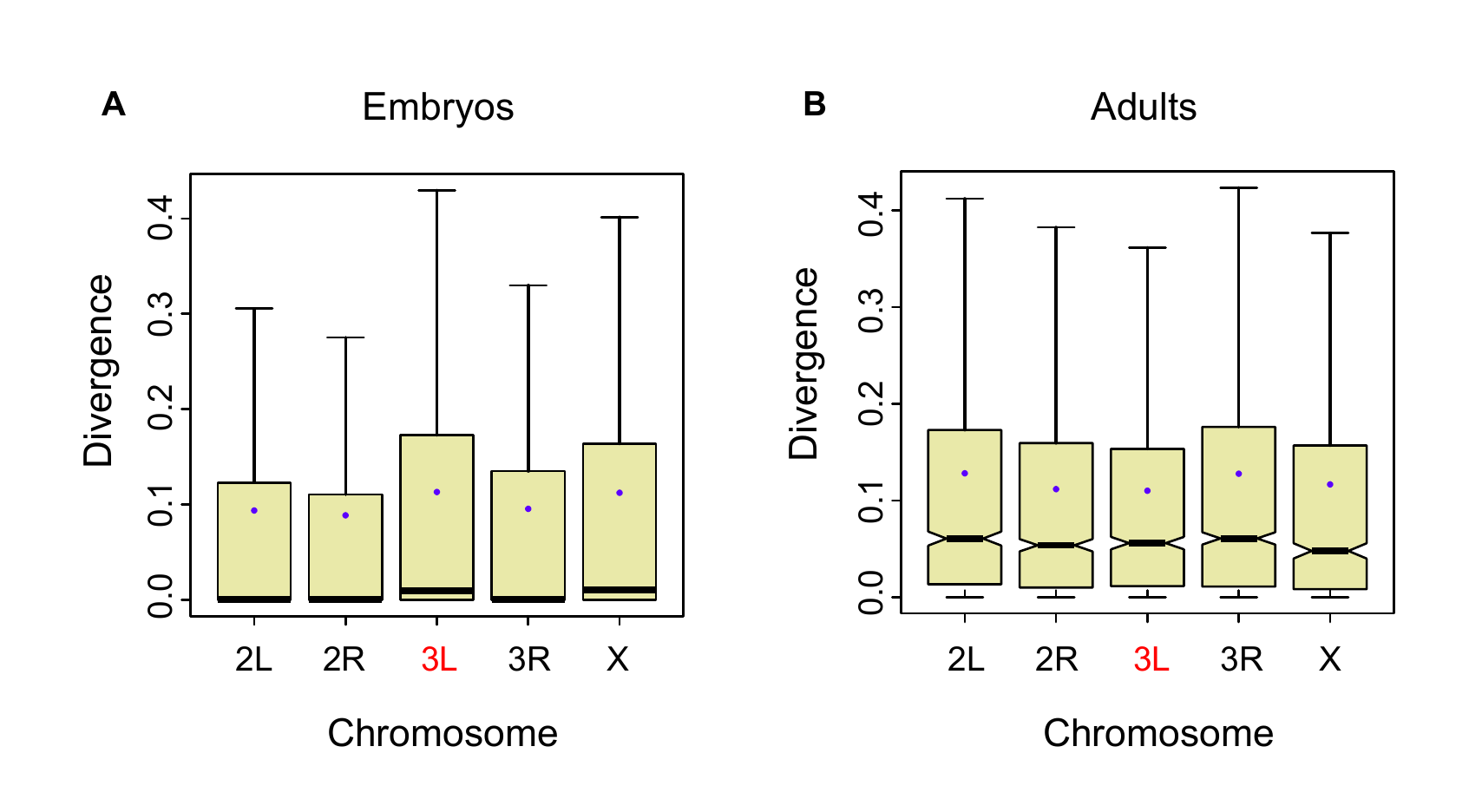}
\end{center}
\caption{
{\bf Expression divergence is higher for the ancestral branch of the neo-X (Muller element D).}\\
{\bf A}, Per-gene, per-chromosome distributions of the length of the ancestral branch leading to the obscura sub-group (\emph{D. persimilis} and \emph{D. pseudoobscura}; see Figure S\ref{phylogbranches}) in the embryonic data divided by the sum of all branch lengths (3L is the neo-X chromosome in the obscura sub-group). {\bf B}, Per-gene, per-chromosome distributions of the length of the branch leading to \emph{D. pseudoobscura} in the adult data divided by the sum of all branch lengths.
}
\label{ancneox}
\end{figure}

In the obscura sub-group, Muller's element D (3L in \emph{D. melanogaster}) has become X-linked and is referred to as a neo-X chromosome. If X-linkage were the cause of increased expression divergence, then we would expect to see accelerated evolution of gene expression on this chromosome relative to the remaining autosomes in this lineage \cite{thorntonetal2006}. As with the global X-effect, we see a small but significant increase in divergence on the ancestral branch of the obscura sub-group in the between-species embryonic dataset ($P = 0.0012$, Wilcoxon one-tailed test; Figure \ref{ancneox}A). While the ancestral branch shows an excess of divergence (Figure \ref{ancneox}A), the terminal branches do not (Figure S\ref{termneox}). In the adult dataset, there is only one species in the obscura sub-group, and the branch leading to this species does not show an excess of divergence (Figure \ref{ancneox}B). An excess of gene expression divergence on the ancestral branch leading to the obscura 
sub-group for 
the neo-X suggests that evolution of 
this chromosome was accelerated more 
after its formation. More generally, this finding lends independent support to the notion that the X evolves more rapidly than the autosomes.

\subsection*{Lower mutational heritability on the \emph{Drosophila} X}

The discovery that \emph{Drosophila} embryos have both an excess of divergence on the X chromosome between species (Figure \ref{embryodiv}A) and significantly lower levels of gene expression differentiation between strains of a single species (Figure \ref{embryodiv}B) is a pattern consistent with what we would expect to be driven by adaptive evolutionary processes. However, such a pattern could also be explained by random genetic drift since lower effective population sizes limit the amount of genetic variance a species can harbour \cite{vicoso&charlesworth2009} while simultaneously leading to the divergence of separate species through the accumulation of chance variations along separate lineages.

To determine whether it is likely that the X chromosome in \emph{Drosophila} could accumulate mutations at a faster rate than the autosomes simply by virtue of being in a hemizygous state in males, we analysed data from mutation accumulation lines of \emph{D. melanogaster} \cite{rifkinetal2005}. Twelve lines of \emph{D. melanogaster} were allowed to accumulate mutations over a period of 200 generations. Since selection is relaxed in these lines, mutations are free to accumulate in the population and if the X has a biased accumulation of mutations due to its hemizygosity, we would expect an excess of gene expression variation between mutation accumulation lines for genes expressed on the X than for those on the autosomes. Gene expression was measured genome-wide at the late larval and puparium formation stages of the life-cycle. After fitting linear models to the data, the authors extracted the variance attributable to mutations and scaled it by the residual variance to give a measure of mutational 
heritability \cite{rifkinetal2005}. Mutational heritability is a dimensionless quantity, defined as the variance in a trait which is attributable to new mutations in each generation divided by the variance attributable to environmental variance (in an initially homozygous population) \cite{houleetal1996}. Thus, this measure captures the rate of increase in the heritability of a trait due to mutations. The trait of interest for us is gene expression, and this metric allows us to infer how quickly different mutation accumulation lines diverge from one another in terms of the accumulation of mutations affecting gene expression at individual genes.

\begin{figure}[!ht]
\begin{center}
\includegraphics[width=6in]{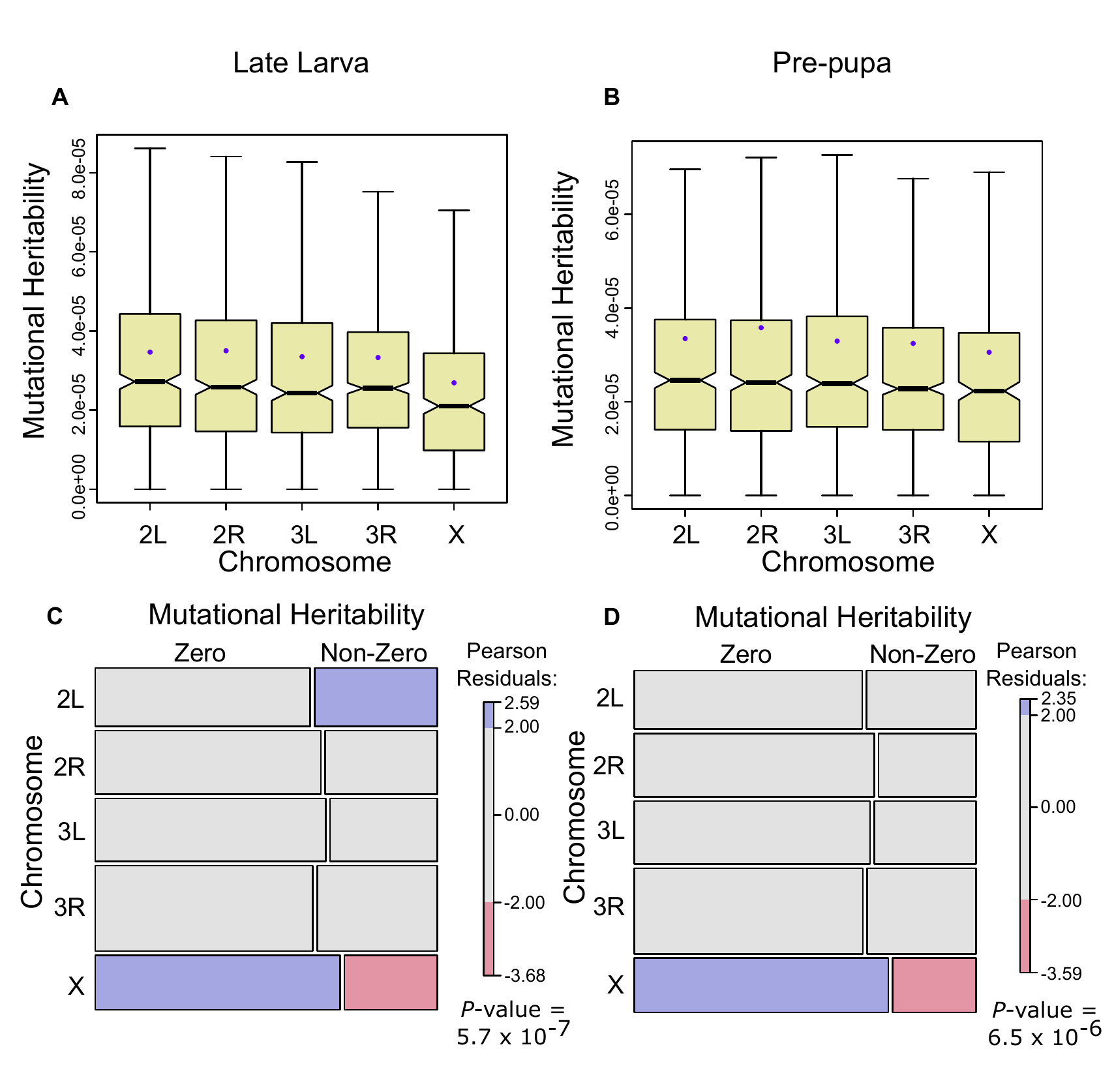}
\end{center}
\caption{
{\bf Gene expression mutational heritabilities are lower for the \emph{Drosophila} X chromosome.}\\
Gene expression mutational heritabilities, estimated from mutation accumulation lines of \emph{D. melanogaster} \cite{rifkinetal2005}, separated onto chromosomes. Genes with measurable mutational heritabilities are shown for the late larva ({\bf A}) and the pre-pupa ({\bf B}). In {\bf C} and {\bf D} genes are categorized as displaying zero or non-zero mutational heritabilities for late larva and pre-pupa respectively and depicted using mosaic plots where the area in the rectangles is proportional to the number in that category combination. Pearson residual shading is used to depict deviations from null expectations -- blue (excess) and red (paucity) colours indicate deviations from the expectation under the null hypothesis that the two variables, mutational heritability and chromosome, are independent \cite{zeileisetal2007}. $P$-values refer to the probability of independence (Chi-squared test).
}
\label{mutherit}
\end{figure}

The results show that, when we restrict the genes to those that have a measurable mutational heritability, the X has the lowest mutational heritability at both life-cycle stages ($P = 5.7 \times 10^{-8}$, Figure \ref{mutherit}A; $P = 0.0143$, Figure \ref{mutherit}B, Wilcoxon one-tailed tests). In addition, when we include those genes that do not have a measurable mutational heritability, we find that the X has both more genes with zero mutational heritability and less genes with a measurable mutational heritability than would be expected by chance (Figures \ref{mutherit}C,D). These results suggest that, for these developmental stages at least, the fixation by random drift of mutations influencing gene expression is not biased on the X chromosome and hence is unlikely to be driving higher gene expression divergence on this chromosome. We note, however, that the mutation accumulation lines do not necessarily perfectly capture the conditions experienced by wild populations of \emph{Drosophila} and so we believe 
it 
is important to conduct further studies designed to answer the question of whether the X fixes more mutations due to its hemizygosity.

\subsection*{A paucity of genes expressed in the cellular blastoderm on the \emph{Drosophila} X}

It was recently discovered that there is a paucity of adult tissue-specific gene expression on the \emph{Drosophila} X chromosome \cite{mikhaylova&nurminsky2011}. This result suggests that the distribution of genes across chromosomes may influence observed differences in chromosomal rates of evolution. To test whether X chromosome genes have unusual embryonic tissue expression patterns, we used a controlled vocabulary of embryonic expression terms based on \emph{in situ} expression data \cite{tomancaketal2007} to ask if there is under- or over-representation of expression terms for genes on the X relative to the whole genome. After correcting for multiple testing, just one term showed a significant departure from its null expectation; genes expressed in the cellular blastoderm are significantly under-represented on the \emph{Drosophila} X ($P_{adj} = 9.5 \times 10^{-5}$; Table S\ref{CVenrich}).

This result makes sense when we consider that dosage compensation of X-expressed zygotic genes in male embryos via the MSL (Male-specific lethal) complex is not fully active until after the blastoderm stage \cite{frankeetal1996, lottetal2011}. The lag in activation of MSL-mediated dosage compensation may disfavour cellular blastoderm expressed genes from residing on the X, especially as they would need to evolve an alternative dosage compensation mechanism \cite{lottetal2011}. More generally, the absence of strong tissue-expression biases on the X chromosome suggests that an unusual chromosomal distribution of tissue-specific embryonic genes is unlikely to be driving the higher gene expression divergence that we find on the X chromosome.

\subsection*{The multi-locus faster-X effect with epistasis and linkage}


\noindent Recent evidence suggests that epistatic interactions between genes constitutes a substantial fraction of the variation of quantitative traits in \emph{Drosophila} \cite{huangetal2012}. Therefore, to determine the relative benefits of chromosomal location and multi-locus co-evolution for beneficial alleles sweeping to fixation in a population, we analysed several diploid population genetic models of the faster-X effect. To compare evolution in equivalent genetic scenarios, we used the ratio of the selection gradient for X-linked versus autosomal cases (see Methods).

\begin{figure}[!ht]
\begin{center}
\includegraphics[width=6in]{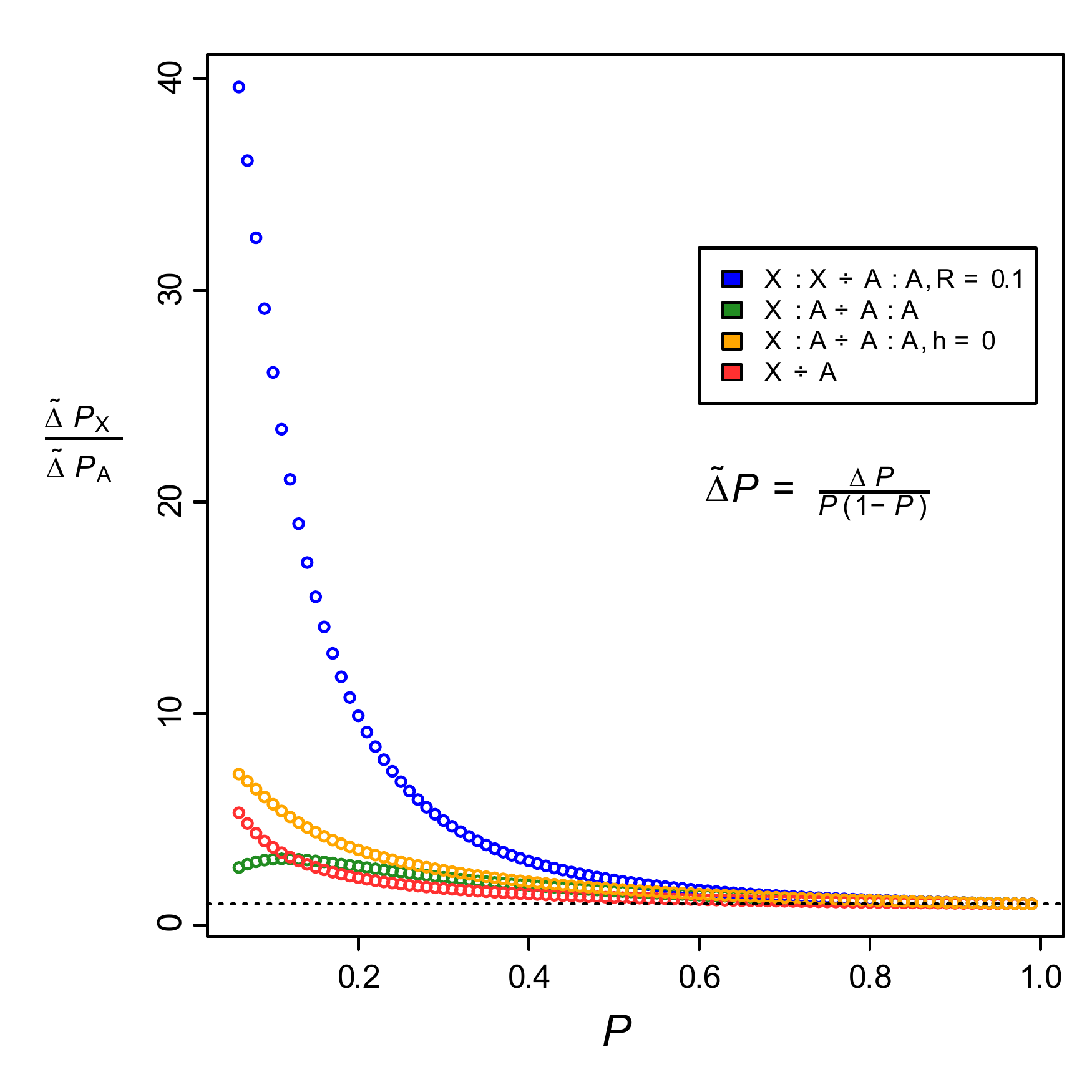}
\end{center}
\caption{
{\bf The faster-X effect is greatest when beneficially-interacting loci are linked on the same chromosome.}\\
The ratio of selection gradients for X-linked models versus their equivalent autosomal cases as a function of allele frequency. Blue points represent the case where both loci are linked on the same chromosome, orange and green points represent the case where the loci are on different chromosomes, and the red points are for the one-locus scenario. Unless otherwise stated in the legend, recombination rates, $R$, are equal to 0.5 (free recombination) and the dominance coefficient, $h$, is 0.01 ($h = 0$ is close to identical to $h = 0.01$ in the one-locus case and hence is not shown). The dashed line indicates a ratio of 1.
}
\label{twolocusmodels}
\end{figure}

The results show that, although a faster-X effect exists in all the cases studied, by far the greatest advantage of X-linkage occurs when both epistatically interacting loci are linked on the same chromosome (Figure \ref{twolocusmodels}, blue circles; Table S\ref{Xlinkedfitness}). When both loci are X-linked there will be no recombination in the heterogametic sex, and this will contribute to an increase in the rate of build-up of linkage disequilibrium between the loci. However, in species such as \emph{D. melanogaster} there is also no recombination occurring between pairs of homologous autosomes in males, and therefore such an effect would contribute to increased evolution on the autosomes. To quantify the magnitude of this effect, we compared the X-linked case to a scenario in which there is no recombination between autosomally linked loci in males. The results show that the effect of a lack of recombination in males cannot account for the advantage enjoyed by X-linked loci, which when compared against 
the autosomal case in 
which there is 
male recombination shows that the advantage in this case is weak and dependent upon high-levels of genetic variance (Figure S\ref{nomalerecomb}). Thus, the benefit of X-linkage in the multi-locus case accrues almost entirely from the increased efficacy of selection when acting on hemizygous males.

When positively-interacting alleles are located on separate chromosomes, it is extremely unlikely that they will sweep to fixation within a plausible time period because recombination will very effectively decay the linkage disequilibrium that is built up by selection in each generation \cite{connallon&clark2010}. When located on the same chromosome, interactions between loci could be considered to be either \emph{cis}-\emph{trans} or \emph{cis}-\emph{cis} interactions \cite{connallon&clark2010}, thereby broadening the scope of possible genetic scenarios that are consistent with faster-X evolution. It remains possible, however, that beneficial \emph{trans}-acting variants located on the autosomes, and interacting with fixed \emph{cis} alleles on the X, are responsible for the excess of divergence that we find on the X. However, there are no reasons to suppose that such interactions ought to be biased in the direction of \emph{trans}-autosomal to \emph{cis}-X, since, due to symmetry, the opposite scenario 
of \emph{trans}-X to \emph{cis}-autosomal appears to be just as likely. Indeed, in a recent study of gene expression in hybrids of \emph{D. yakuba} and \emph{D. santomea}, hybrid male mis-expression was found to be greater for autosomal genes, most likely as a result of faster evolution of X-linked \emph{trans}-acting factors \cite{llopart2012}. Thus, the available evidence suggests that if there is a bias in positive species-specific interactions between the X and the autosomes, it is in the direction of \emph{trans}-X to \emph{cis}-autosomal. Overall, both theory and data support the notion that during adaptive evolution, X-linked alleles have a capacity to sweep to fixation faster than their autosomal equivalents, and this effect is greatly enhanced when there are beneficial interactions between two or more loci.

\subsection*{Higher co-ordination of gene expression in embryos relative to adults}

In a recent study of gene expression evolution in mammals, evidence was reported for a faster-X effect \cite{brawandetal2011} (although a separate study found no evidence for a faster-X effect for gene expression in two species of mice \cite{goodetal2010}). The authors correlated gene expression across homologous chromosomes in species pairs and used one minus Spearman's correlation coefficient as a measure of divergence. The same approach has also been used recently to find an excess of divergence on the X in adult males and females of \emph{Drosophila} species \cite{meiseletal2012a}. Thus, we can ask why this correlation-based measure of divergence uncovers an X-effect in adults when our per-gene expression-level measure of divergence does not (at least not globally -- see Figure S\ref{melsimmalebias}).

\begin{figure}[!ht]
\begin{center}
\includegraphics[width=6in]{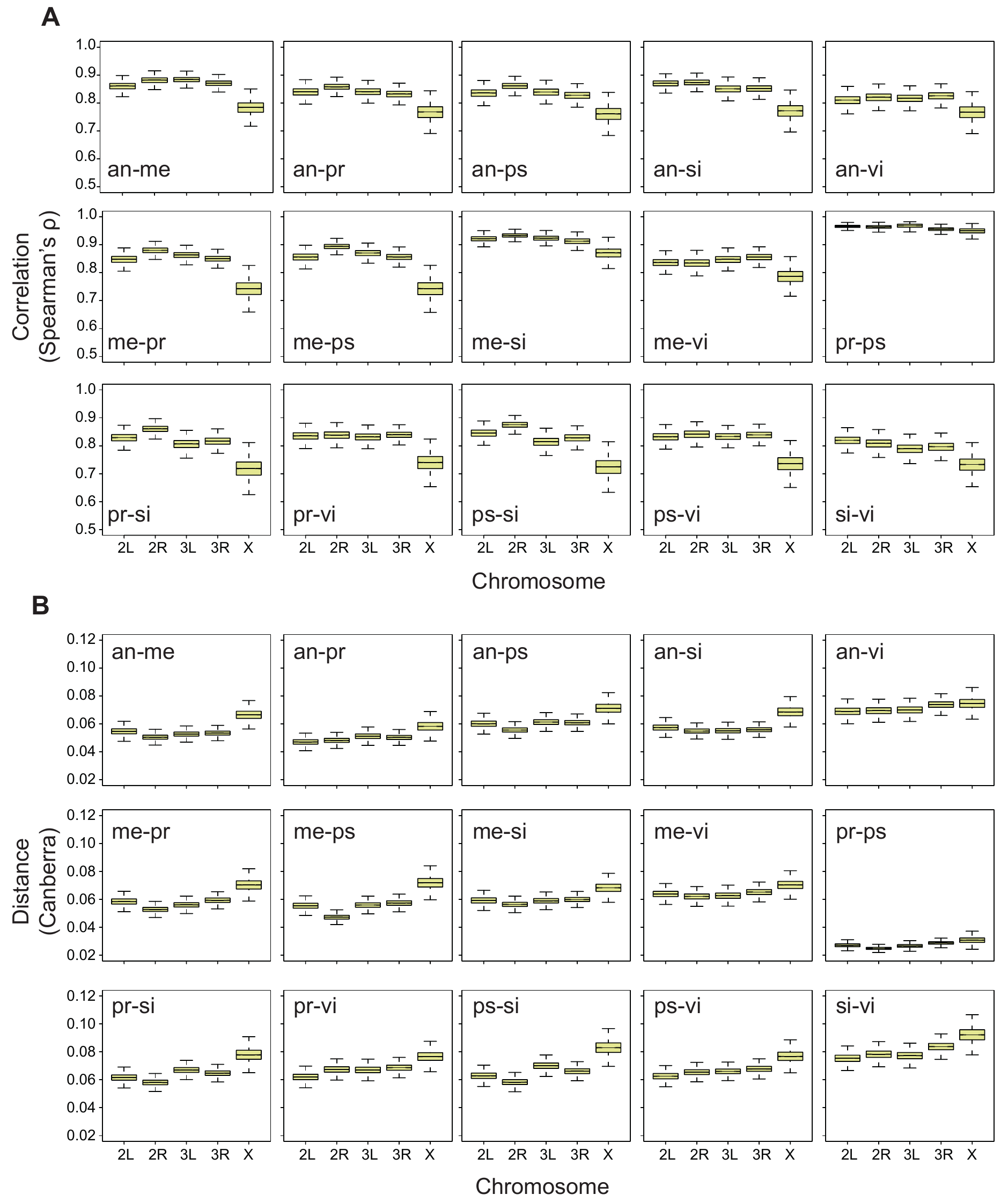}
\end{center}
\caption{
{\bf Divergence on the X in embryos is greater using both Spearman's $\rho$ and the Canberra distance.}\\
Bootstrapped distributions of {\bf A}, Spearman's $\rho$ (divergence is $1-\rho$) and {\bf B}, the mean Canberra distance across chromosomes in \emph{Drosophila} embryos for all pair-wise species comparisons.
}
\label{corrdistembryos}
\end{figure}

\begin{figure}[!ht]
\begin{center}
\includegraphics[width=6in]{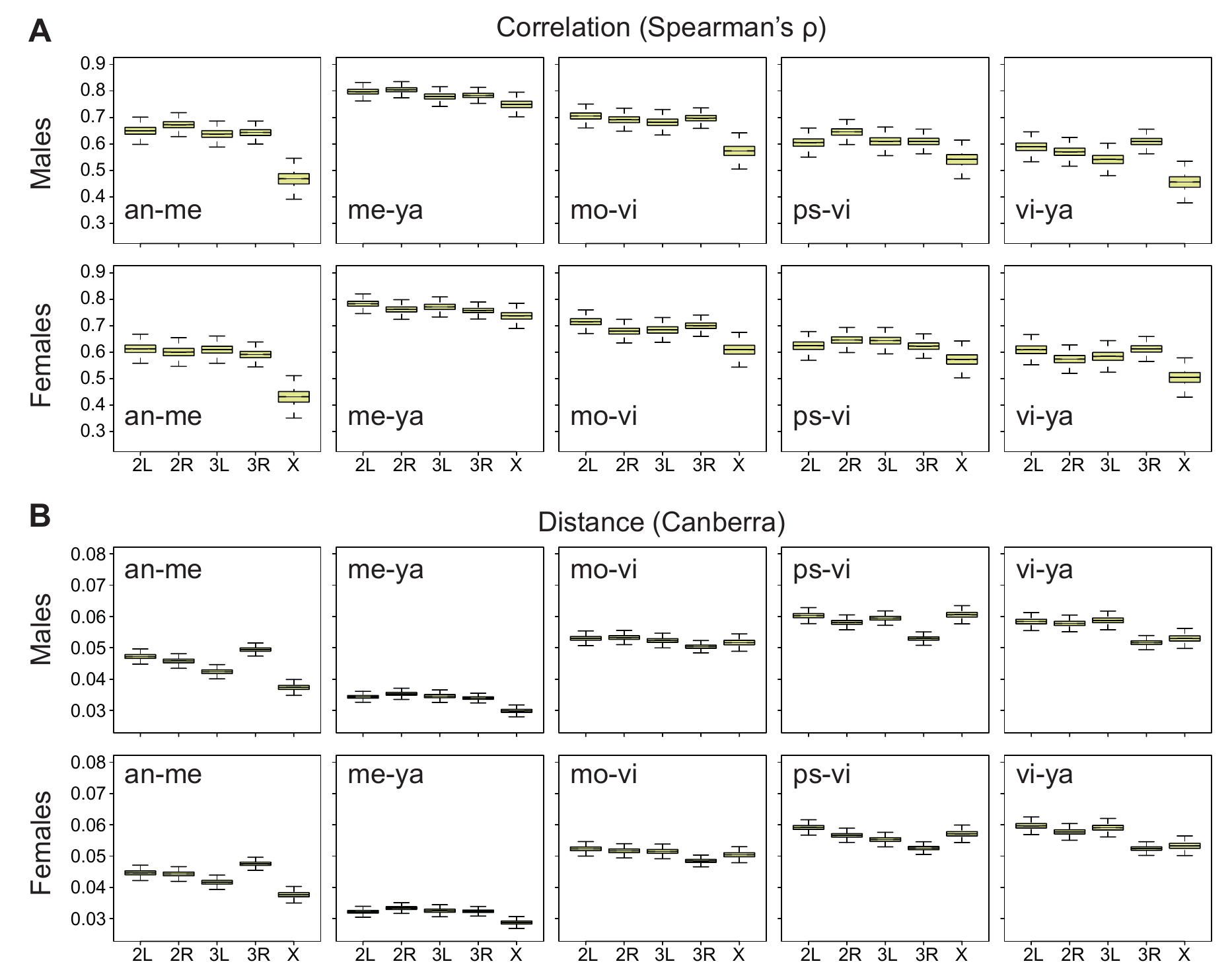}
\end{center}
\caption{
{\bf Divergence on the X in adults is greater using Spearman's $\rho$, but not the Canberra distance.}\\
Bootstrapped distributions of {\bf A}, Spearman's $\rho$ (divergence is $1-\rho$) and {\bf B}, the mean Canberra distance across chromosomes in \emph{Drosophila} males and females for a selection of pair-wise species comparisons (all pair-wise comparisons are shown in Figures S\ref{allcomps1},S\ref{allcomps2}).
}
\label{corrdistadults}
\end{figure}

To aid our search for an answer to this question, we first applied the correlation method to both embryos and adult males and females in the datasets that we have used. The results show that the X chromosome has a reduced cross-species correlation relative to the autosomes in the embryos (Figure \ref{corrdistembryos}A), just as it has in both adult males and females (Figure \ref{corrdistadults}A,B; all pair-wise comparisons are shown in Figure S\ref{allcomps1}) \cite{meiseletal2012a}. However, when we use an absolute distance metric to determine the per-chromosome differences between species, we find that, while the X consistently displays a greater distance between species in embryos (Figure \ref{corrdistembryos}B), in adults the X chromosome is largely equivalent to the autosomes (Figure \ref{corrdistadults}C,D; Figure S\ref{allcomps2}). Thus, the question arises as to why the X chromosome appears more divergent in terms of correlations but not in terms of distances?

The answer must be sought in the component of gene expression divergence that each measure is capturing. Spearman's rank correlation coefficient is a dimensionless number that in the context of gene expression in two species, determines the extent to which expression relationships between genes are retained across the two species, and the strength of the correlation is insensitive to absolute expression differences (Figure S\ref{corrschematic}). Thus, this measure of divergence captures how co-ordinated expression is across a specific set of genes in two different species. In contrast, absolute distances, and per-gene expression changes, measure to what extent individual genes differ in expression level in two species, and these metrics are insensitive to how co-ordinated expression is between different genes. This suggests, therefore, that gene expression on the X chromosome in adults is weakly co-ordinated relative to expression on the autosomes even though absolute expression differences are not 
significantly greater on the X (Figure S\ref{corrschematic}).

Furthermore, when we compare the chromosomal correlations in embryos and adults, we find that embryos have much higher correlations overall than the adults even when we reduce them both to a common set of genes and species (Figure S\ref{corrcomparison}). This suggests that gene expression is generally more highly co-ordinated in \emph{Drosophila} embryos relative to adults.

\clearpage

\section*{Discussion}

We have presented evidence that gene expression in \emph{Drosophila} embryos evolves faster on the X chromosome between species, but slower on the X chromosome within species (Figure \ref{embryodiv}). The salience of this result is substantially strengthened by the discovery that the Muller D element has a significantly longer ancestral branch leading to the obscura sub-group in the embryonic data (Figure \ref{ancneox}A). The Muller D element segregates as a neo-X chromosome in the obscura sub-group (\emph{D. persimilis} and \emph{D. pseudoobscura} in our data), and therefore provides a powerful, independent test for faster evolution of the X chromosome. In addition, we find that gene expression evolves faster on the X chromosome in embryos when we employ a more global measure of expression divergence (Figure \ref{corrdistembryos}A), a measure which we find can vary independently of per-gene expression level divergence (Figures \ref{corrdistadults},S\ref{corrschematic}). In what follows, we discuss 
different potential interpretations of these results.

\subsection*{Adaptive versus non-adaptive evolution}

The excess of gene expression divergence that we find in the embryonic data could be driven by a relaxation of selective constraints acting on X-linked gene expression. We would predict that relaxed selective constraints would lead to an elevation of within-species gene expression variation on the X, and, contrary to this prediction, we find that gene expression variation within inbred strains of \emph{D. melanogaster} is significantly lower on the X relative to the autosomes (Figure \ref{embryodiv}B,D) suggesting that X-linked gene expression is not evolving under a relaxation of selective constraint. In support of this finding, we find a corresponding reduction in gene expression variation on the X in both adult males and females (Figure \ref{adultdiv}B,D,F,H) \cite{meiseletal2012a}.

Nonetheless, it remains possible that elevated between-species variance coupled with diminished within-species variance is a consequence of random genetic drift, or demographic effects such as bottlenecks \cite{avery1984, singhetal2007}. If the hemizygosity of the X chromosome in males, and the resulting potentially diminished effective population size of the X, were resposible for the lower within-species variance in X-linked gene expression, then we would expect to find an excess of fixation of X-linked gene expression mutations in separate mutation accumulation lines. However, we find the opposite pattern, that mutation accumulation lines display less gene expression variation for X-linked genes (Figure \ref{mutherit}). Part of the reason for this could be due to the X chromosome presenting a smaller mutational target than the autosomes as a result of being in a hemizygous state in males, but this effect of hemizygosity will be present in wild populations of \emph{Drosophila} as much as in lab-reared 
lines. It is also possible that, while the experimenters made every effort to neutralise the effects of mutations, selective effects remained in the accumulated mutations and that purifying selection is stronger on the X relative to the autosomes.

Prior studies have found that the X chromosome in \emph{Drosophila} experiences more effective purifying selection against weakly deleterious and recessive mutations \cite{singhetal2005, singhetal2008, vicosoetal2008, takahashietal2009}, and in non-recombining chromosomal regions, the X has been shown to experience the smallest reduction in the efficacy of selection \cite{camposetal2012}. In addition, studies of nucleotide diversity on the X in both coding and non-coding regions in \emph{Drosophila} species suggest that adaptive processes best explain the observed variance on the X \cite{andolfatto2005, singhetal2007, andolfattoetal2011}, including recent data showing that there is an absence of X-autosomal differences for putatively neutral sites \cite{huetal2012}. Overall, our findings are consistent with there being an excess of adaptive evolution of X-linked gene expression, although this does not mean that drift or demographic effects are not involved in shaping gene expression 
evolution.

\subsection*{\emph{cis} versus \emph{trans} effects}

Gene expression is influenced by both \emph{cis}-acting regulatory sequences, and by \emph{trans}-acting factors, such as transcription factors. Thus, while we observe an excess of X-linked divergence of gene expression, this could be the result of either \emph{trans}-acting factors potentially located on other chromosomes, X-linked \emph{cis}-acting variants, or a combination of both. Several studies have found evidence for both \emph{cis} and \emph{trans} effects influencing gene expression differences both within and between \emph{Drosophila} species \cite{wittkoppetal2004, lemosetal2008, wittkoppetal2008, wangetal2008a, wittkoppetal2008b, mcmanusetal2010}. Thus far, however, the evidence suggests that there is an excess of \emph{cis}-acting variants influencing divergence between species \cite{wittkoppetal2004, lemosetal2008, wittkoppetal2008, grazeetal2009}, and that \emph{cis}-regulatory divergence increases with the divergence time between species \cite{lemosetal2008, mcmanusetal2010}. One study 
reported an excess of \emph{trans}-acting variation influencing gene expression in a comparison of \emph{D. melanogaster} and \emph{D. sechellia}, although as noted by the authors this could be related to the unusual demographic history and life-history evolution of \emph{D. sechellia} \cite{mcmanusetal2010}.

It's possible that the excess of X chromosome divergence that we see is the result of a bias in the direction of autosomal \emph{trans}-acting factors impacting the X chromosome more than the reverse situation of X-linked \emph{trans}-acting factors affecting the autosomes. Current evidence suggests, however, that the opposite is the case -- that there is a bias towards \emph{trans}-acting factors on the X impacting autosomal \emph{cis}-elements resulting in an excess of autosomal mis-expression in \emph{Drosophila} hybrids \cite{llopart2012}, including a study of mis-expression in hyrbid \emph{D. simulans} males carrying an X-linked allele introgressed from \emph{D. mauritiana} \cite{luetal2010}. Therefore, if there are species-specific interactions between the X and the autosomes, it seems unlikely that they would be biased in such a way as to account for our results.

Theoretical considerations also do not favour the notion that \emph{trans}-acting factors could be driving the majority of the divergence that we find, assuming that a substantial fraction of this divergence is adaptive. Mutations in \emph{trans}-acting factors are more likely to be pleiotropic, and so should have less scope to influence adaptive evolution than the more modular effects of mutations in \emph{cis}-regulatory regions \cite{stern2000, prudhommeetal2007, wray2007, rebeizetal2009, connallon&clark2010}. Furthermore, population genetic models of the faster-X effect show that if there are two or more interacting loci with beneficial interactions between them, then X-linked loci enjoy a far greater benefit than autosomal loci (Figure \ref{twolocusmodels}). Whether adaptive changes occur in \emph{cis} or in \emph{trans} also has important consequences for the scope of mutations to have recessive or partially recessive effects on fitness, which in turn is of central importance for the 
faster-X phenomenon \cite{charlesworthetal1987}. We address these issues towards the end of the Discussion.

\subsection*{Embryos versus adults}

In the embryonic between-species data, we found evidence for faster evolution of gene expression on the X chromosome using two different measures of divergence (Figures \ref{embryodiv}A,\ref{corrdistembryos}A). The first measure captures the change in expression levels on a per-gene basis (Figure \ref{embryodiv}A), and the second captures the extent to which gene expression relationships between genes have changed in pairs of species, and hence how co-ordinated expression is across a subset of genes (Figures \ref{corrdistembryos}A,S\ref{corrschematic}). In contrast, in the adults, we see evidence for higher divergence on the X chromosome using only the second measure of divergence (Figure \ref{corrdistadults}A) and not the first (Figure \ref{adultdiv}A). This suggests that, while the X displays lower levels of co-ordinated expression in pairs of species in the adult, it does not exhibit significant differences in expression level on a per-gene basis. Then we must ask, why does the embryo diverge more on the 
X in terms of per-gene expression levels than the adults?

Embryogenesis is a highly dynamic process, driven by a cascade of gene expression unraveling through a highly co-ordinated developmental network leading to large batteries of genes being switched on and off at precise moments during development \cite{peter&davidson2011}. In contrast, in a fully developed adult, cells are largely fully differentiated, and gene expression is to a much lesser degree responding to a pre-determined developmental program, and is freer to respond to changes in the environment. Thus, it makes sense that we find gene expression to be overall much more highly co-ordinated in the embryo relative to the adults (Figure S\ref{corrcomparison}). But it is precisely because of the broad dynamic range of embryonic gene expression, with a large fraction of the zygotic genome being activated in a series of waves as embryogenesis proceeds (Figure S\ref{karolinafig}), that even subtle shifts in timing could potentially produce large differences in expression levels. In a whole adult fly, however, 
genes are likely expressed in subsets of tissues and organs such that we will not find extremely low or high expression levels for most genes when we extract RNA from all of the tissues simultaneously, thereby diminishing the dynamic range of the data. Therefore, our results highlight the need to perform more precise organ-by-organ comparisons of gene expression in future between-species studies of adult flies. In addition, our analysis draws attention to the different components of divergence that are captured by different measures of gene expression divergence.

\subsection*{The faster-X hypothesis}

Taking the above considerations and all of our results into account, we believe that the X effect we find in the embryos is best explained within the framework of the faster-X hypothesis. This does not mean that all of the divergence we see is driven by adaptive substitutions in \emph{cis}-regulatory regions on the X chromosome, but rather that the excess of X chromosomal divergence that we find together with the reduction of expression variation in inbred strains of \emph{D. melanogaster} is most consistent within an adaptive evolutionary scenario. In support of this interpretation, researchers found an excess of adaptive substitutions on the X chromosome in a long-term evolution experiment involving lines of \emph{D. melanogaster} selected for increased rates of egg-to-adult development \cite{burkeetal2010}. An interesting theoretical corollary of the fast-X interpretation is that it suggests that adaptive substitutions are more likely to occur via new mutations than from standing genetic 
variation \cite{orr&betancourt2001}.

If we adopt a faster-X interpretation of the data, then we must provide some explanation as to why beneficial \emph{cis}-regulatory mutations have recessive or partially recessive effects on fitness, in keeping with the original model \cite{haldane1924}. Current evidence in adult \emph{Drosophila} species suggest the opposite, that \emph{cis}-acting variants have largely additive effects relative to \emph{trans}-acting factors, which show more deviations from additivity towards dominance and recessiveness \cite{lemosetal2008, mcmanusetal2010}. However, these experiments determine the additivity of the phenotype of a \emph{cis} variant (where the phenotype is its gene expression level), and not necessarily its effect on fitness. Theory suggests that mutations could have fitness consequences that are non-linear even if they have additive phenotypic effects \cite{sellisetal2011}. Therefore, it is possible that phenotypic measures of \emph{cis}-acting elements fail to capture their effects on fitness.

To understand the fitness effect of a mutation in an organismal context, we must focus on the biology of the organism, and not just on its genetics. One potential route towards non-additive intra-locus effects on fitness is canalisation. The canalisation of embryonic development, such that it is resistant to environmental or genetic perturbations, has long been recognized as a crucial element contributing to the evolution of robustness in developmental systems \cite{waddington1942}. The evolution of dominance is a means by which the components of a network could become canalised \cite{fisher1928, bourguet1999, proulx&phillips2005, billiard&castric2011}. While selection acting on modifiers of dominance will typically be weak (of the order of the mutation rate), it can be substantially stronger in non-equilibrium populations where genetic variation is maintained at high levels by processes such as migration and hybridisation \cite{bourguet1999, billiard&castric2011}. The notion that the evolution of robustness 
(i.e., an attempt to prevent change of the phenotype) could lead to faster evolution of the X may seem counter-intuitive. However, the relationship between robustness and evolvability is well established, and suggests that the evolution of phenotypic robustness can often facilitate adaptive evolution \cite{wagner2008, masel&trotter2010, draghietal2010}. We present this scenario partly to illustrate that the biological details of an individual species, such as species range and migratory pressures, might play a significant role in determining how its chromosomes evolve.

\subsection*{Outlook}

We report evidence that gene expression evolves faster on the X chromosome in \emph{Drosophila} embryos. While our results are consistent with adaptive evolutionary processes, more work is required to unravel the details underpinning this excess of divergence at the genetic, phenotypic, and fitness levels. We contend that variations in biological and life-history details, such as differences in dosage compensation menchanisms, can strongly impact how the chromosomes of different species evolve. We therefore stress the importance of appreciating biological context when attempting to understand chromosomal evolution. Deciphering the relationship between species-specific biology and chromosomal patterns of evolution promises to provide fertile ground for future research.

\clearpage

\section*{Methods}

\subsection*{Embryo collections and RNA isolation and labeling}

We used inbred strains of \emph{D. melanogaster}, originally collected from farmer's markets in North Carolina and provided as a resource by the Drosophila Genetic Reference Panel (DGRP; http://dgrp.gnets.ncsu.edu/) \cite{mackayetal2012}. Seventeen strains were selected for the collection of 0-2 hour old embryos.

Populations of healthy adults from 3-7 days of age, were reared at 25$^{\circ}$C and used for embryo collections. To synchronize the age of the embryos in each sample, we pre-laid the flies three times for 1 hour with a fresh apple juice plate with yeast paste before every collection. Another fresh plate with yeast was used to collect the embryos. After collection, embryos were rinsed with distilled water and then dechorionated in 100\% bleach for 2 minutes before being washed in desalinated water. The embryos were then transferred into a 1.5-ml tube and snap-frozen in liquid nitrogen and stored at $-80^{\circ}$C. Three biological replicates were collected for each strain.

To isolate RNA, embryos were thawed on ice and homogenized with a pellet pestle and a pellet pestle cordless motor (Kontes). RNA was isolated with the RNeasy Mini kit (Qiagen) and eluted with 30 ml of distilled water. The RNA concentration was measured with the NanoDrop spectrophotometer and RNA quality was assessed with Bioanalyser using the Agilent RNA 6000 Nano kit.

To prepare samples for hybridization to the chip, we followed the Agilent One-Colour Microarray-Based Gene Expression Analysis protocol version 6.5 (Low Input Quick Amp Labeling). The starting amount of RNA was normalized to 100 ng for all samples.

\subsection*{Gene expression data sets}

Embryonic expression in \emph{Drosophila} was taken from a species-specific microarray data set, in which eight time-points were sampled for the duration of embryogenesis of \emph{D. melanogaster}, \emph{D. simulans}, \emph{D. ananassae}, \emph{D. pseudoobscura}, \emph{D. persimilis}, and \emph{D. virilis} \cite{kalinkaetal2010}. Adult \emph{Drosophila} expression was collected from a microarray experiment that measured the gene expression of whole flies sorted into males and females and taken from \emph{D. melanogaster}, \emph{D. ananassae}, \emph{D. mojavensis}, \emph{D. pseudoobscura}, \emph{D. simulans}, \emph{D. virilis}, and \emph{D. yakuba} \cite{zhangetal2007a}. Gene expression mutation accumulation data was taken from a microarray study of mutation accumulation lines of \emph{D. melanogaster} \cite{rifkinetal2005}. Adult \emph{D. melanogaster} strain data was taken from a whole-genome microarray study of gene expression in whole adult flies from 40 inbred strains separated into males and 
females \cite{ayrolesetal2009}.


\subsection*{Measures of chromosomal expression divergence and differentiation}

To quantify gene expression divergence in a chromosomal context, we fitted the following linear model \cite{kerretal2000} to $log_{2}$ gene expression measures, $y_{ijkl}$,

\[ y_{ijkl} = \mu + S_{j} + C_{k} + GC_{i(k)} + SC_{jk} + GCS_{i(k)j} + e_{ijkl} \]

\noindent where $S_{j}$ is the effect of the $j{\text{'th}}$ species, $C_{k}$ is the effect of the $k{\text{'th}}$ chromosome, and $GC_{i(k)}$ is the effect of the $i{\text{'th}}$ gene nested in the $k{\text{'th}}$ chromosome. The interaction between the $j{\text{'th}}$ species and the $i{\text{'th}}$ gene nested in the $k{\text{'th}}$ chromosome, $GCS_{i(k)j}$, provides information about species-specific chromosomal expression of a gene and is given by

\[ GCS_{i(k)j} = \bar{y}_{ijk.} - \bar{y}_{i.k.} - \bar{y}_{.jk.} + \bar{y}_{..k.}  \]

\noindent where values are averaged over missing subscripts indicated by dots. Thus, the effect of the $i{\text{'th}}$ gene in the $j{\text{'th}}$ species is the excess that cannot be explained by the expression of the $i{\text{'th}}$ gene across species, the expression of the $k{\text{'th}}$ chromosome in the $j{\text{'th}}$ species, and the overall expression on the $k{\text{'th}}$ chromosome. When there are multiple expression measures over a time-course, our measure of divergence is designed to detect translations up or down in expression level across the time course as a whole (see Figure S\ref{vinc}).

Differentiation of gene expression between inbred strains was determined using the R package `limma' \cite{smyth2005}. Limma fits linear regression models to each gene separately. The differentiation of each gene was then scored as the mean log fold change of the gene across all pairwise strain comparisons.

\subsection*{Branch length analysis}

Absolute pairwise species contrasts of the $GCS_{i(k)j}$ values were transformed into branch lengths using the Fitch–Margoliash least squares method (implemented in the PHYLIP program ‘fitch’) \cite{felsenstein1989}. Negative branch lengths were set to zero, and for all genes the topology of the known phylogeny was used \cite{markow&ogrady2007}. Per-gene expression divergence was then expressed as the sum of all of the branch lengths in each gene tree separately.

To test for acceleration on one lineage, for each gene we expressed the branch length of the focal lineage as a proportion of the total of all branch lengths. In the embryonic dataset we chose the ancestral branch leading to the common ancestor of \emph{D. pseudoobscura} and \emph{D. persimilis} but not including the terminal branches (Figure \ref{ancneox}A). For the adult dataset, which does not have data for \emph{D. persimilis}, we used the terminal branch leading to \emph{D. pseudoobscura} (Figure \ref{ancneox}B).

\subsection*{Resampling branch lengths}

Mean summed branch lengths were bootstrapped by resampling the genes on each chromosome 10,000 times with replacement and in each bootstrap replicate calculating the mean summed branch lengths for the genes on each chromosome (Figure 1C,D). Individual branches in the embryonic and adult datasets were tested for an excess of divergence on the X chromosome using the number of bootstrap replicates in which mean autosomal branch lengths were greater than the mean on the X chromosome (Figure S1). All resampling was carried out using the R statistical  programming environment \cite{R2012}.

In both of the \emph{Drosophila} between-species data sets, the smallest sample of genes was on the X chromosome (Table S\ref{chromsummary}). To determine whether the differences between the X and the autosomes could have been caused by a sampling bias on the X, we resampled the number of genes present on the X from the autosomes 10,000 times without replacement and each time recalculated the mean divergence. The distributions of these resampled means are shown in Figure S\ref{resampnumbergenes}.

\subsection*{Accounting for sex-biased expression in adults}

Expression of genes in the adults can be biased towards one of the sexes \cite{zhangetal2007a}, and it's possible that sex-biased genes might exhibit stronger differences in divergence across the chromosomes. We focused on male and female-biased genes identified in \cite{zhangetal2007a} in each of the species. Genes that show a male-bias in at least one species show a significant excess of divergence in both males and females ($P_{male} = 1.57 \times 10^{-11}$; $P_{female} = 6.33 \times 10^{-6}$; Figures S\ref{biasmales},S\ref{biasfemales}) \cite{ranzetal2003, meiklejohnetal2003}, and conversely female-biased genes are significantly more conserved in both males and females ($P_{male} = 6.27 \times 10^{-7}$; $P_{female} = 3.42 \times 10^{-10}$; Figures S\ref{biasmales},S\ref{biasfemales}). When we look at divergence across chromosomes, however, we find that sex-biased genes are not significantly more divergent on the X in either sex (Figure S\ref{allsexbias}). Interestingly, when we restrict male-biased genes 
to 
those in \emph{D. melanogaster} and \emph{D. simulans} we do find a weak but significant excess of divergence on the X ($P = 0.0022$; Figure S\ref{melsimmalebias}), which is absent for the same genes expressed in females ($P = 0.117$; Figure S\ref{melsimmalebias}). The biological function of these genes is enriched for carbohydrate metabolism ($P_{adj} = 2.7 \times 10^{-6}$) and alcohol metabolism ($P_{adj} = 1.1 \times 10^{-6}$), which might suggest that these are genes that have evolved rapidly and relatively recently, thus preserving the signal of an excess of divergence on the X. Indeed, we find that these genes are significantly more divergent than average ($P = 1.0 \times 10^{-4}$; Figure S\ref{melsimdiv}).

\subsection*{The X-effect during embryogenesis}

In the between-species embryonic data, our measure of divergence is designed to detect translations in expression up or down in different species across the embryonic time course as a whole (Figure S\ref{vinc}). However, it remains possible that much of the difference that we detect between the X and the autosomes is driven by a subset of the time points. To test this, we extracted divergence measures from each time point separately. We then bootstrap resampled divergence measures for the X chromosome and the autosomes and in each bootstrap replicate calculated the ratio of mean X to mean autosomal divergence. The results show that at every time point the X chromosome displays an excess of divergence relative to the autosomes (X/A ratio $> 1$; Figure \ref{timediv}). Furthermore, all of the resampled time point distributions heavily overlap with one another indicating that higher expression divergence on the X is not driven solely by one or a subset of time points.

\subsection*{Resampling according to gene expression level}

Differences in gene expression divergence across chromosomes could be influenced by consistent differences in expression levels across chromosomes. In the between-species embryo data, the X chromosome has the weakest mean expression level (Figure S4), whereas in the adults, the X chromosome has the highest mean expression level (Figure S\ref{explevel}). Higher expression in the adults could be a reflection of a paucity of adult tissue-specific expression on the X chromosome \cite{mikhaylova&nurminsky2011}. To elucidate the relationship between expression level and divergence in these data sets, we ranked genes by their expression level (lowest to highest), binned them into groups of 50 genes, and measured the deviation of each group's mean divergence from the global mean divergence.

The results show that for the embryos, the relationship is non-linear, with groups of the weakest expressed genes diverging less than the global average (Figure S\ref{explevelbins}). Thus, although an increasing expression level does predict less divergence, divergence cannot be attributed simply to stochastic fluctuations of the weakest expressed genes. In the adults, the relationship is more linear, with the weakest expressed genes showing the highest divergence (Figure S\ref{explevelbins}). Thus, higher expression on the X in adults may at least partly explain the lower levels of divergence relative to the embryos.

To clarify the relationship between expression level and chromosomal divergence, we bootstrap sampled genes from each chromosome while weighting their probability of being sampled according to their expression level. To sample genes according to expression level we weighted the probability of being sampled according to the cumulative distribution function of a normal distribution with a specified mean expression level and standard deviation. We defined the standard deviation as the standard deviation of the whole expression level distribution divided by the number of mean expression levels that were being sampled. Genes were then sampled with replacement 10,000 times for each mean expression level for each chromosome in both the embryonic and adult datasets. Fewer mean expression levels were taken for the adult data due to its lower expression level variance.

The results show that, in the embryo, divergence on the X is greater than the autosomes for intermediate gene expression levels, but not when expression is high or low (Figure S\ref{explevelboot}A). In contrast to this result, in the adult data the X shows higher expression divergence when gene expression is low or high (Figure S\ref{explevelboot}B). Thus, the higher expression divergence of the X in the embryos is not driven by expression levels at the extremes of the distribution.

\subsection*{Testing for sex ratio effects}

While divergence on the X is not driven by particular periods during development, it is possible that there is a bias in the direction of expression differences between species. For example, if there was a persistent skew towards a male-biased sex ratio in one species relative to another and if dosage compensation in males was incomplete, then we would expect X-linked genes to show a skew towards lower expression in this species as the male-biased population would amplify the incomplete dosage compensation. To test this, we contrasted normalized expression in pairs of species and scored genes as up or down in one species relative to the other. We then asked if the X-chromosome showed significant skews in the number of genes scored as up or down in these species pairs relative to the autosomes. The results show this is not the case for any species pair (Figure S\ref{expskewall}), and this is shown in more detail for the \emph{D. persimilis} versus \emph{D. pseudoobscura} contrast (Figure S\ref{expskewprps}), 
which is pertinent given that there is an excess of X chromosome divergence in this species comparison ($P = 0.0042$; Figures S\ref{phylogbranches}, S\ref{prpschrom}). Therefore, there do not appear to be systematic biases in the direction of expression differences between species and hence this is unlikely to be a factor driving the higher divergence of the X chromosome.

\subsection*{Uncovering the relationship between expression evolution and excess chromosomal divergence}

The discovery that different groups of genes exhibit differences in their chromosomal divergence in adults suggested that there may be a relationship between excess chromosomal divergence and the rate of gene expression evolution. To test this, we scored the ratio of mean divergence of genes belonging to each percentile of each chromosome's divergence distribution relative to the same percentile of the other chromosomes. The results show that in both the embryos and the adult males, excess divergence on the X chromosome increases as the genes become more divergent while such a pattern is not seen consistently on any of the other chromosomes (Figure S\ref{percentilediv}). In addition we find that while in the embryos most of the genes on the X exhibit an excess of divergence relative to the autosomes, in adult males these genes are restricted to a subset of those on the X. The top enriched biological functions for these genes are primary sex determination, secondary metabolic process, and adult behavior 
(Table 
S\ref{XAadultmales}), all likely to be fast-evolving traits and processes. It is interesting to note that in both cases, the fastest evolving genes do not display an excess of divergence on the X. Overall, however, we find that fast-evolving genes tend to diverge more on the X in both embryos and adult males.

\subsection*{Correcting for non-expressed/weakly expressed genes}

In the embryonic time course, an initially bimodal gene expression distribution gradually becomes unimodal as the zygotic genome is switched on during embryogenesis (Figure S\ref{karolinafig}). If the X chromosome happened to be over-represented for genes in the lower mode of this bimodal distribution, then it is possible that much of the excess divergence we find on the X could be driven by spurious divergence between non-expressed genes. Therefore, to test for this we used the expectation-maximisation algorithm to determine a cutoff expression level (based on time point 1) below which a gene could be considered as non-expressed at any time point ($log_{2}$ expression of 8.513).

We then defined three gene sets based on increasingly more stringent criteria for being thrown out from the analysis. The first set (termed ``Two'') consists of genes that are not expressed in at least two species in at least one time point (1502 genes). The second set (``Six'') consists of genes that are not expressed in at least six species in at least one time point (849 genes), and the final set (``Six-Eight'') consists of genes that are not expressed in at least six species at every time point (536 genes). Expression distributions for these gene sets shows that they increasingly capture more weakly expressed genes as the criteria for exclusion becomes more stringent (Figure S\ref{dropgenesExp}). When we compare gene expression divergence for the data set after removing these gene sets, we find that the excess of divergence on the X is not affected (Figure S\ref{dropgenesXeffect}) showing that this effect is not driven by spurious divergence between non-expressed or weakly expressed genes.

\subsection*{Mutation accumulation analysis}

To determine whether the lower effective population size of the X chromosome might increase the chance that it fixes weakly deleterious mutations, we used gene expression mutation accumulation data to assess potential chromosomal biases in the accumulation of gene expression differences. We used jack-knifed mutational variance estimates scaled by residual variances to provide estimates of the mutational heritability of gene expression changes between lines \cite{rifkinetal2005}. As a large fraction of the genes at both the late larval and puparium formation stages did not exhibit measurable mutational heritabilities, we separated the genes with measurable estimates (Figure \ref{mutherit}A,B). In addition, we categorized genes as having measurable mutational heritabilities from those without and compared the ratios of these two categories across chromosomes using contingency tables. The results were visualized using residual-based shading with the R package `vcd' \cite{zeileisetal2007} (Figure \ref{mutherit}
C,D).

\subsection*{Embryonic tissue expression enrichment analysis}

A hierarchically-arranged controlled vocabulary (CV) of embryonic tissue expression terms based on an in situ expression data set \cite{tomancaketal2007} was used for assessing under- or over-representation of expression patterns for genes on the Drosophila X chromosome. Enrichment of terms was carried out in the R package 'topGO' \cite{alexaetal2006} using custom-written code. The parent-child algorithm was employed to control for the inheritance bias between parent and child terms in the CV hierarchy \cite{grossmannetal2007} (Table S\ref{CVenrich}). The resulting P-values were adjusted using the Benjamini-Hochberg correction in the R package `multtest' \cite{pollardetal2003}.





\subsection*{Multi-locus population genetic models of the faster-X effect}

In all of our models, we assume that selection coefficients are equal in the two sexes, which corresponds to the assumption of complete dosage compensation in \cite{charlesworthetal1987}, and, in the case of the two-locus models, that there is a beneficial epistatic interaction between one of the alleles at each locus. In addition, we assume that viability selection operates on the diploid zygotes, that mating is random, and that double heterozygotes experience half of the fitness benefit of single heterozygotes (Tables S\ref{Xlinkedfitness},\ref{fitnesses}).

We derived genotype frequency recurrence equations to describe the evolutionary dynamics in our models and then solved the equations numerically. To compare evolution in the equivalent X versus autosomal scenarios, we extracted the change in allele frequency of the \emph{cis}-acting beneficial allele between generations, $\Delta P$. We used the ratio of selection gradients in the equivalent models as a comparative statistic. The selection gradient describes the change in relative fitness as the allele frequency of the beneficial variant changes. Using the Robertson-Price identity \cite{robertson1966,price1970} to describe the change in allele frequency, $P$, in terms of relative fitness, $\tilde{w}$,

\[\Delta P = Cov(\tilde{w},P),\]

\noindent and replacing with the regression coefficient, $Cov(\tilde{w},P) = \beta_{\tilde{w},P}\sigma^{2}_{P}$,

\[\Delta P = \beta_{\tilde{w},P}\sigma^2_P = \frac{d\tilde{w}}{dP}P(1-P),\]

\noindent then the selection gradient, $\frac{d\tilde{w}}{dP}$, is equal to the change in allele frequency divided by its variance, $\tilde{\Delta} P = \frac{\Delta P}{P(1-P)}$. We plot the ratio of selection gradients in the X versus autosomal cases (Figures \ref{twolocusmodels},S\ref{nomalerecomb}).

\subsection*{Correlation-based measures of divergence}

Spearman's $\rho$ was measured for pairs of chromosomes in pairs of species for both the embryonic and adult data. Correlation coefficients were bootstrapped by resampling the genes 10,000 times on each chromosome separately (Figures \ref{corrdistembryos}A,\ref{corrdistadults}A). For the embryos, we used expression averaged across time, and found that correlations derived from this measure agreed very well with correlations derived from expression within single time points in terms of a reduction of correlation on the X chromosome. In addition, we took the mean Canberra distance across chromosomes for pairs of species, averaging it by dividing by the number of genes on each chromosome separately (Figures \ref{corrdistembryos}B,\ref{corrdistadults}B).

The correlation approach captures the extent to which chromosomal subsets of genes tend to conserve their expression relationships in pairs of species. However, this approach fails to capture the level of conservation of gene expression in a chromosomal subset relative to a separate chromosomal subset across pairs of species. For example, we might wish to ask whether the expression relationship of genes on the X chromosome relative to the autosomal arm 2L shares a conserved pattern in a pair of species. To answer questions of this nature, we introduce a variant of Spearman's correlation coefficient which allows us to rank genes in a chromosomal subset relative to genes in a separate chromosomal subset for pairs of species. For the correlation of subset $A$ relative to subset $B$ in two species we have

\[\tilde{\rho}_{A:B} = \frac{\sum\nolimits_{i}^{n}(x_{i_{A:B}}-\bar{x}_{A})(y_{i_{A:B}}-\bar{y}_{A})}{\sqrt{\strut\sum\nolimits_{i}^{n}(x_{i_{A:B}}-\bar{x}_{A})^{2}\sum\nolimits_{i}^{n}(y_{i_{A:B}}-\bar{y}_{A})^{2}}},\]

\noindent where $x_{i_{A:B}}$ and $y_{i_{A:B}}$ are the ranks of the $i$'th gene's expression level (from the $n$ genes that belong to subset $A$) relative to gene expression in subset $B$ for species $x$ and species $y$ respectively. Thus, this relative measure captures whether expression in subset $A$ is co-ordinated relative to subset $B$ in pairs of species. 

As it is established that correlation coefficients within subsets can vary, sometimes dramatically, from correlation at the level of aggregates (known as the Yule-Simpson effect \cite{yule1903, simpson1951, blyth1972, wagner1982, wilcox2001}), we believe that it is necessary to account for possible discrepancies when measuring correlation within subsets drawn from a larger population (Figure S\ref{simpsonsparadox}). When we measure relativised correlations for chromosomal subsets in the embryonic and adult data, we find that the X chromosome displays a significantly higher correlation when correlating against an autosomal background in adult females (Figure S\ref{relativecorrs}). This suggests that in adult females the X is generally more co-ordinated in relation to the autosomes than in relation to itself ($P = 0.015$; Wilcoxon two-tailed test), a pattern that could be driven, in part, by gene interactions between the X and the autosomes. More generally, this result highlights the importance of considering 
cross-chromosome relationships when 
using correlation-based measures of divergence.

\clearpage

\section*{Acknowledgments}
We gratefully acknowledge Julia Jarrells, Britta Jedamzik, and Nicola Gscheidel at the MPI-CBG Microarray Facility for their help with processing RNA samples for microarray hybridization. We thank Casey Bergman for helpful discussion and for proposing the analysis of differential rates of evolution on Muller's elements, and the analysis of mutation accumulation data. We also thank Nick Barton, Michael Hiller, and four anonymous reviewers for helpful comments on the manuscript, and Iva Kelava for preparing Figures \ref{corrdistembryos}, \ref{corrdistadults}, and S\ref{allcomps1}--S\ref{corrschematic}.

\section*{Author Contributions}
DTG first discovered higher gene expression divergence on the X chromosome in \emph{Drosophila} embryos and conceived the branch length analysis. MAK, PT, and ATK conceived the embryonic inbred strain collections, and MAK conducted the experiments. ATK conceived and conducted the gene expression and statistical analyses, and analysed the population genetics models. ATK wrote the manuscript with support from co-authors.

\clearpage

\bibliography{/home/alexk/Documents/refs/evol_genet}

\clearpage

\section*{Figures}

\clearpage

\section*{Supplementary Figures}

\setcounter{figure}{0}

\renewcommand{\figurename}{Supplementary Figure}

\begin{figure}[!ht]
\begin{center}
\includegraphics[width=6in]{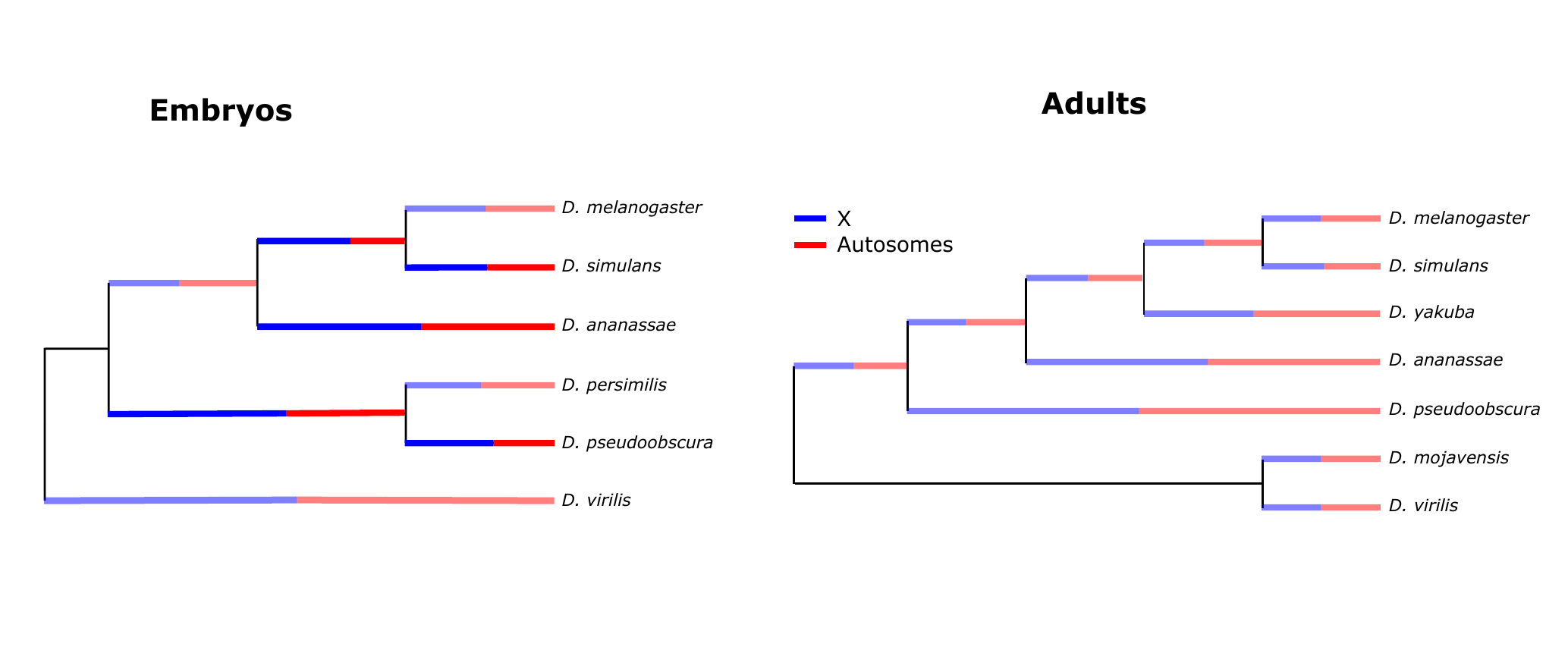}
\end{center}
\caption{
Phylogenies of the species analyzed with the relative mean lengths of each branch for genes on the X vs genes on the autosomes depicted in blue and red respectively. Bold branches are significantly longer for genes on the X chromosome based on 10,000 bootstrap replicates at the 5\% level.
}
\label{phylogbranches}
\end{figure}

\begin{figure}[!ht]
\begin{center}
\includegraphics[width=6in]{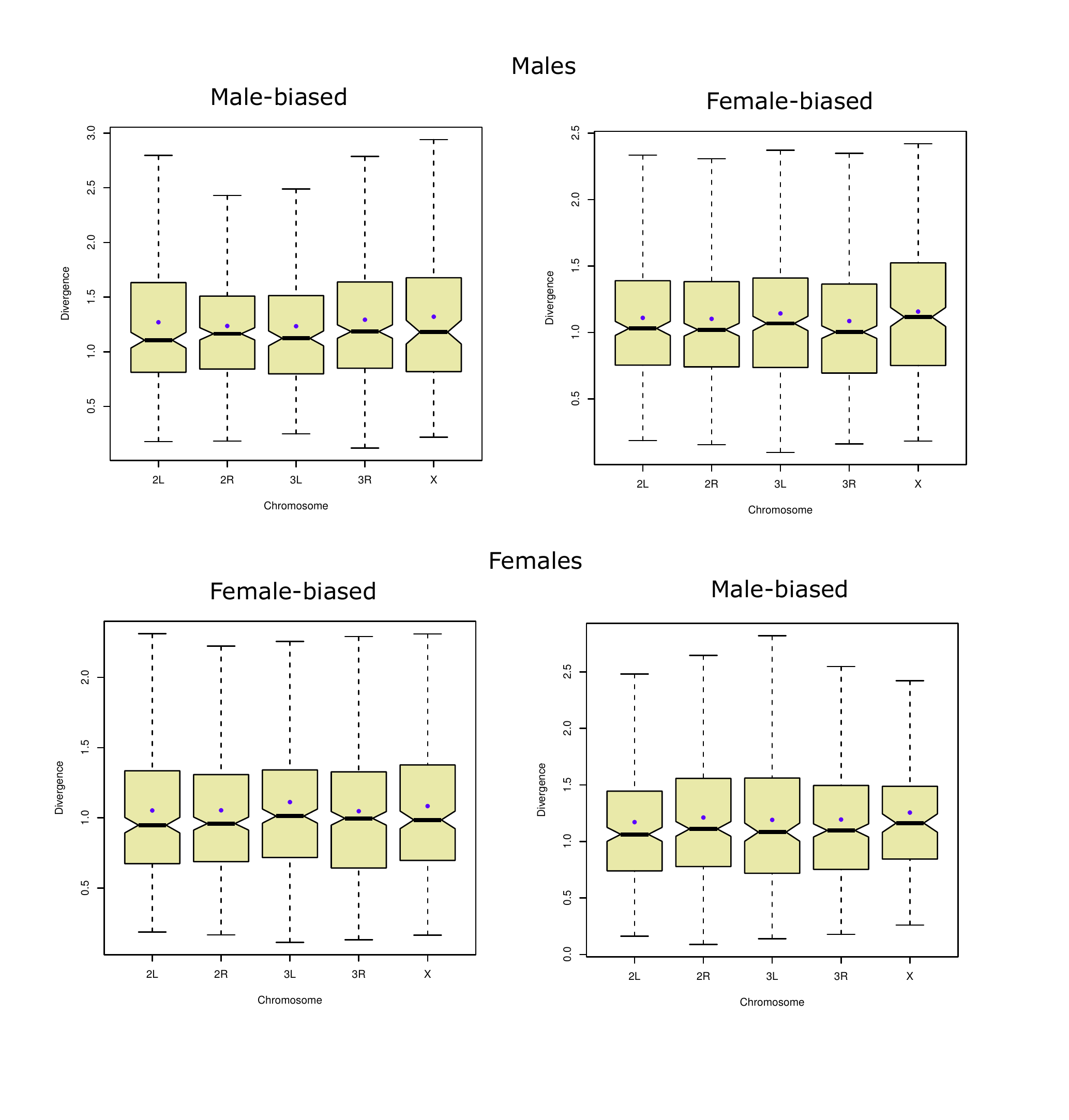}
\end{center}
\caption{
Divergence of gene expression across chromosomes in both adult males and females for genes with sex-biased expression patterns.
}
\label{allsexbias}
\end{figure}

\begin{figure}[!ht]
\begin{center}
\includegraphics[width=6in]{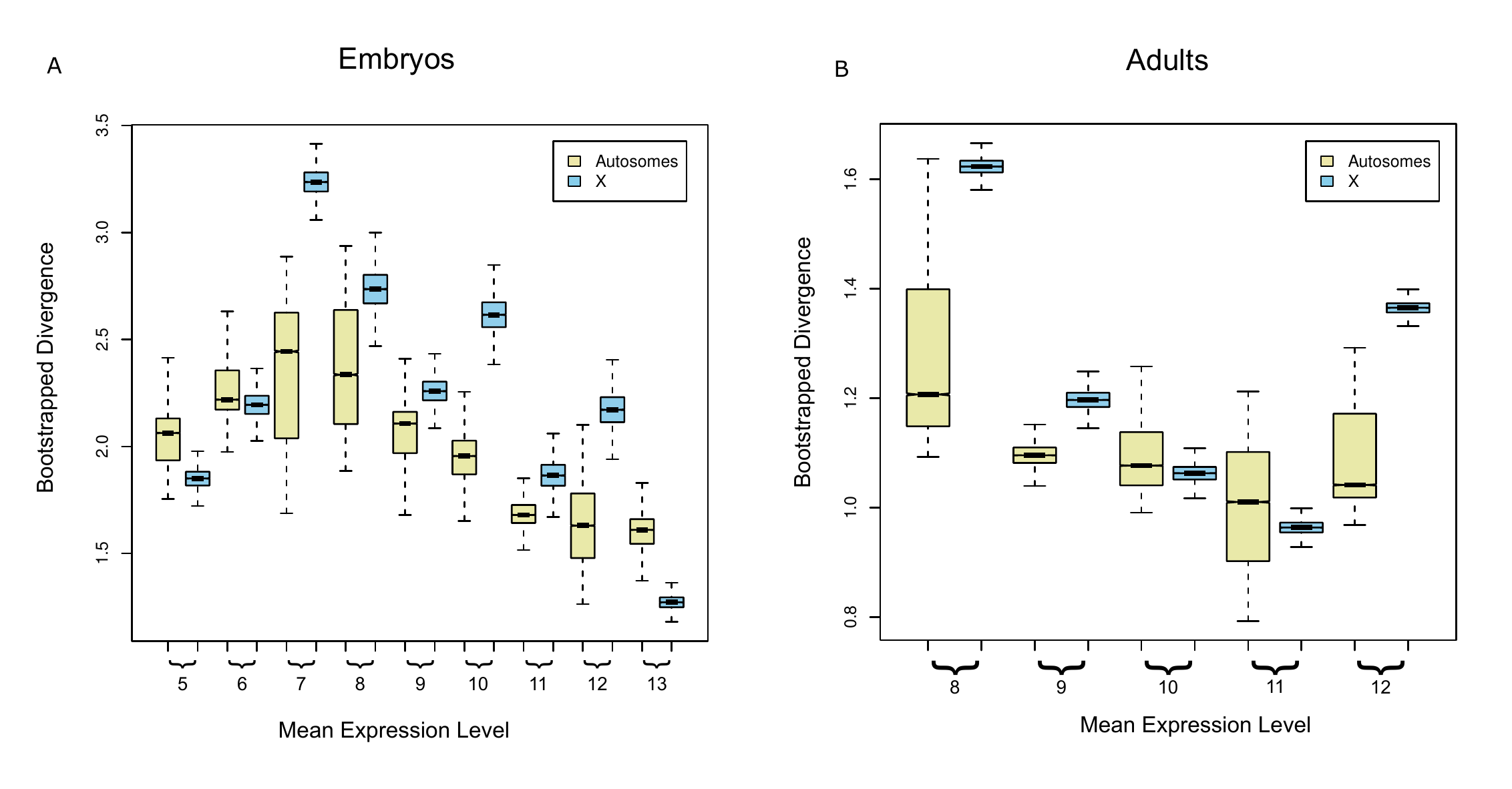}
\end{center}
\caption{
{\bf Embryonic expression divergence on the X is not driven by extreme expression levels.}\\
Bootstrapped divergence measures generated by resampling genes according to their expression levels. Genes were resampled per chromosome using 10,000 bootstrap replicates for both embryos, {\bf A}, and adults, {\bf B}. There are more expression levels sampled for embryos because they have a broader gene expression level distribution than the adults.
}
\label{explevelboot}
\end{figure}

\begin{figure}[!ht]
\begin{center}
\includegraphics[width=6in]{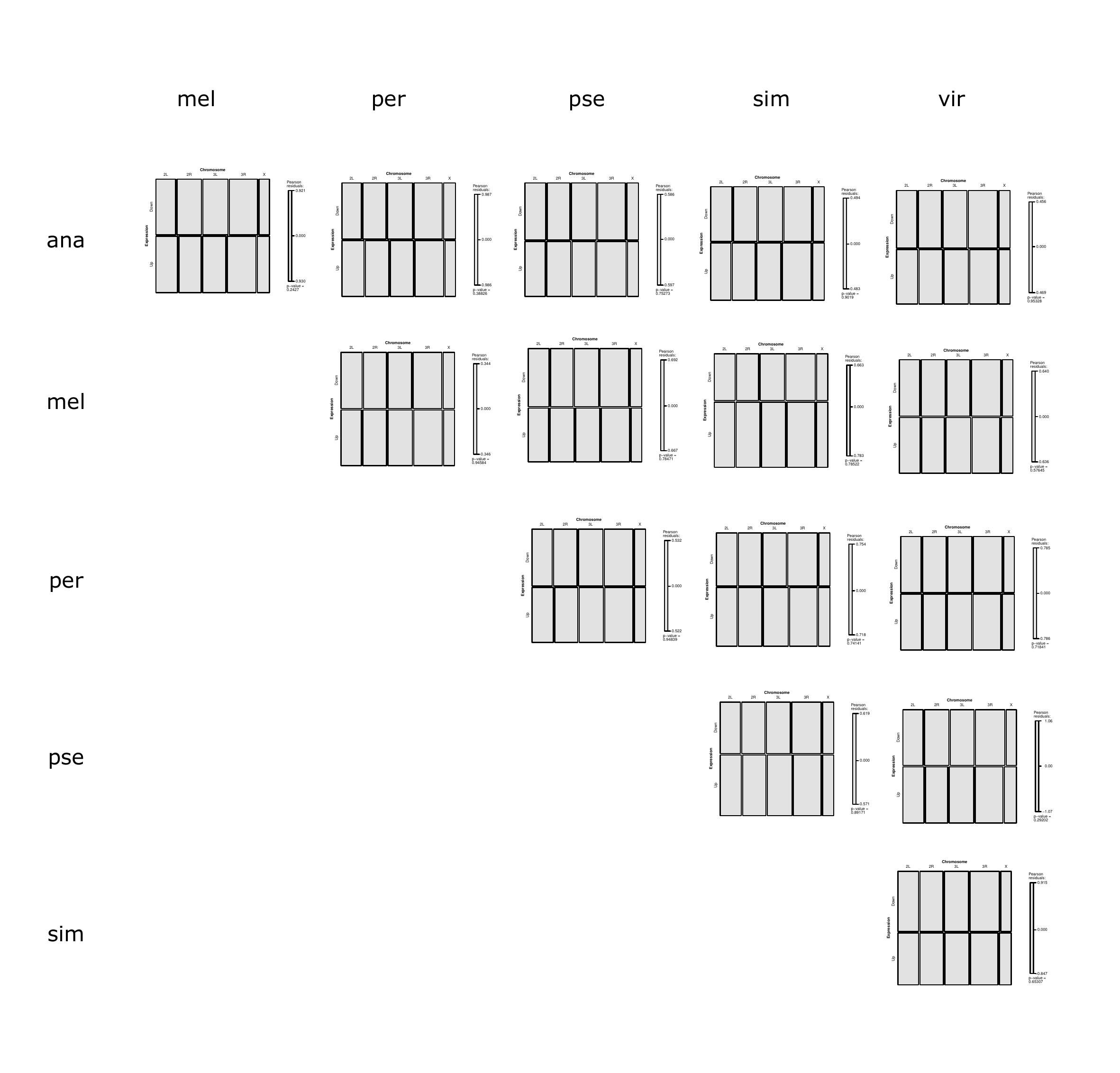}
\end{center}
\caption{
Mosaic plots for all pair-wise species comparisons of normalized gene expression categorised as up or down relative to one of the species. Mosaic plots visualize categorical data (contingency table) using rectangles that are proportional to the number of counts in each row-column combination, and highlight in red variable combinations that have less than expected numbers and in blue those that have more than expected based on Pearson residuals \cite{zeileisetal2007}. $P$-values are based on Chi-squared tests, which test whether the two main variables, Expression and Chromosome, are independent.
}
\label{expskewall}
\end{figure}

\begin{figure}[!ht]
\begin{center}
\includegraphics[width=6in]{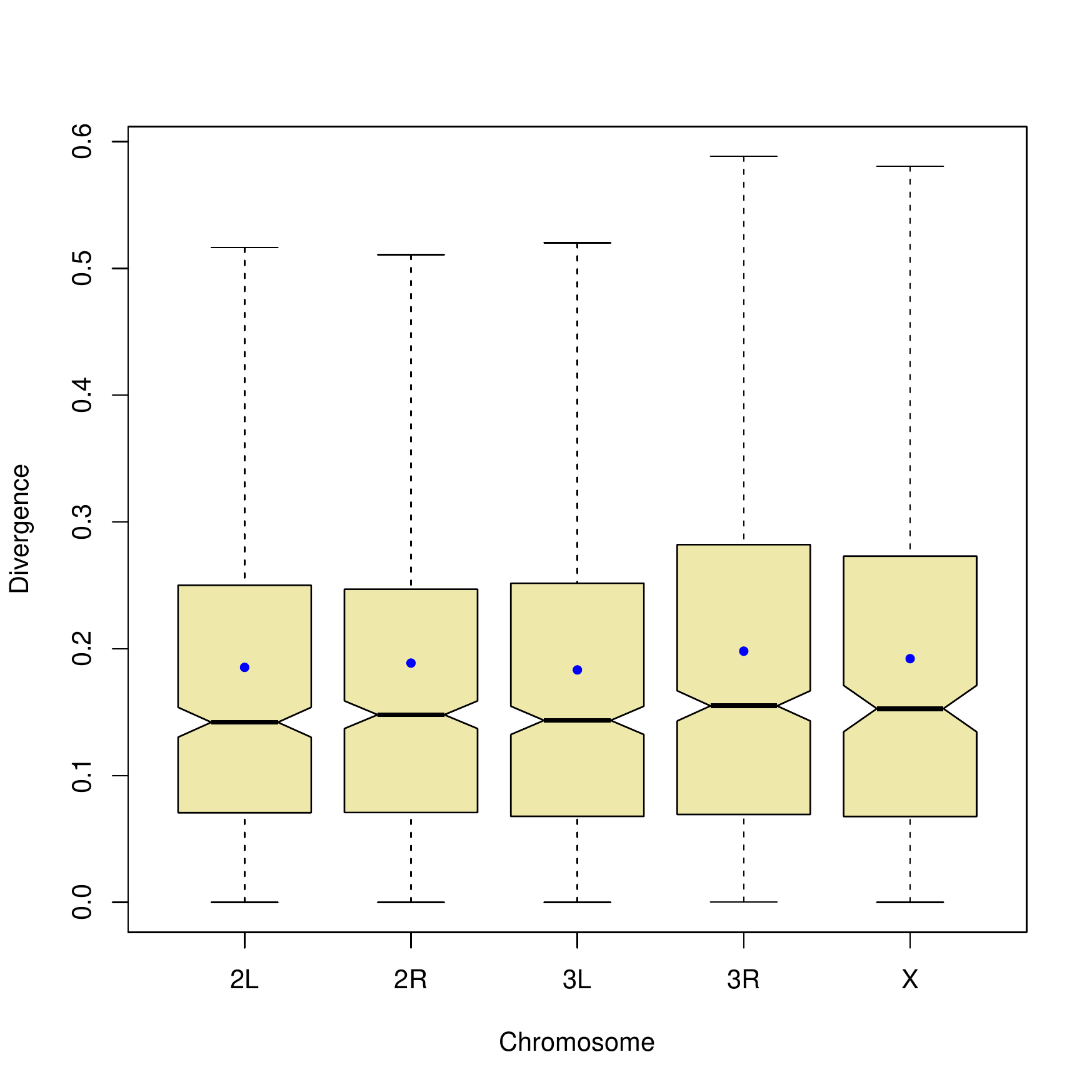}
\end{center}
\caption{
The lengths of the summed terminal branches leading to \emph{D. persimilis} and \emph{D. pseudoobscura} as a fraction of the total branch length for \emph{Drosophila} embryos.
}
\label{termneox}
\end{figure}

\begin{figure}[!ht]
\begin{center}
\includegraphics[width=7in]{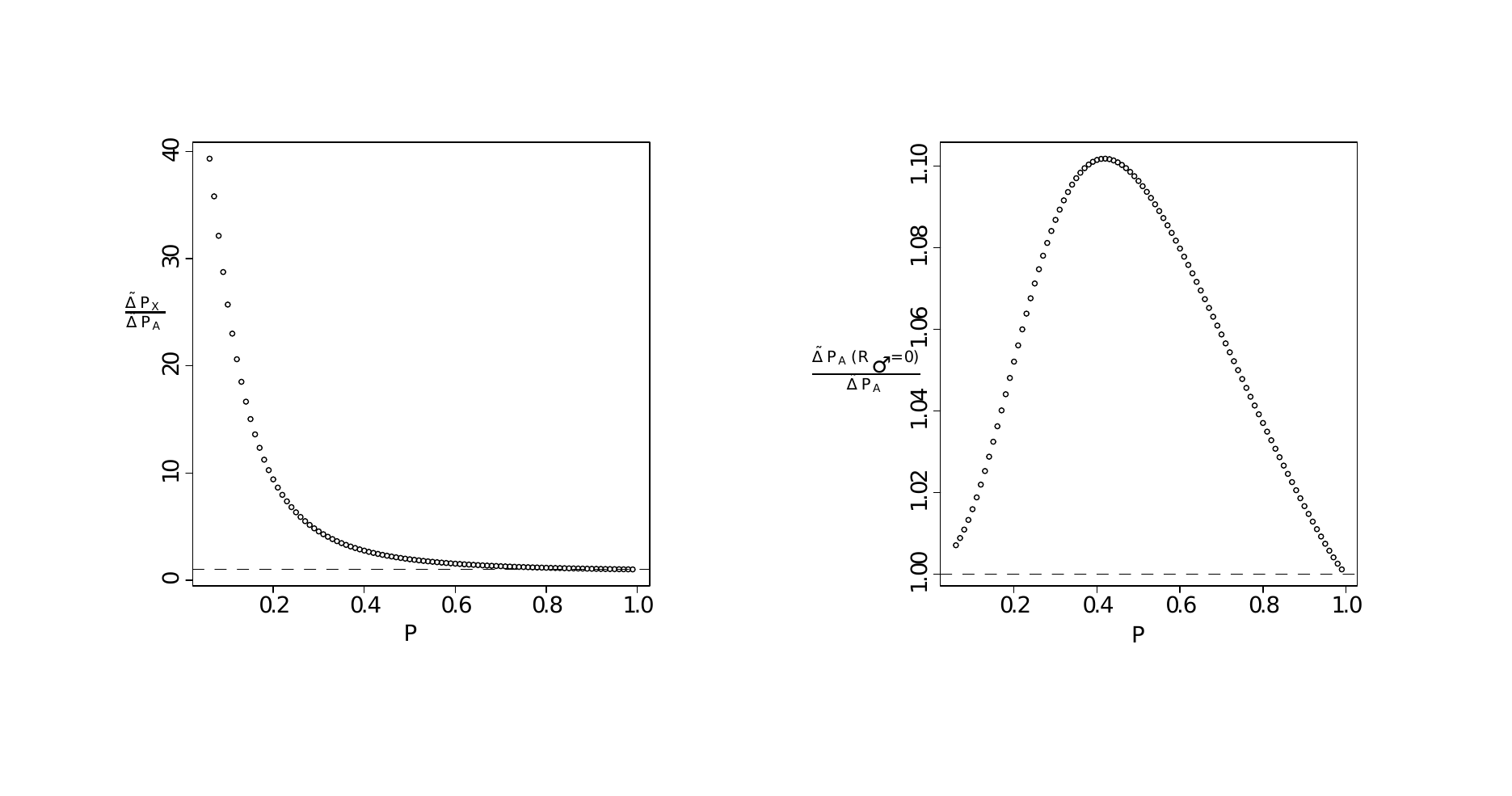}
\end{center}
\caption{
Selection gradient ratios when there is no recombination between homologous pairs of male autosomes. The left panel shows the ratio when both loci are X-linked versus both loci being linked on the same autosome but with no male recombination. The right panel shows the ratio for autosomes when there is no recombination in males versus the case when there is. Parameter values: recombination rates, $R$, are equal to 0.5 (free recombination) and the dominance coefficient, $h$, is 0.01. The dashed line indicates a ratio of 1.
}
\label{nomalerecomb}
\end{figure}

\begin{figure}[!ht]
\begin{center}
\includegraphics[width=6in]{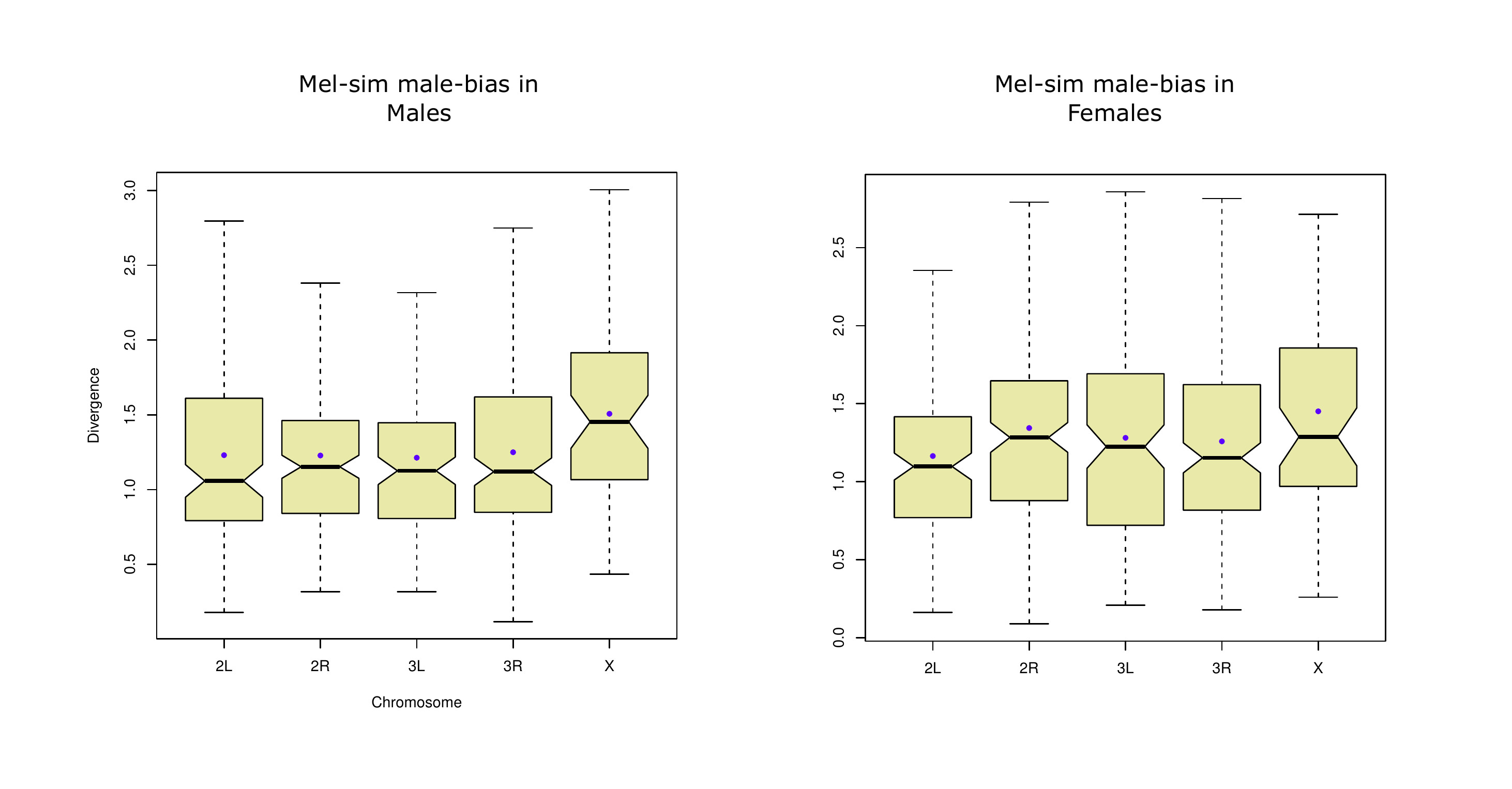}
\end{center}
\caption{
Divergence of gene expression across chromosomes in both adult males and females for 656 genes with male-biased expression in either \emph{D. melanogaster} or \emph{D. simulans}.
}
\label{melsimmalebias}
\end{figure}

\begin{figure}[!ht]
\begin{center}
\includegraphics[width=4in]{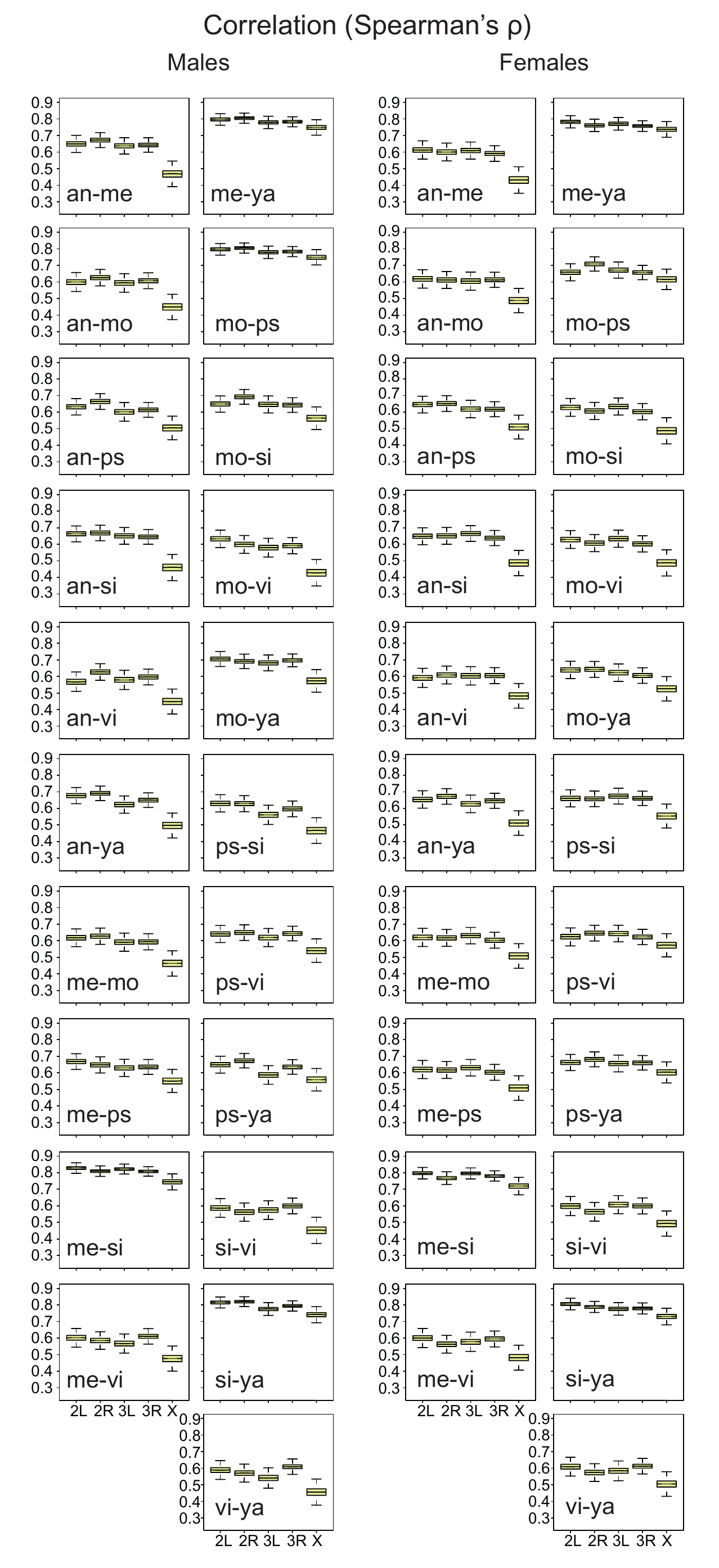}
\end{center}
\caption{
Bootstrapped (10,000 replicates) Spearman's $\rho$ correlation coefficients for adult males and females for all pair-wise species comparisons.
}
\label{allcomps1}
\end{figure}

\begin{figure}[!ht]
\begin{center}
\includegraphics[width=4in]{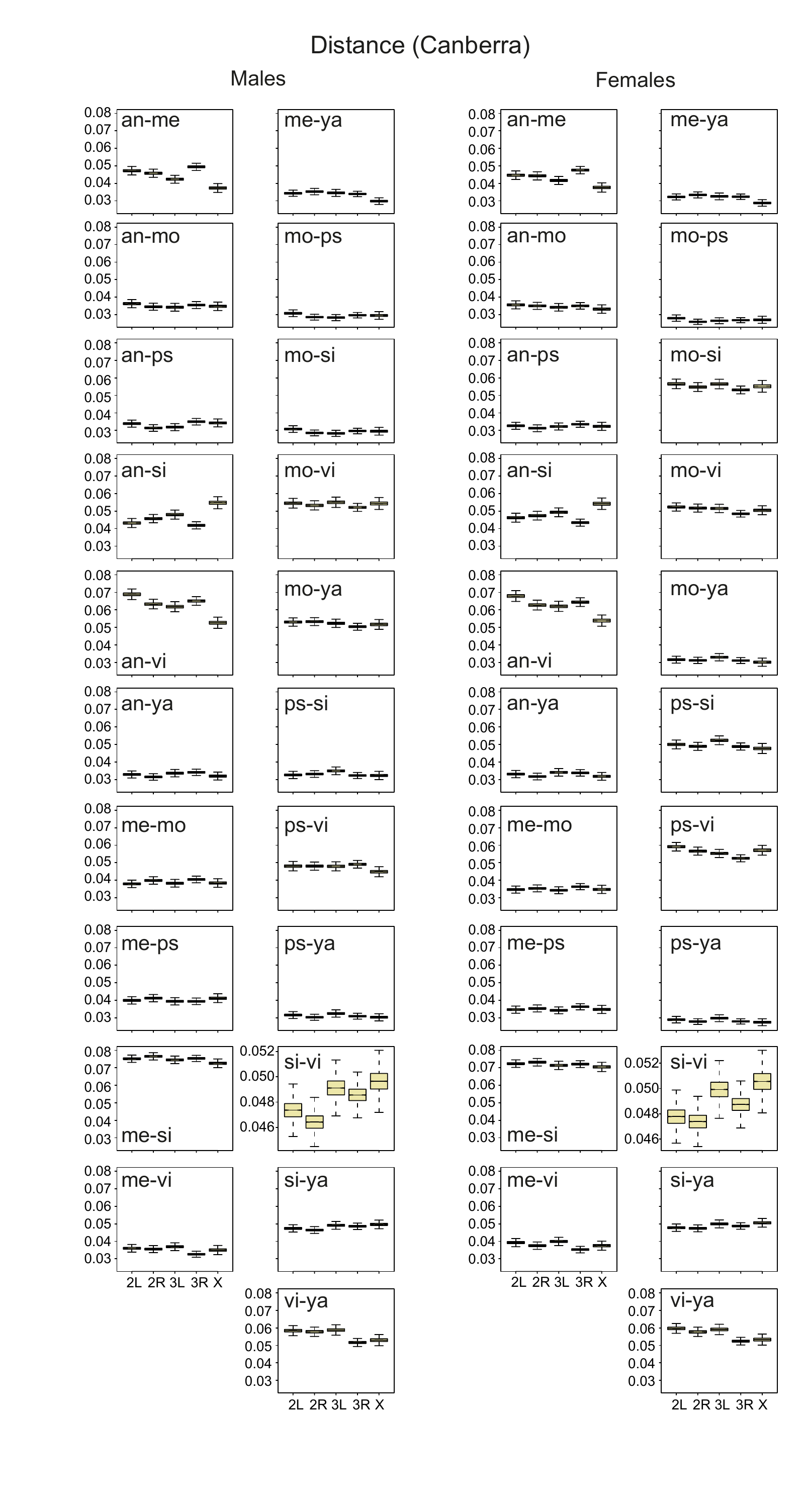}
\end{center}
\caption{
Bootstrapped (10,000 replicates) Mean Canberra distances for adult males and females for all pair-wise species comparisons.
}
\label{allcomps2}
\end{figure}

\begin{figure}[!ht]
\begin{center}
\includegraphics[width=6in]{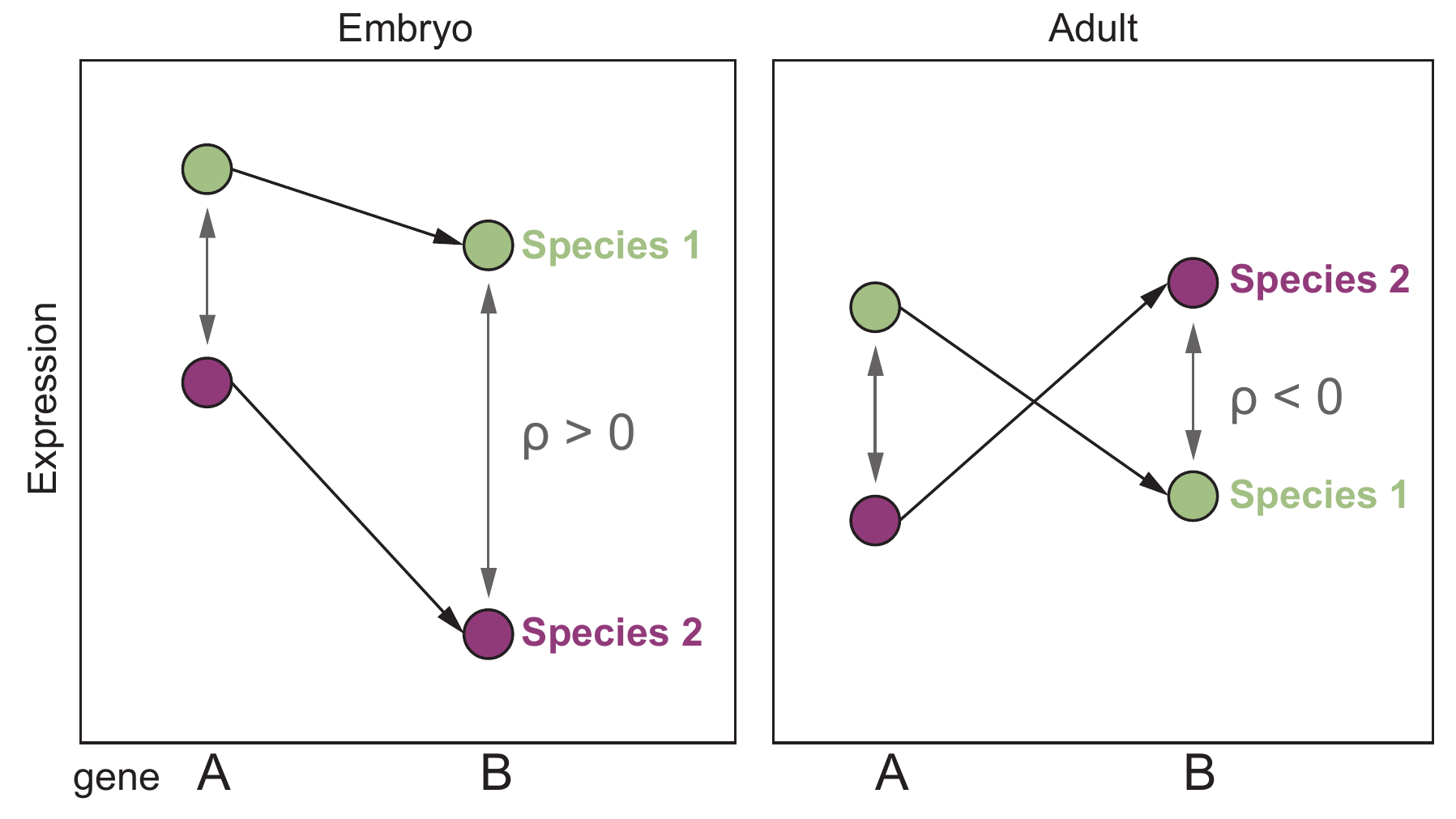}
\end{center}
\caption{
A schematic depicting gene expression in two genes showing why Spearman's $\rho$ would produce a positive correlation despite large differences in expression level and a negative correlation when expression co-ordination between genes is diminished regardless of how much absolute gene expression levels have changed.
}
\label{corrschematic}
\end{figure}

\begin{figure}[!ht]
\includegraphics[width=6in]{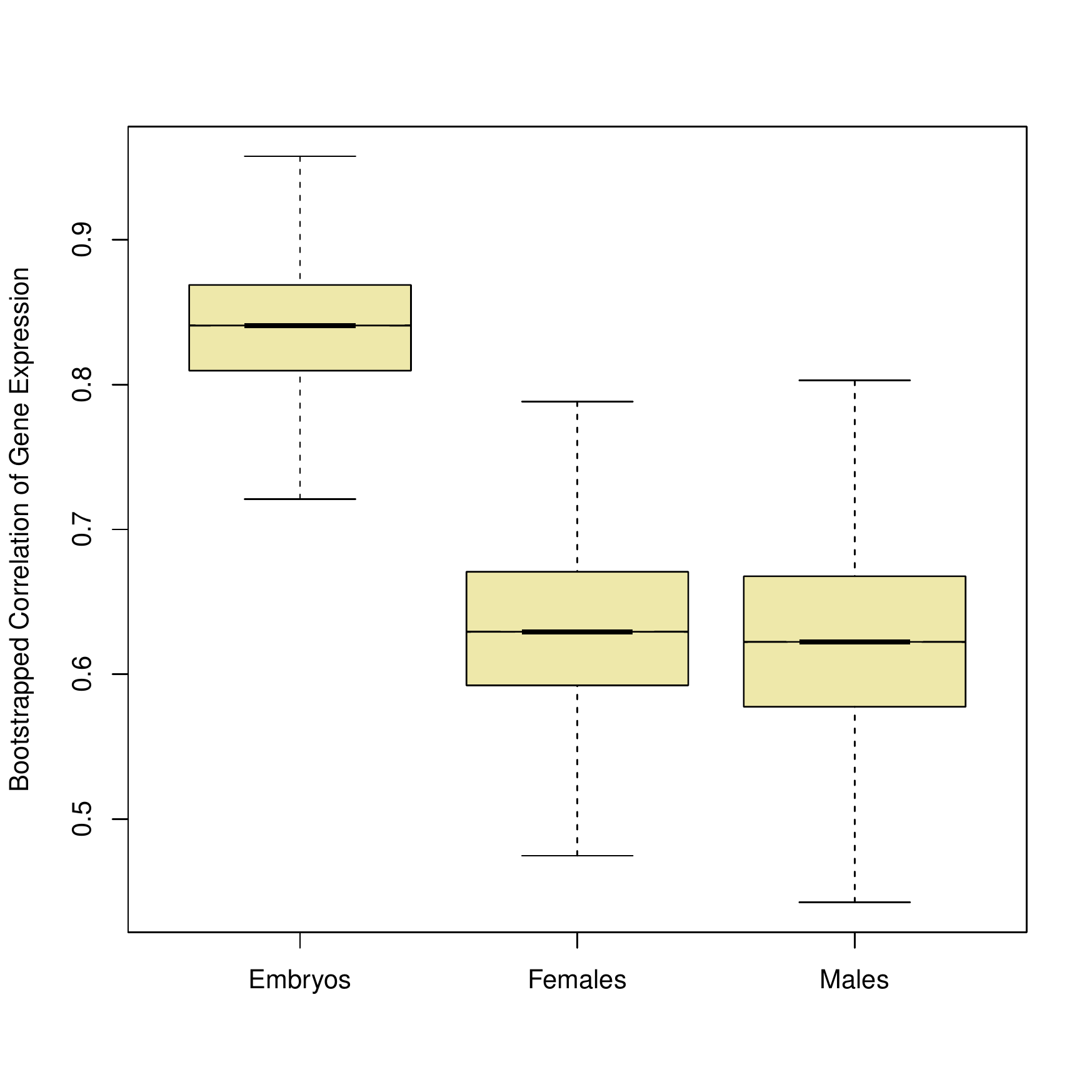}
\caption{
All bootstrapped Spearman's $\rho$ correlations across all chromosomes for embryos and adult males and females.
}
\label{corrcomparison}
\end{figure}

\begin{figure}[!ht]
\begin{center}
\includegraphics[width=6in]{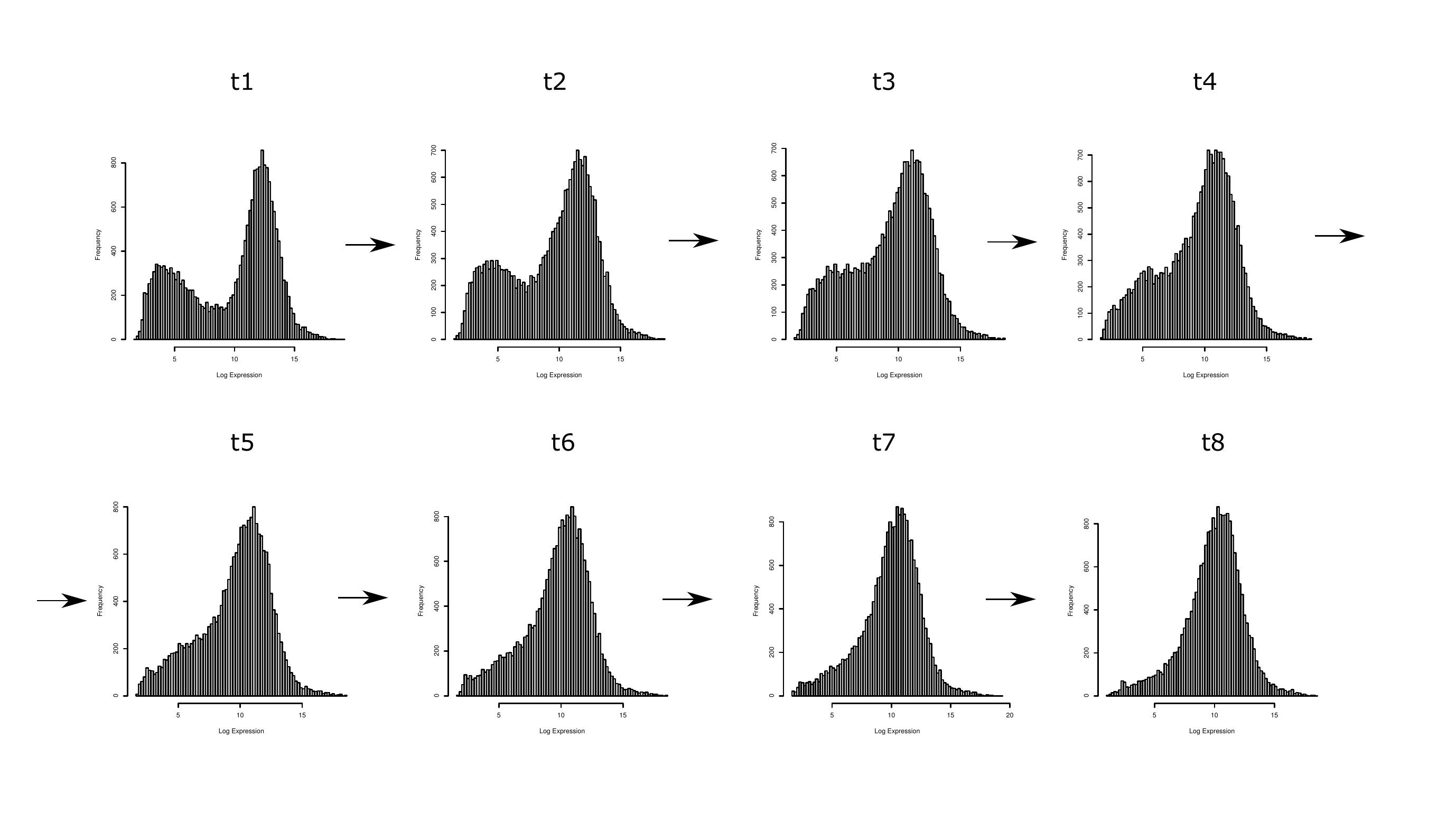}
\end{center}
\caption{
The distribution of gene expression levels during embryogenesis of \emph{D. melanogaster} showing that an initially bimodal distribution, where the lower mode represents unexpressed zygotic genes, becomes a unimodal distribution through time as the zygotic genome is activated.
}
\label{karolinafig}
\end{figure}

\begin{figure}[!ht]
\begin{center}
\includegraphics[width=6in]{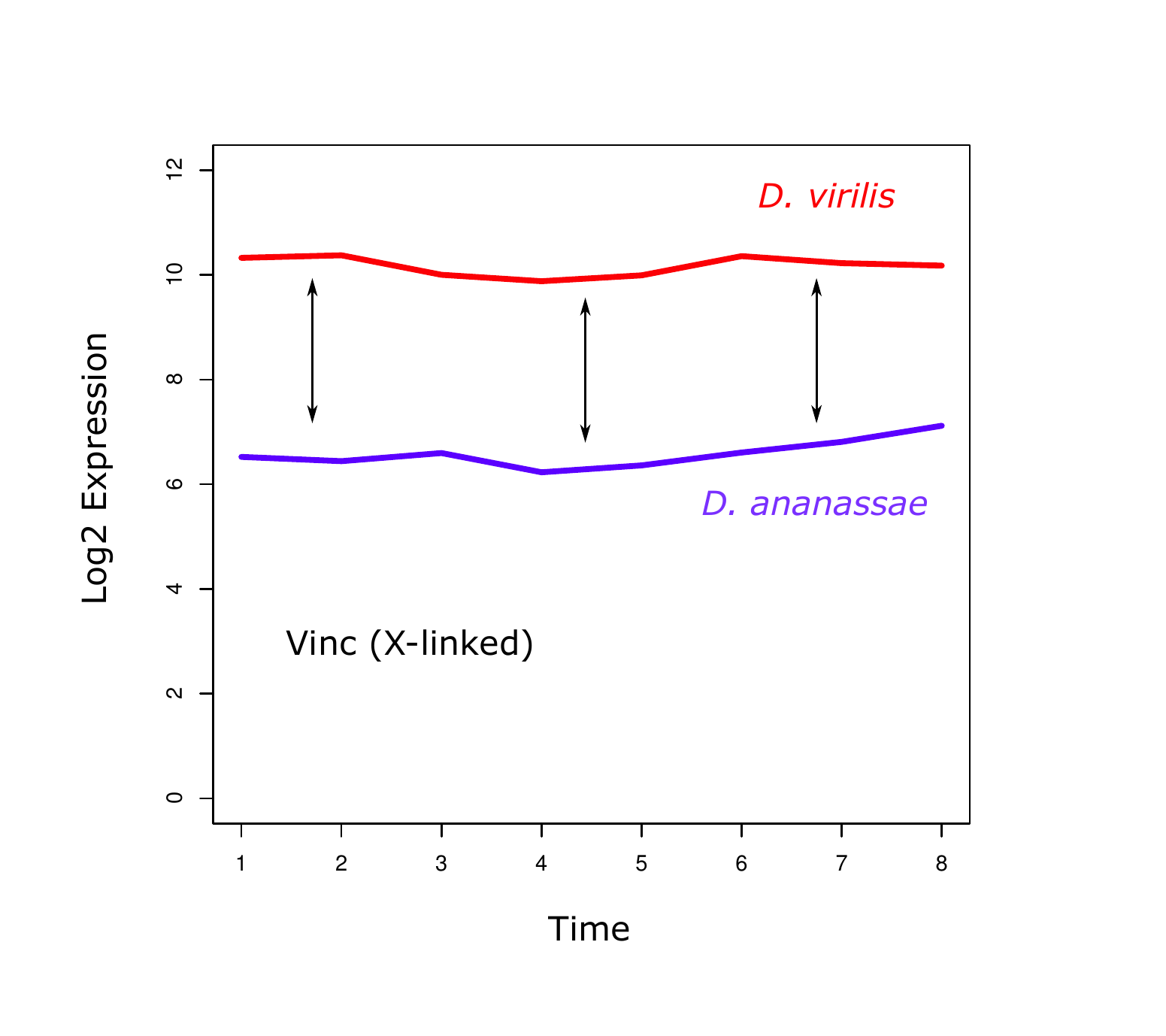}
\end{center}
\caption{
Log$_{2}$ gene expression time course for the X-linked gene Vinculin (\emph{Vinc}) for \emph{D. ananassae} and \emph{D. virilis} showing divergence across the whole time course.
}
\label{vinc}
\end{figure}

\begin{figure}[!ht]
\begin{center}
\includegraphics[width=6in]{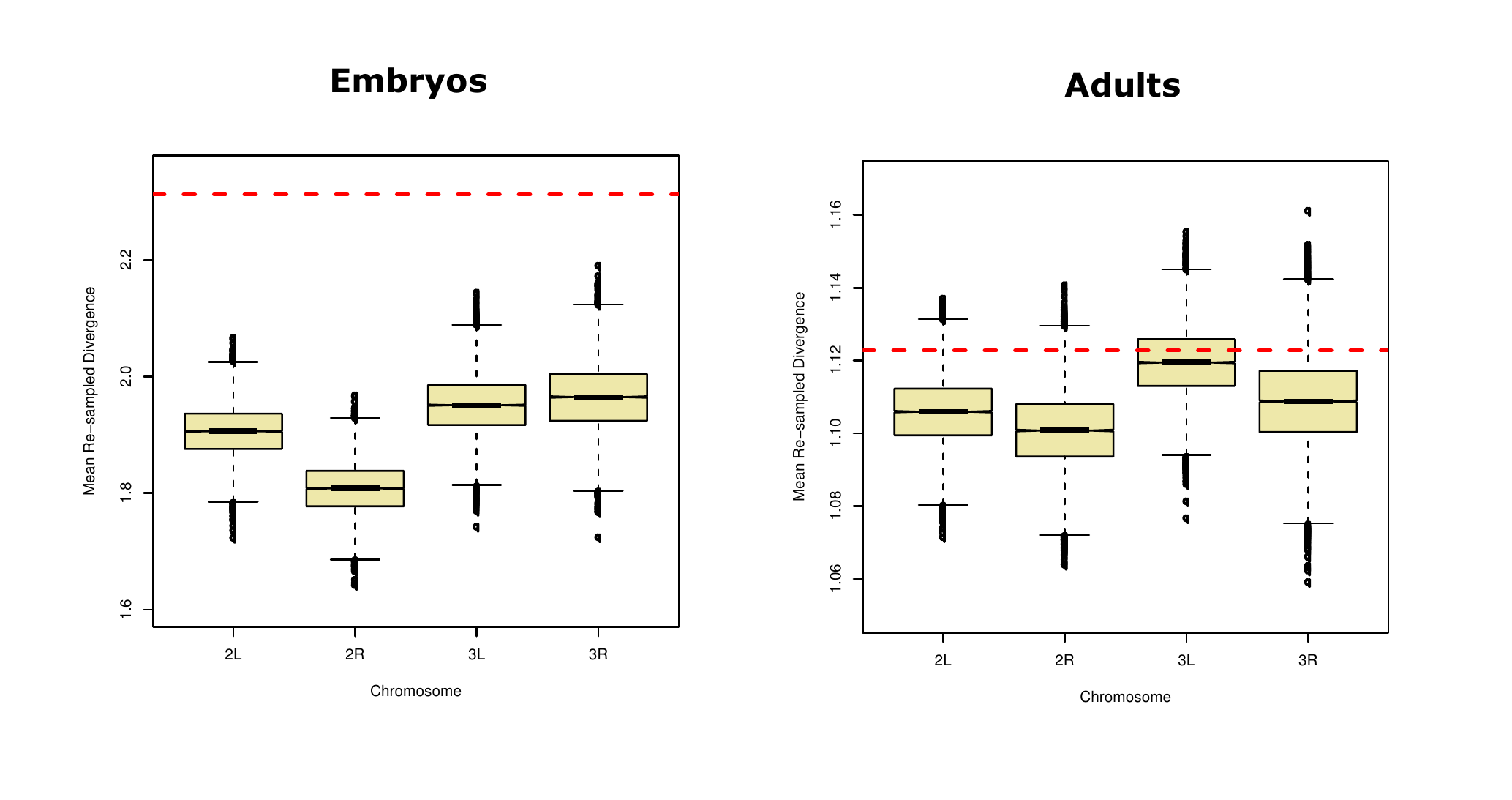}
\end{center}
\caption{
The distributions of resampled mean divergences for each autosome with the mean of the X chromosome indicated by a dashed red line for embryos and adults. Autosomal genes were resampled so that they matched the number of genes on the X chromosome and in each of 10,000 resamples the mean divergence per chromosome was recorded.
}
\label{resampnumbergenes}
\end{figure}

\begin{figure}[!ht]
\begin{center}
\includegraphics[width=6in]{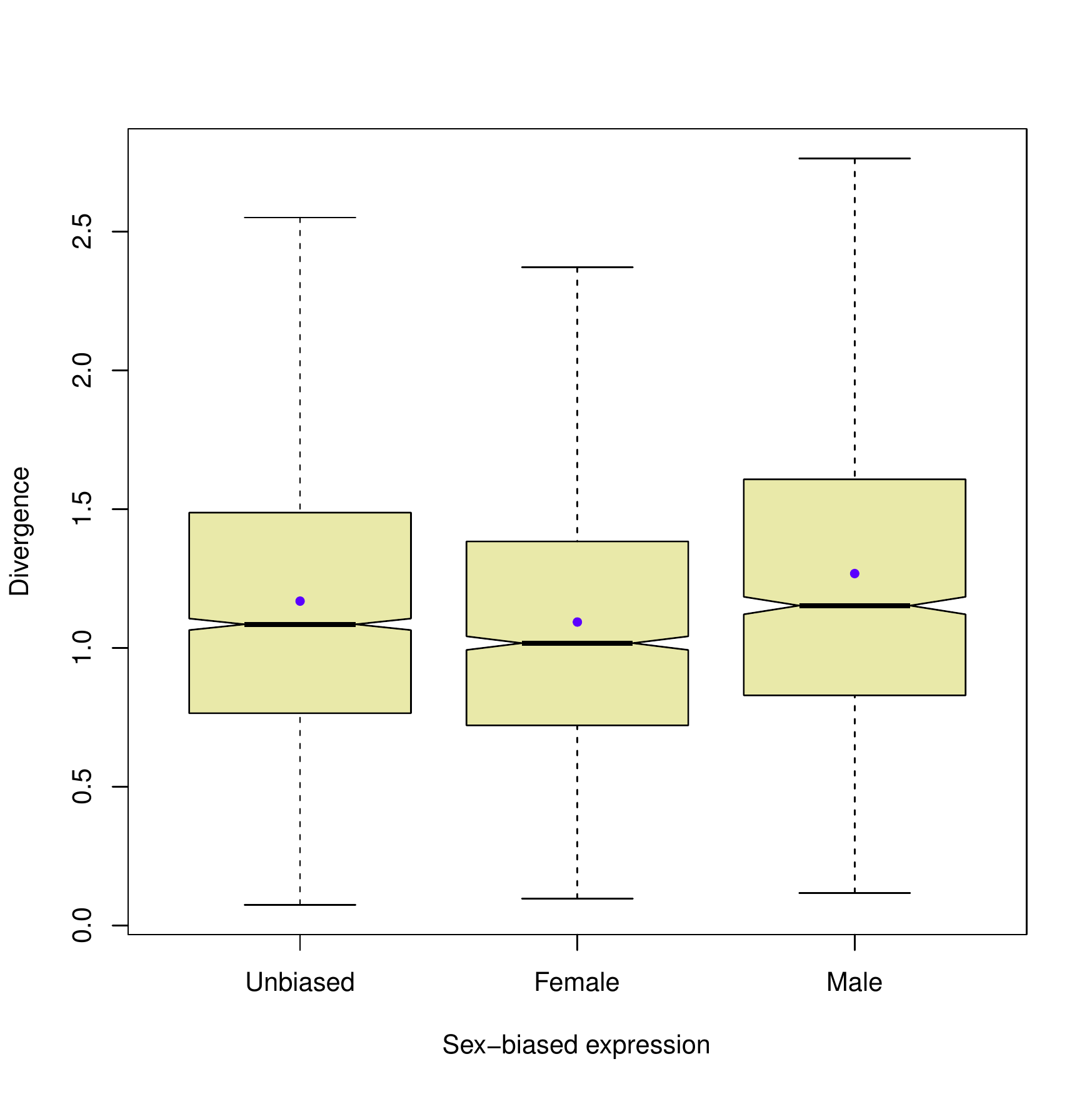}
\end{center}
\caption{
Divergence of gene expression in adult males for genes that show unbiased, male-biased, and female-biased expression patterns.
}
\label{biasmales}
\end{figure}

\begin{figure}[!ht]
\begin{center}
\includegraphics[width=6in]{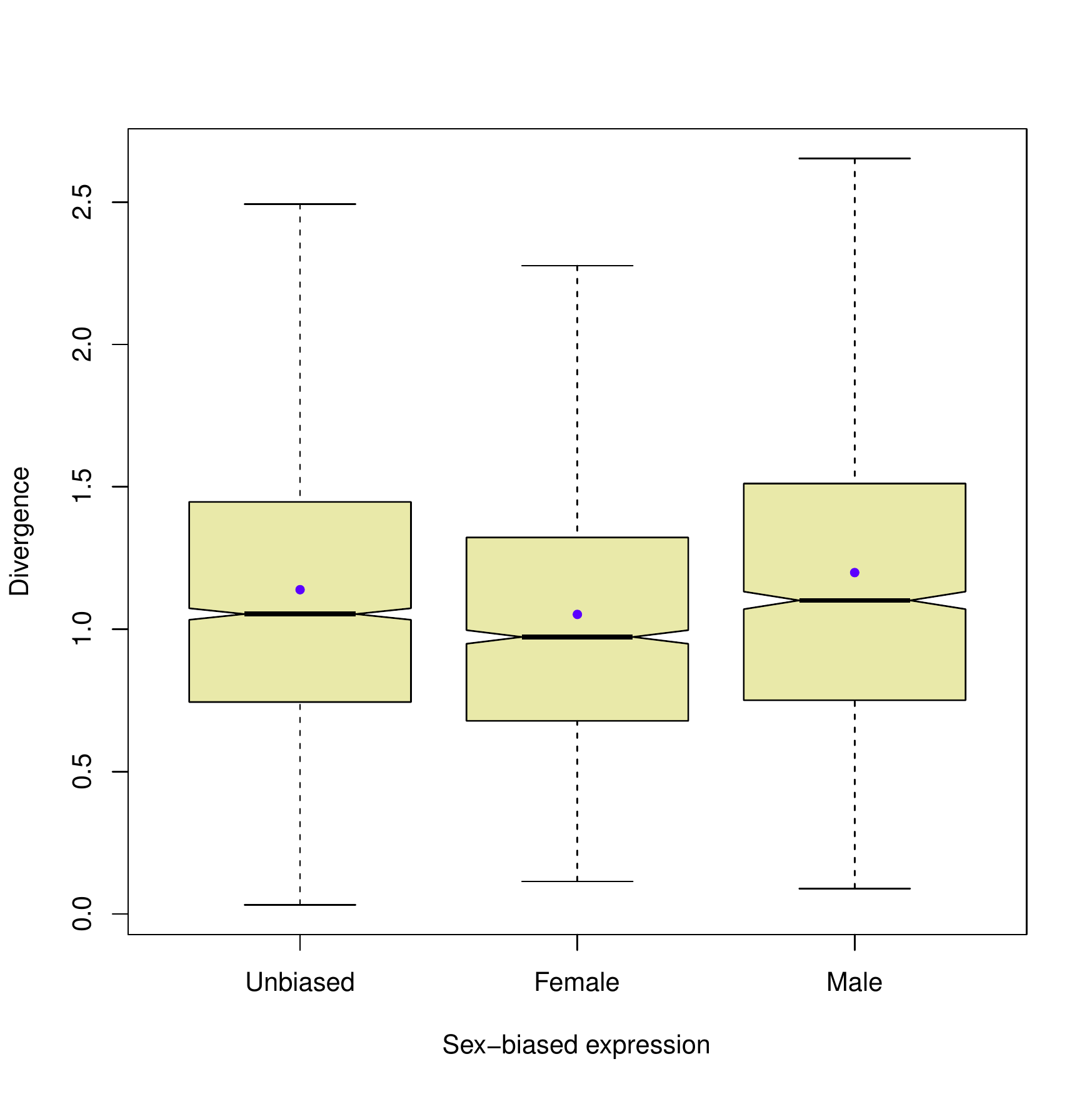}
\end{center}
\caption{
Divergence of gene expression in adult females for genes that show unbiased, male-biased, and female-biased expression patterns.
}
\label{biasfemales}
\end{figure}

\begin{figure}[!ht]
\begin{center}
\includegraphics[width=6in]{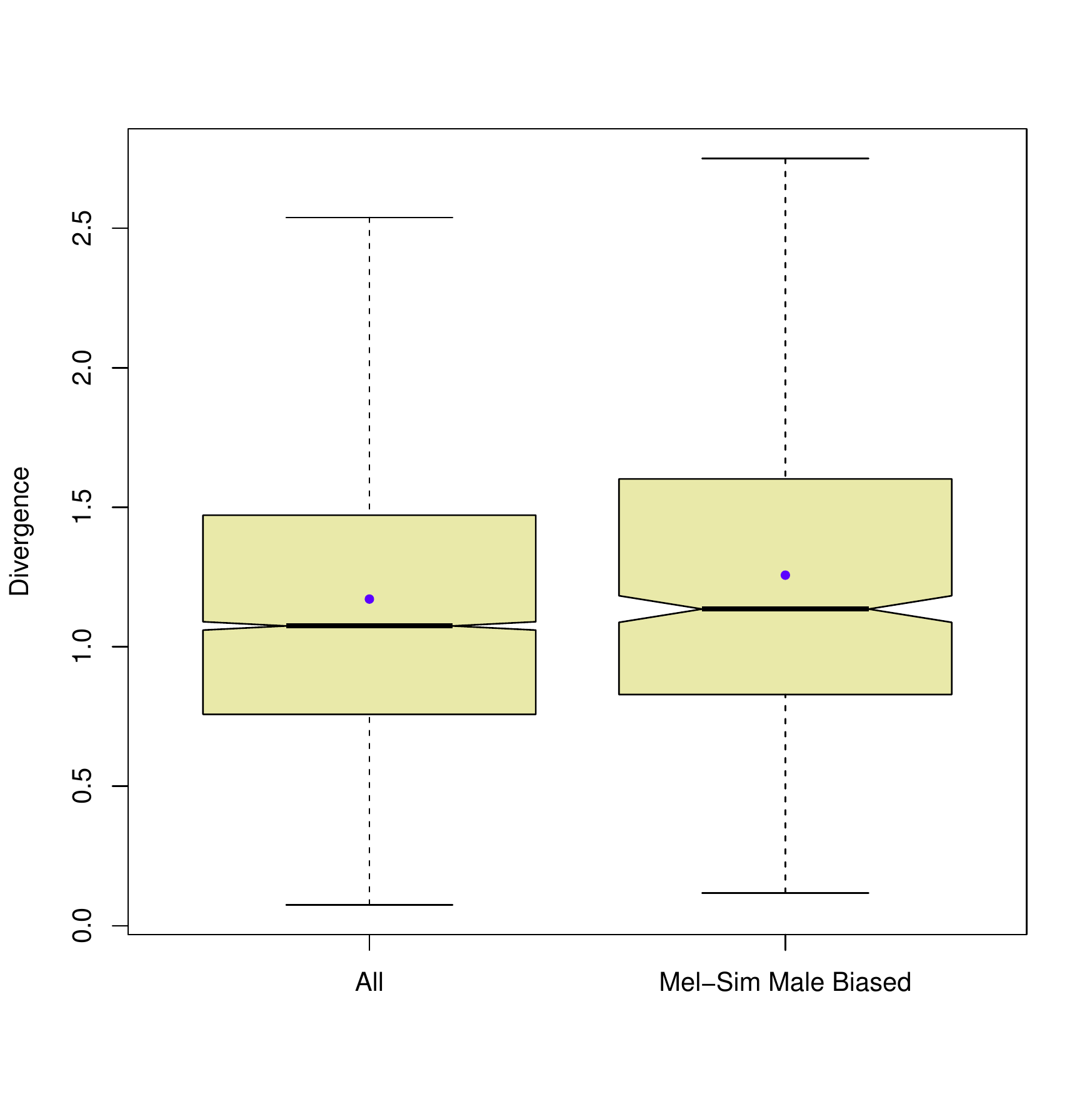}
\end{center}
\caption{
Divergence of gene expression in adult males for 656 genes with male-biased expression in either \emph{D. melanogaster} or \emph{D. simulans} relative to all genes in the dataset.
}
\label{melsimdiv}
\end{figure}

\begin{figure}[!ht]
\begin{center}
\includegraphics[width=6in]{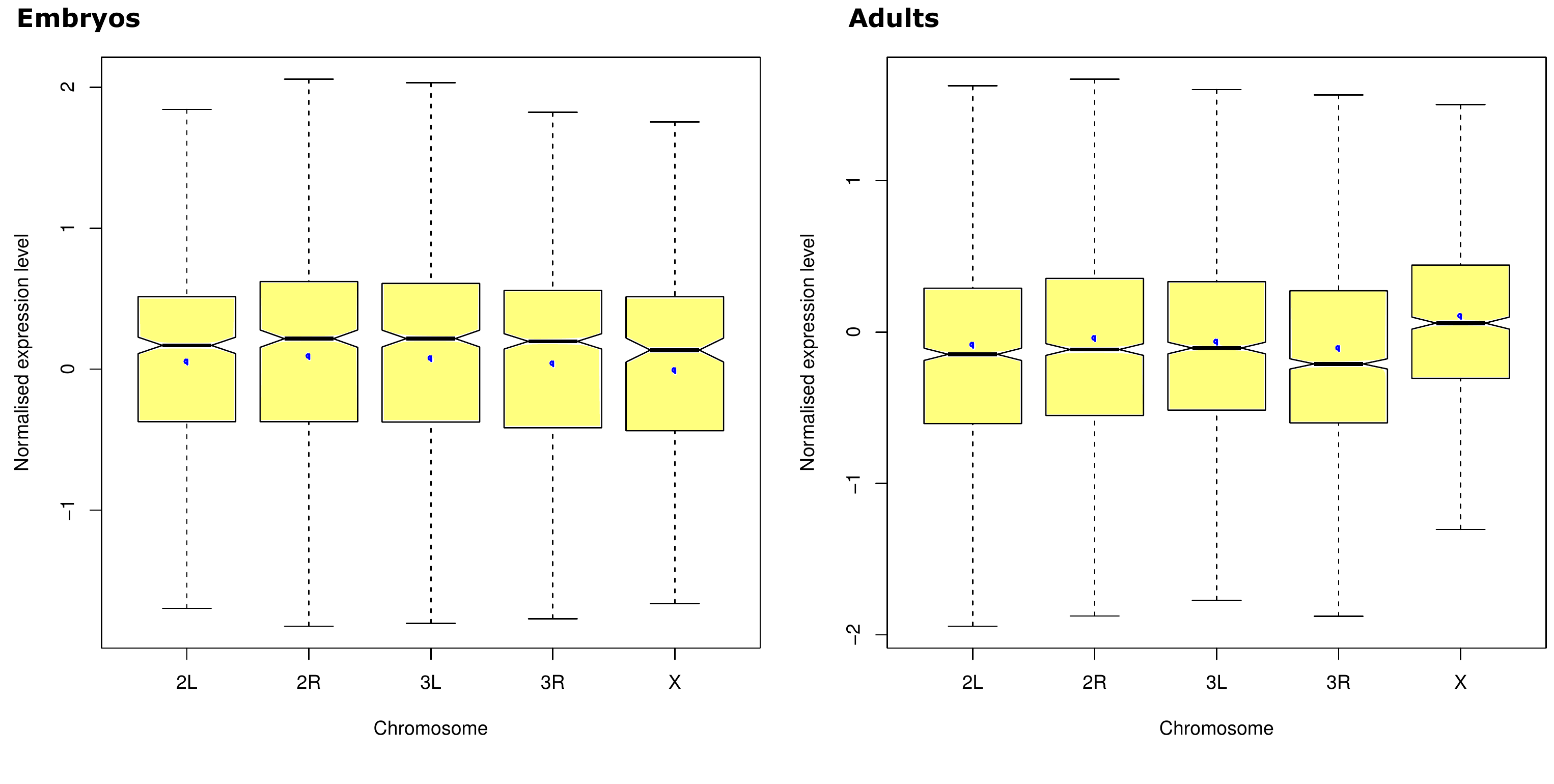}
\end{center}
\caption{
Gene expression level by chromosome for embryos and adults in the Drosophila data sets. Expression level is shown as the deviation of each gene's mean log$_{2}$ expression level from the global mean.
}
\label{explevel}
\end{figure}

\clearpage

\begin{figure}[!ht]
\begin{center}
\includegraphics[width=6in]{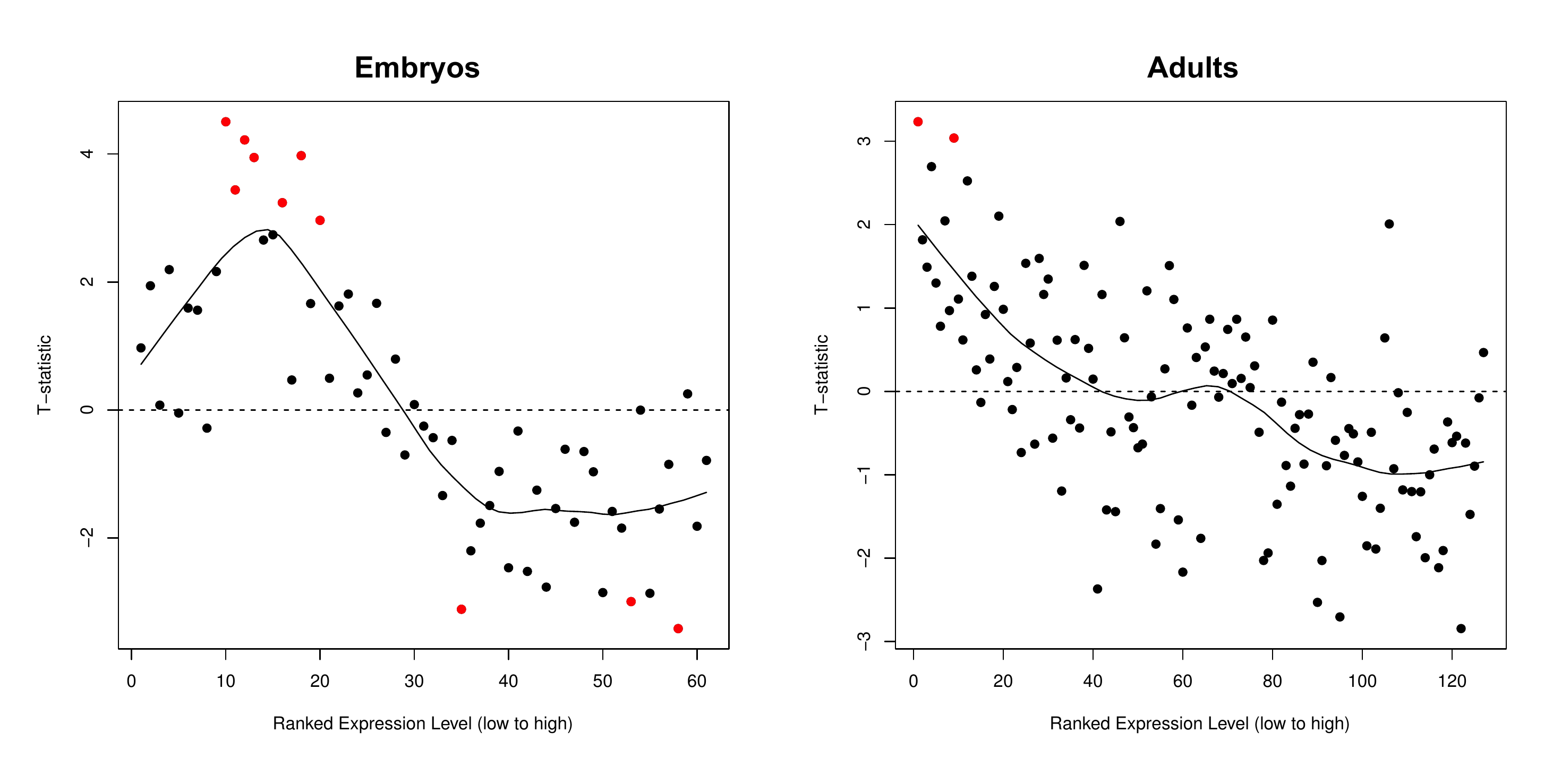}
\end{center}
\caption{
The relationship between expression level and divergence for embryos and adults in the Drosophila data sets. Genes are ranked by expression, from lowest to highest, binned into groups of 50, and their mean divergence deviation from the global mean (log divergence) is shown as a T-statistic, with significant values highlighted in red. A LOESS curve is fitted to the data.
}
\label{explevelbins}
\end{figure}

\begin{figure}[!ht]
\begin{center}
\includegraphics[width=6in]{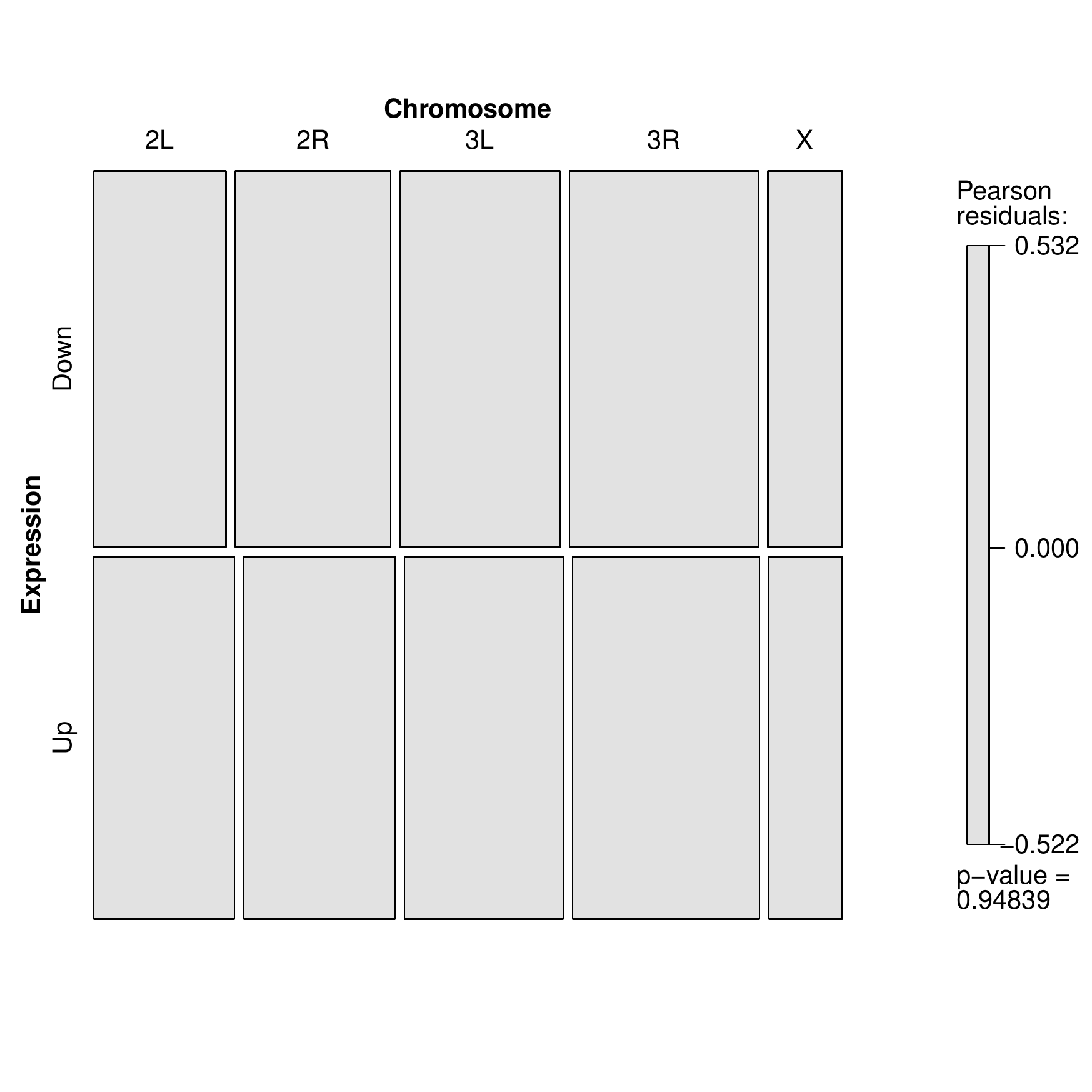}
\end{center}
\caption{
Mosaic plots for the \emph{D. persimilis}-\emph{D. pseudoobscura} species comparison of normalized gene expression categorised as up or down relative to one of the species. Mosaic plots visualize categorical data (contingency table) using rectangles that are proportional to the number of counts in each row-column combination, and highlight in red variable combinations that have less than expected numbers and in blue those that have more than expected based on Pearson residuals \cite{zeileisetal2007}. $P$-values are based on Chi-squared tests, which test whether the two main variables, Expression and Chromosome, are independent.
}
\label{expskewprps}
\end{figure}

\begin{figure}[!ht]
\begin{center}
\includegraphics[width=6in]{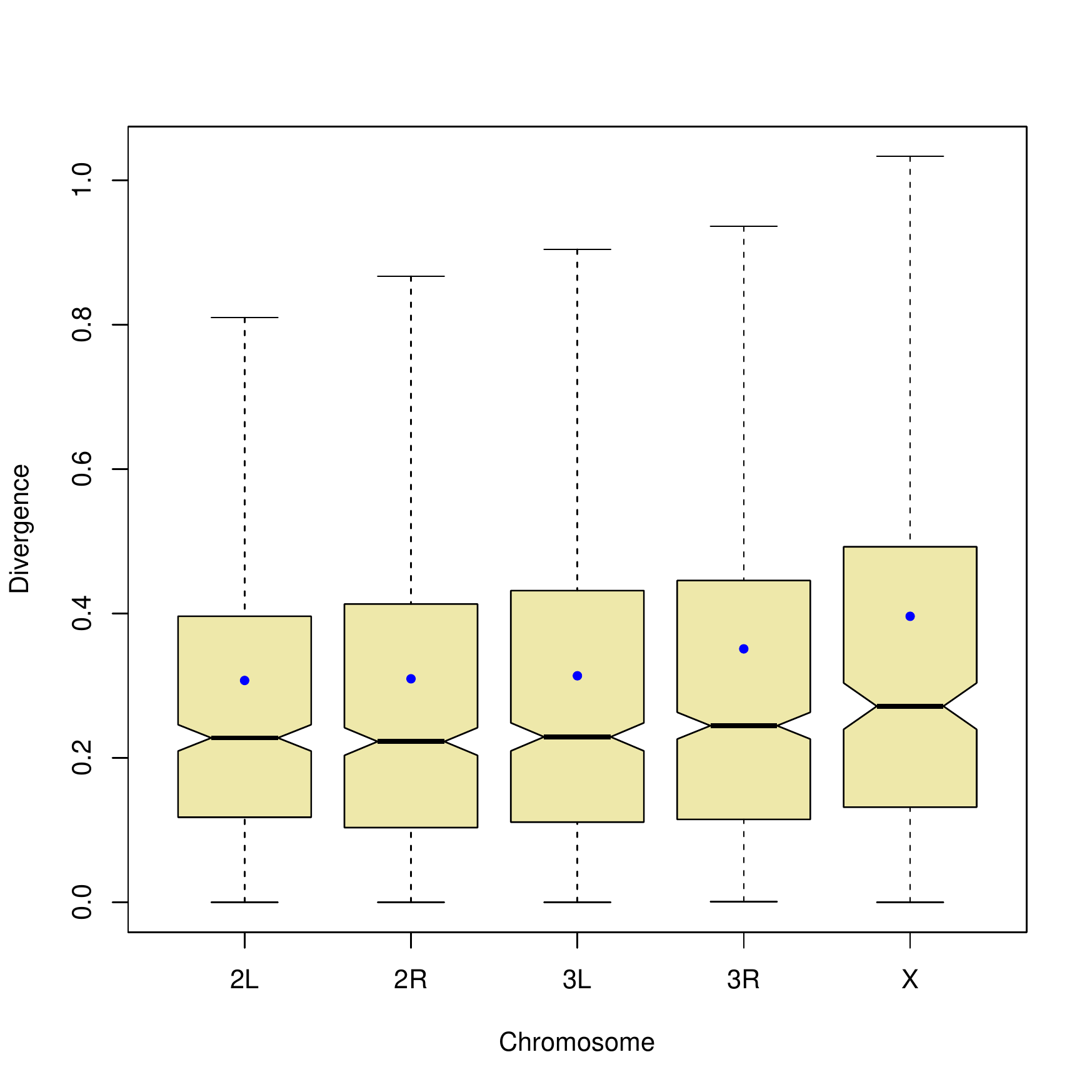}
\end{center}
\caption{
Gene expression divergence per chromosome along the branches leading to \emph{D. persimilis} and \emph{D. pseudoobscura}.
}
\label{prpschrom}
\end{figure}

\begin{figure}[!ht]
\begin{center}
\includegraphics[width=6in]{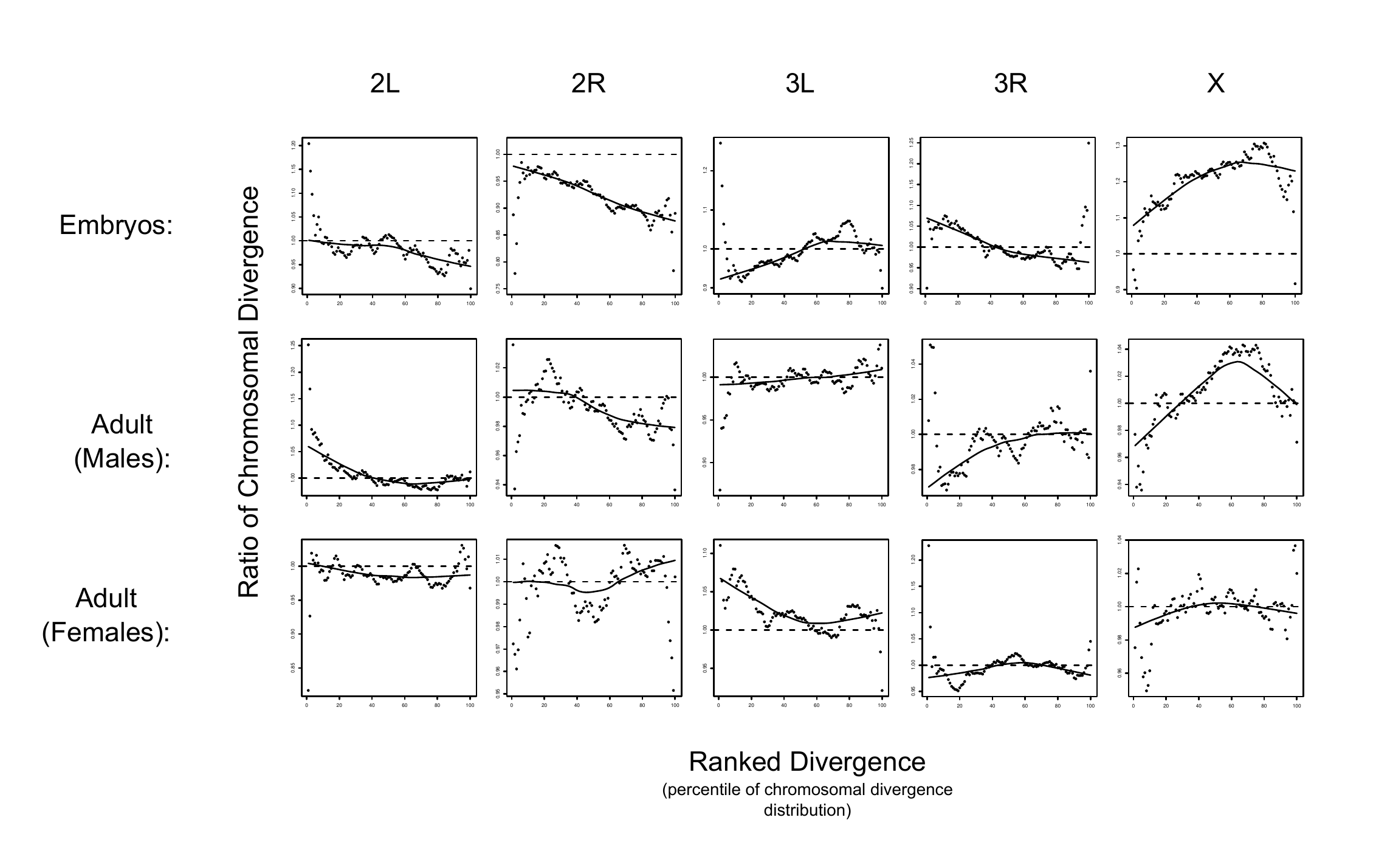}
\end{center}
\caption{
{\bf Fast-evolving genes tend to diverge more on the X in embryos and adult males.}\\
The mean ratio of chromosomal divergence to divergence in the rest of the genome. Mean divergence is plotted for genes belonging to each percentile of a particular chromosome's divergence distribution (separately for 2L, 2R, etc) relative to genes in the same percentile of the divergence distribution of the the rest of the genome (all other chromosomes). The results show that, for the X chromosome, the excess of X/A divergence is higher for faster-evolving genes in both embryos and adult males. Lines are LOESS fits to the data and dashed lines indicate ratios of 1.
}
\label{percentilediv}
\end{figure}

\begin{figure}[!ht]
\begin{center}
\includegraphics[width=6in]{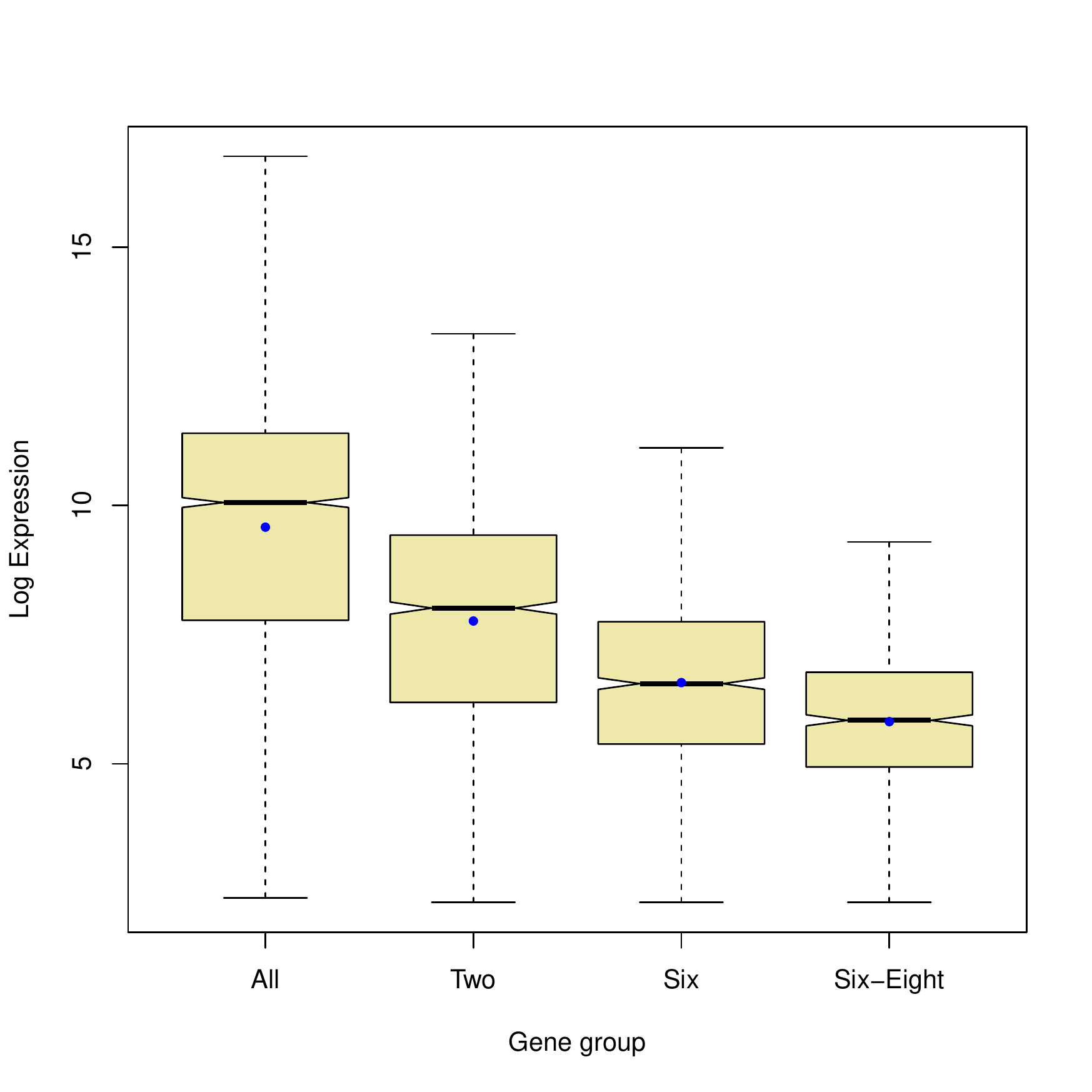}
\end{center}
\caption{
Log expression distributions for gene sets excluded for being non-expressed in at least two species in at least one time point (``Two''), in at least six species in at least one time point (``Six''), and in all species at all time points (``Six-Eight''). See Methods.
}
\label{dropgenesExp}
\end{figure}

\begin{figure}[!ht]
\begin{center}
\includegraphics[width=6in]{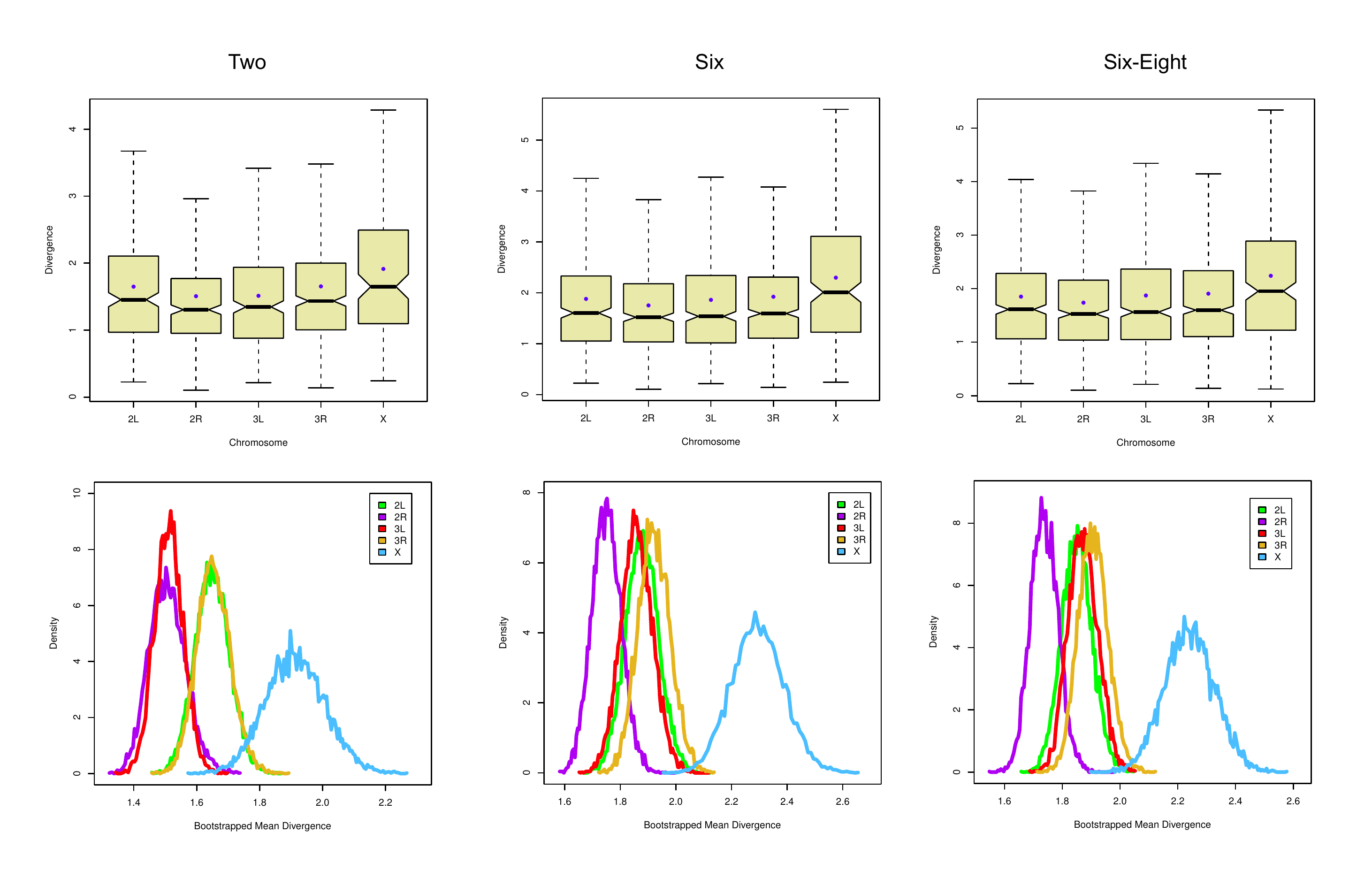}
\end{center}
\caption{
Gene expression divergence on the X chromosome relative to the autosomes for sets of genes with groups of non-expressed genes removed using various different criteria: non-expressed in at least two species in at least one time point (``Two''), non-expressed in at least six species in at least one time point (``Six''), and non-expressed in all species at all time points (``Six-Eight''). See Methods.
}
\label{dropgenesXeffect}
\end{figure}

\begin{figure}[!ht]
\begin{center}
\includegraphics[width=6in]{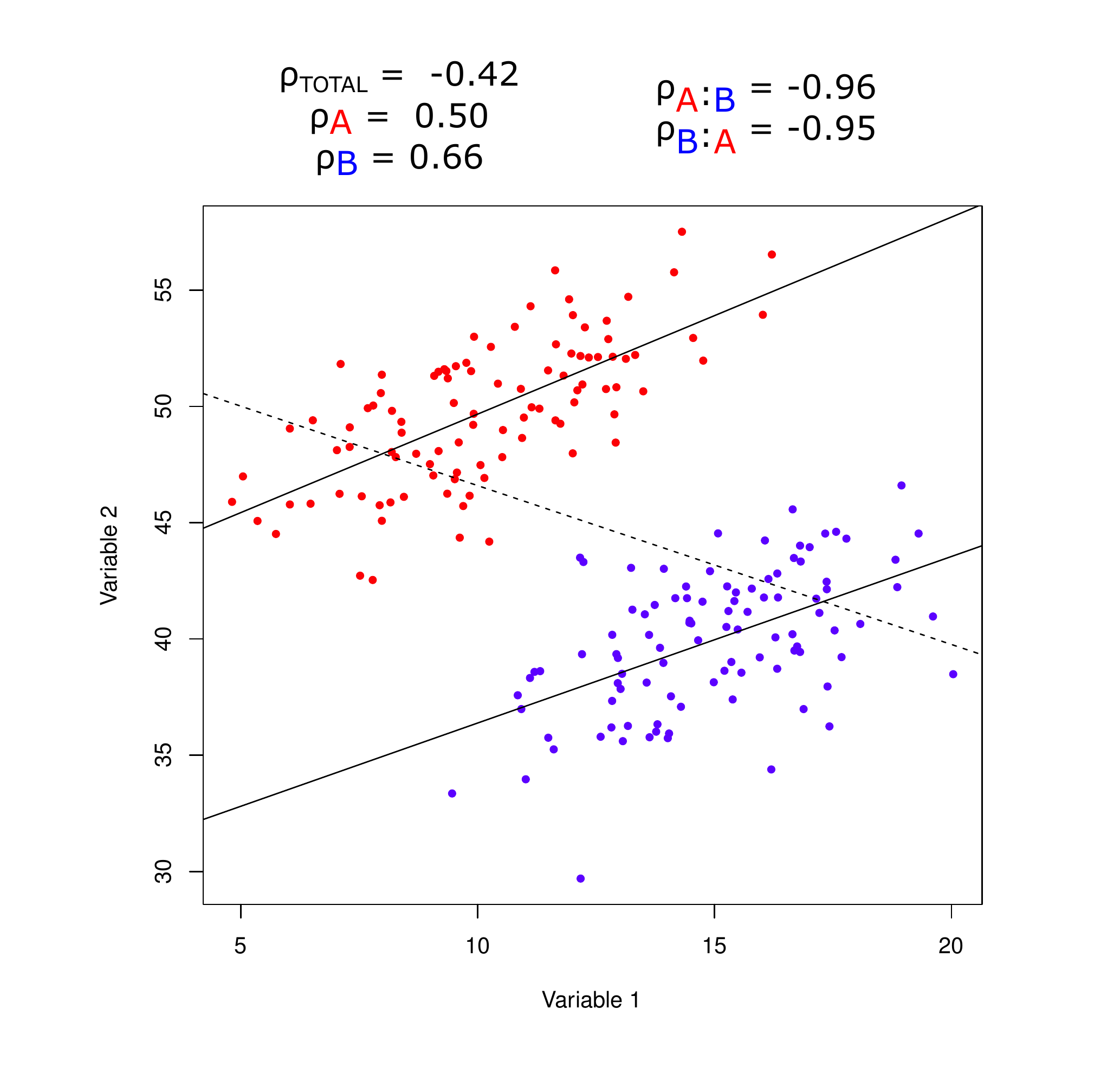}
\end{center}
\caption{
Simulated bivariate data illustrating the Yule-Simpson effect \cite{yule1903, simpson1951, blyth1972, wagner1982, wilcox2001} when correlating subsets that belong to a larger aggregate. The red and blue points represent two subsets within the total population which display positive correlations when correlated as subsets (unbroken lines) yet a negative correlation when taken as a total population (dashed line). When we use a relativised Spearman's correlation (see Methods), however, we find that these subsets display negative correlations relative to each other thereby explaining why there is a negative correlation for the total population.
}
\label{simpsonsparadox}
\end{figure}

\begin{figure}[!ht]
\begin{center}
\includegraphics[width=6in]{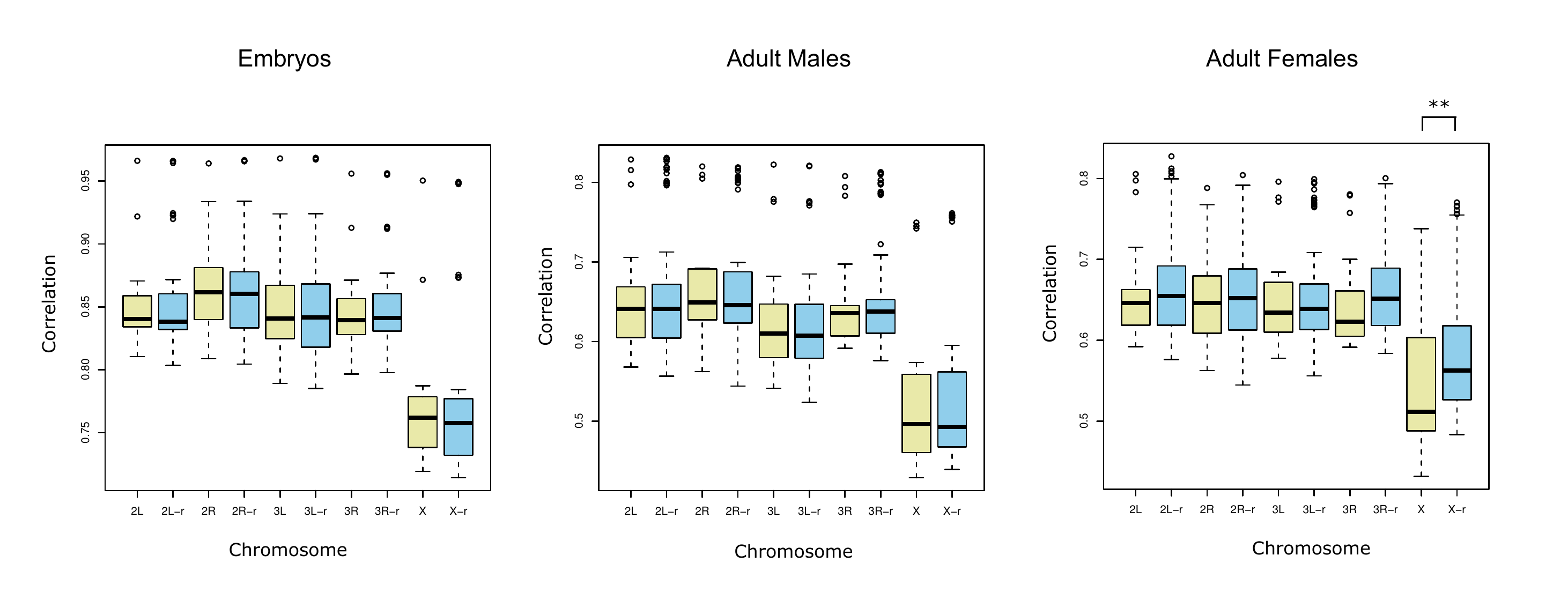}
\end{center}
\caption{
Distributions of pairwise species chromosome correlations for embryos, adult males, and adult females. In light blue are the distributions of a relativised Spearman's rank correlation coefficient (see Methods). The suffix ``\_r'' indicates that these are the relative correlation coefficients for a particular chromosome in relation to the other chromosomes.
}
\label{relativecorrs}
\end{figure}

\clearpage

\section*{Supplementary Tables}

\renewcommand{\tablename}{Supplementary Table}

\begin{table}[!h]
\caption{
\bf{The chromosomal distribution of genes in the expression datasets}}
\begin{center}
\begin{tabular}{|l|l|l|l|l|l|l|l|}
\hline
Stage & Comparison & 2L & 2R & 3L & 3R & X & Total \\
\hline\hline
Embryos & species & 579 & 651 & 677 & 798 & 314 & 3019 \\
Embryos & strains & 2267 & 2521 & 2437 & 3005 & 1998 & 12228 \\
Adults & species & 1214 & 1360 & 1215 & 1662 & 881 & 6332 \\
Adults & strains & 1694 & 1843 & 1721 & 2217 & 1275 & 8750 \\

\hline
\end{tabular}
\end{center}
\label{chromsummary}
\end{table}

\begin{table}[!h]
\caption{
\bf{Contrasts for \emph{Drosophila} embryo species comparisons}}
\begin{center}
\begin{tabular}{|l|l|l|l|l|l|}
\hline
Contrast & Mean 1st & Mean 2nd & W-stat & $P$-value & $P_{adj}$-value\\
\hline\hline
Aut-X & 1.912150 & 2.312695 & 331955.5 & $2.19$ x $10^{-7}$ & - \\
2L-X & 1.906623 & 2.312695 & 74430 & $3.8$ x $10^{-6}$ & $1.9$ x $10^{-5}$\\
2R-X & 1.808078 & 2.312695 & 79352 & $8.8$ x $10^{-9}$ & $8.8$ x $10^{-8}$\\
3L-X & 1.951095 & 2.312695 & 89006 & $1.8$ x $10^{-5}$ & $4.7$ x $10^{-5}$\\
3R-X & 1.963467 & 2.312695 & 104232.5 & $6.3$ x $10^{-6}$ & $2.1$ x $10^{-5}$\\
2L-2R & 1.906623 & 1.808078 & 196983.5 & $0.171$ & $0.244$\\
2L-3L & 1.906623 & 1.951095 & 193720 & $0.723$ & $0.803$\\
2L-3R & 1.906623 & 1.963467 & 227856 & $0.664$ & $0.803$\\
2R-3L & 1.808078 & 1.951095 & 208468 & $0.089$ & $0.148$\\
2R-3R & 1.808078 & 1.963467 & 244404 & $0.053$ & $0.106$\\
3L-3R & 1.951095 & 1.963467 & 269368 & $0.926$ & $0.926$\\

\hline
\end{tabular}
\end{center}
\vspace{0.25cm}
\begin{flushleft}{
Aut - all autosomes. W - Wilcoxon rank sum test statistic. P-values adjusted according to Benjamini-Hochberg correction.}
\end{flushleft}
\label{contr1}
\end{table}

\clearpage

\begin{table}[!h]
\caption{
\bf{Contrasts for \emph{D. melanogaster} embryo strain comparisons}}
\begin{center}
\begin{tabular}{|l|l|l|l|l|l|}
\hline
Contrast & Mean 1st & Mean 2nd & W-stat & $P$-value & $P_{adj}$-value\\
\hline\hline
Aut-X & 1.400482 & 1.287024 & 11273798 & $1.16$ x $10^{-9}$ & - \\
2L-X & 1.411071 & 1.287024 & 2496381 & $7.8$ x $10^{-9}$ & $7.8$ x $10^{-8}$\\
2R-X & 1.395688 & 1.287024 & 2731109 & $1.1$ x $10^{-6}$ & $3.5$ x $10^{-6}$\\
3L-X & 1.422667 & 1.287024 & 2667918 & $3.8$ x $10^{-9}$ & $1.9$ x $10^{-7}$\\
3R-X & 1.380755 & 1.287024 & 3205879 & $4.6$ x $10^{-5}$ & $1.2$ x $10^{-4}$\\
2L-2R & 1.411071 & 1.395688 & 2905198 & $0.318$ & $0.424$\\
2L-3L & 1.411071 & 1.422667 & 2771018 & $0.852$ & $0.852$\\
2L-3R & 1.411071 & 1.380755 & 3514421 & $0.048$ & $0.096$\\
2R-3L & 1.395688 & 1.422667 & 3032346 & $0.433$ & $0.481$\\
2R-3R & 1.395688 & 1.380755 & 3844265 & $0.339$ & $0.424$\\
3L-3R & 1.422667 & 1.380755 & 3764470 & $0.074$ & $0.123$\\

\hline
\end{tabular}
\end{center}
\vspace{0.25cm}
\begin{flushleft}{
Aut - all autosomes. W - Wilcoxon rank sum test statistic. P-values adjusted according to Benjamini-Hochberg correction.}
\end{flushleft}
\label{contr2}
\end{table}

\clearpage

\begin{table}[!h]
\caption{
\bf{Contrasts for \emph{Drosophila} adult female species comparisons}}
\begin{center}
\begin{tabular}{|l|l|l|l|l|l|}
\hline
Contrast & Mean 1st & Mean 2nd & W-stat & $P$-value & $P_{adj}$-value\\
\hline\hline
Aut-X & 1.128154 & 1.129415 & 2409026 & $0.994$ & - \\
2L-X & 1.118199 & 1.129415 & 528305.5 & $0.636$ & -\\
2R-X & 1.127064 & 1.129415 & 599021 & $0.997$ & -\\
3L-X & 1.141319 & 1.129415 & 543405.5 & $0.549$ & -\\
3R-X & 1.126056 & 1.129415 & 729981 & $0.904$ & -\\
2L-2R & 1.118199 & 1.127064 & 815877.5 & $0.609$ & -\\
2L-3L & 1.118199 & 1.141319 & 717046 & $0.237$ & -\\
2L-3R & 1.118199 & 1.126056 & 999595.5 & $0.675$ & -\\
2R-3L & 1.127064 & 1.141319 & 812860.5 & $0.479$ & -\\
2R-3R & 1.127064 & 1.126056 & 1133633 & $0.884$ & -\\
3L-3R & 1.141319 & 1.126056 & 1028575 & $0.390$ & -\\

\hline
\end{tabular}
\end{center}
\vspace{0.25cm}
\begin{flushleft}{
Aut - all autosomes. W - Wilcoxon rank sum test statistic. P-values adjusted according to Benjamini-Hochberg correction.}
\end{flushleft}
\label{contr3}
\end{table}

\clearpage

\begin{table}[!h]
\caption{
\bf{Contrasts for \emph{Drosophila} adult male species comparisons}}
\begin{center}
\begin{tabular}{|l|l|l|l|l|l|}
\hline
Contrast & Mean 1st & Mean 2nd & W-stat & $P$-value & $P_{adj}$-value\\
\hline\hline
Aut-X & 1.168791 & 1.184388 & 2361976 & $0.3552$ & - \\
2L-X & 1.170921 & 1.184388 & 525043.5 & $0.477$ & -\\
2R-X & 1.159897 & 1.184388 & 584673 & $0.336$ & -\\
3L-X & 1.172074 & 1.184388 & 525263 & $0.467$ & -\\
3R-X & 1.169593 & 1.184388 & 717711.5 & $0.413$ & -\\
2L-2R & 1.170921 & 1.159897 & 830314 & $0.800$ & -\\
2L-3L & 1.170921 & 1.172074 & 738058 & $0.975$ & -\\
2L-3R & 1.170921 & 1.169593 & 1011397 & $0.907$ & -\\
2R-3L & 1.159897 & 1.172074 & 822211.5 & $0.832$ & -\\
2R-3R & 1.159897 & 1.169593 & 1125957 & $0.860$ & -\\
3L-3R & 1.1195 & 1.169593 & 1011382 & $0.938$ & -\\

\hline
\end{tabular}
\end{center}
\vspace{0.25cm}
\begin{flushleft}{
Aut - all autosomes. W - Wilcoxon rank sum test statistic. P-values adjusted according to Benjamini-Hochberg correction.}
\end{flushleft}
\label{contr4}
\end{table}

\clearpage

\begin{table}[!h]
\caption{
\bf{Contrasts for \emph{D. melanogaster} female adult strain comparisons}}
\begin{center}
\begin{tabular}{|l|l|l|l|l|l|}
\hline
Contrast & Mean 1st & Mean 2nd & W-stat & $P$-value & $P_{adj}$-value\\
\hline\hline
Aut-X & 0.6113846 & 0.5431882 & 5192587 & $7.28$ x $10^{-6}$ & - \\
2L-X & 0.6399925 & 0.5431882 & 1219771 & $1.46$ x $10^{-9}$ & $1.68$ x $10^{-8}$\\
2R-X & 0.6033502 & 0.5431882 & 1244176 & $0.0051$ & $0.0078$\\
3L-X & 0.6187143 & 0.5431882 & 1189794 & $3.78$ x $10^{-5}$ & $0.00020$\\
3R-X & 0.5943979 & 0.5431882 & 1494048 & $0.0025$ & $0.0046$\\
2L-2R & 0.6399925 & 0.6033502 & 1663789 & $7.1$ x $10^{-4}$ & $0.0046$\\
2L-3L & 0.6399925 & 0.6187143 & 1520440 & $0.029$ & $0.040$\\
2L-3R & 0.6399925 & 0.5943979 & 2010076 & $1.6$ x $10^{-4}$ & $0.0019$\\
2R-3L & 0.6033502 & 0.6187143 & 1548129 & $0.219$ & $0.240$\\
2R-3R & 0.6033502 & 0.5943979 & 2049155 & $0.868$ & $0.868$\\
3L-3R & 0.6187143 & 0.5943979 & 1959594 & $0.143$ & $0.175$\\

\hline
\end{tabular}
\end{center}
\vspace{0.25cm}
\begin{flushleft}{
Aut - all autosomes. W - Wilcoxon rank sum test statistic. P-values adjusted according to Benjamini-Hochberg correction.}
\end{flushleft}
\label{contr5}
\end{table}

\clearpage

\begin{table}[!h]
\caption{
\bf{Contrasts for \emph{D. melanogaster} male adult strain comparisons}}
\begin{center}
\begin{tabular}{|l|l|l|l|l|l|}
\hline
Contrast & Mean 1st & Mean 2nd & W-stat & $P$-value & $P_{adj}$-value\\
\hline\hline
Aut-X & 0.4768255 & 0.4189356 & 5359535 & $9.89$ x $10^{-11}$ & - \\
2L-X & 0.4877695 & 0.4189356 & 1249111 & $2.52$ x $10^{-13}$ & $2.56$ x $10^{-12}$\\
2R-X & 0.4705672 & 0.4189356 & 1281926 & $1.49$ x $10^{-5}$ & $4.97$ x $10^{-5}$\\
3L-X & 0.4884414 & 0.4189356 & 1255394 & $1.38$ x $10^{-11}$ & $6.89$ x $10^{-11}$\\
3R-X & 0.467385 & 0.4189356 & 1526993 & $7.42$ x $10^{-5}$ & $1.86$ x $10^{-4}$\\
2L-2R & 0.4877695 & 0.4705672 & 1657322 & $0.0015$ & $0.0021$\\
2L-3L & 0.4877695 & 0.4884414 & 1473490 & $0.583$ & $0.583$\\
2L-3R & 0.4877695 & 0.467385 & 2013942 & $9.99$ x $10^{-5}$ & $2.00$ x $10^{-4}$\\
2R-3L & 0.4705672 & 0.4884414 & 1505481 & $0.0088$ & $0.011$\\
2R-3R & 0.4705672 & 0.467385 & 2064474 & $0.563$ & $0.583$\\
3L-3R & 0.4884414 & 0.467385 & 2025011 & $9.20$ x $10^{-4}$ & $0.0015$\\

\hline
\end{tabular}
\end{center}
\vspace{0.25cm}
\begin{flushleft}{
Aut - all autosomes. W - Wilcoxon rank sum test statistic. P-values adjusted according to Benjamini-Hochberg correction.}
\end{flushleft}
\label{contr6}
\end{table}

\clearpage

\begin{table}[!h]
\caption{
\bf{Contrasts for \emph{Drosophila} embryos for a common set of 2072 genes and 5 species.}}
\begin{center}
\begin{tabular}{|l|l|l|l|l|l|}
\hline
Contrast & Mean 1st & Mean 2nd & W-stat & $P$-value & $P_{adj}$-value\\
\hline\hline
2L-X & 1.767753 & 2.023779 & 39029 & $0.009315$ & $0.0232$\\
2R-X & 1.647628 & 2.023779 & 39580 & $5.4$ x $10^{-5}$ & $5.4$ x $10^{-4}$\\
3L-X & 1.759429 & 2.023779 & 40418 & $0.004624$ & $0.0154$\\
3R-X & 1.736327 & 2.023779 & 51914.5 & $8.5$ x $10^{-4}$ & $0.0042$\\
2L-2R & 1.767753 & 1.647628 & 99783 & $0.02553$ & $0.051$\\
2L-3L & 1.767753 & 1.759429 & 89248 & $0.3774$ & $0.377$\\
2L-3R & 1.767753 & 1.736327 & 119319 & $0.2105$ & $0.263$\\
2R-3L & 1.647628 & 1.759429 & 91380 & $0.06257$ & $0.104$\\
2R-3R & 1.647628 & 1.736327 & 121461.5 & $0.09044$ & $0.129$\\
3L-3R & 1.759429 & 1.736327 & 123275 & $0.3443$ & $0.377$\\

\hline
\end{tabular}
\end{center}
\vspace{0.25cm}
\begin{flushleft}{
W - Wilcoxon rank sum test statistic. P-values adjusted according to Benjamini-Hochberg correction.}
\end{flushleft}
\label{contrcommon1}
\end{table}

\clearpage

\begin{table}[!h]
\caption{
\bf{Contrasts for \emph{Drosophila} adults for a common set of 2072 genes and 5 species.}}
\begin{center}
\begin{tabular}{|l|l|l|l|l|l|}
\hline
Contrast & Mean 1st & Mean 2nd & W-stat & $P$-value & $P_{adj}$-value\\
\hline\hline
2L-X & 0.9187558 & 0.9741846 & 40504.5 & $0.04798$ & $0.164$\\
2R-X & 0.90749 & 0.9741846 & 44751.5 & $0.04946$ & $0.164$\\
3L-X & 0.9049189 & 0.9741846 & 41350 & $0.01444$ & $0.144$\\
3R-X & 0.9461049 & 0.9741846 & 57753.5 & $0.1443$ & $0.288$\\
2L-2R & 0.9187558 & 0.90749 & 92572 & $0.4904$ & $0.490$\\
2L-3L & 0.9187558 & 0.9049189 & 90367.5 & $0.2641$ & $0.293$\\
2L-3R & 0.9187558 & 0.9461049 & 112693 & $0.2353$ & $0.293$\\
2R-3L & 0.90749 & 0.9049189 & 99597 & $0.2614$ & $0.293$\\
2R-3R & 0.90749 & 0.9461049 & 124054 & $0.2174$ & $0.293$\\
3L-3R & 0.9049189 & 0.9461049 & 115402 & $0.08814$ & $0.220$\\

\hline
\end{tabular}
\end{center}
\vspace{0.25cm}
\begin{flushleft}{
W - Wilcoxon rank sum test statistic. P-values adjusted according to Benjamini-Hochberg correction.}
\end{flushleft}
\label{contrcommon2}
\end{table}

\clearpage

\begin{table}[!h]
\caption{
\bf{Characterisation of genes with a percentile X/A divergence ratio greater than 1.015 in adult males.}}
\begin{center}
\begin{tabular}{|l|l|l|l|l|l|l|}
\hline
ID & Term & \# & Sig. & Exp. & $P$-value & $P_{adj}$-value\\
\hline\hline
 GO:0007538 & primary sex determination & 11 & 5 & 0.6 & $8.9$ x $10^{-5}$ & 0.312\\
 GO:0019748 & secondary metabolic process & 24 & 7 & 1.32 & $1.6$ x $10^{-4}$ & 0.312\\
 GO:0030534 & adult behavior & 34 & 7 & 1.86 & $6.2$ x $10^{-4}$ & 0.625\\
 GO:0007362 & terminal region determination & 10 & 3 & 0.5 & 0.0062 & 0.625\\
 GO:0046152 & ommochrome metabolic process & 13 & 4 & 0.71 & 0.0076 & 1.0\\
\hline
\end{tabular}
\end{center}
\vspace{0.25cm}
\begin{flushleft}{
Enrichment is based on the `parent-child' algorithm in the topGO R package and Fisher's exact test applied to 352 genes that have an X/A percentile divergence ratio of $>$ 1.015 against the background of the genes in the dataset. \# - total number of genes with this annotation in the dataset. Sig. - significant, Exp. - expected. $P_{adj}$-value - adjusted according to the Benjamini-Hochberg false discovery rate.}
\end{flushleft}
\label{XAadultmales}
\end{table}

\clearpage

\begin{table}[!h]
\caption{
\bf{Characterisation of the embryonic expression patterns of genes residing on the X chromosome in \emph{Drosophila}.}}
\begin{center}
\begin{tabular}{|l|l|l|l|l|l|l|l|}
\hline
Test & ID & Term & \# & Sig. & Exp. & $P$-value & $P_{adj}$-value\\
\hline\hline
\multirow{3}{*}{Under}& 254 & cellular blastoderm & 3039 & 418 & 477.37 & $2.6$ x $10^{-7}$ & $9.5$ x $10^{-5}$\\
& 273 & visual anlage & 100 & 7 & 15.71 & $0.017$ & $1.0$\\
& 222 & visual primordium & 88 & 6 & 13.82 & $0.022$ & $1.0$\\
\cline{1-8}
\multirow{3}{*}{Over}& 493 & no staining (stage 5) & 2773 & 471 & 435.58 & 0.00029 & 0.10643\\
& 346 & muscle system primordium & 670 & 109 & 105.24 & $0.00392$ & 0.71392\\
& 580 & apically cleared & 74 & 19 & 11.62 & $0.01309$ & $1.0$\\
\hline
\end{tabular}
\end{center}
\vspace{0.25cm}
\begin{flushleft}{
Enrichment is based on the `parent-child' algorithm in the topGO R package and Fisher's exact test applied to 2228 genes that reside on the X chromosome in \emph{Drosophila}, and enrichment is relative to the whole genome. Terms with uncorrected \emph{P}-values below 0.05 are shown. \# - total number of genes with this annotation in the dataset. Sig. - significant, Exp. - expected. $P_{adj}$-value - adjusted according to the Benjamini-Hochberg false discovery rate.}
\end{flushleft}
\label{CVenrich}
\end{table}

\clearpage

\begin{table}[!ht]
\caption{
Fitnesses in a diploid two-locus epistatic model with X-linkage.}
\begin{center}
\begin{tabular}{c c|c|c|c|c|c|}
\cline{3-7}
 & & \multicolumn{5}{|c|}{\male} \\
\cline{3-7}
 & & TC & Tc & tC & tc & 00 \\
\hline
\multicolumn{1}{|c}{} \multirow{4}{*}{\female} & \multicolumn{1}{|c|}{TC} & 1 & 1 & 1 & 1 + $\frac{h}{2}s$ & 1 \\
\multicolumn{1}{|c}{} & \multicolumn{1}{|c|}{Tc} & 1 & 1 & 1 + $\frac{h}{2}s$ & $1 + h s$ & 1 \\
\multicolumn{1}{|c}{} & \multicolumn{1}{|c|}{tC} & 1 & 1 + $\frac{h}{2}s$ & 1 & $1 + h s$ & 1 \\
\multicolumn{1}{|c}{} & \multicolumn{1}{|c|}{tc} & 1 + $\frac{h}{2}s$ & $1 + h s$ & $1 + h s$ & $1 + s$ & $1 + s$ \\
\hline
\end{tabular}
\end{center}
\begin{flushleft}{\small{
Fitnesses of different male-female gametic combinations when both the loci are located on the X chromosome. T/t - trans-acting gene; C/c - cis-acting locus; 00 - indicates a male gamete carrying a Y chromosome; $s$ - selection coefficient; $h$ - dominance coefficient.}}
\end{flushleft}
\label{Xlinkedfitness}
\end{table}

\clearpage

\begin{table}[!ht]
\caption{
Fitnesses in a diploid two-locus epistatic model.}
\begin{center}
\begin{tabular}{c c|c|c|c|c|c|c|}
\cline{3-8}
 & & \multicolumn{6}{|c|}{\male} \\
\cline{3-8}
 & & TC & Tc & tC & tc & T0 & t0 \\
\hline
\multicolumn{1}{|c}{} \multirow{4}{*}{\female} & \multicolumn{1}{|c|}{TC} & 1 & 1 & 1 & 1 + $\frac{h}{2}s$ & 1 & 1 \\
\multicolumn{1}{|c}{} & \multicolumn{1}{|c|}{Tc} & 1 & 1 & 1 + $\frac{h}{2}s$ & $1 + h s$ & 1 & $1 + h s$ \\
\multicolumn{1}{|c}{} & \multicolumn{1}{|c|}{tC} & 1 & 1 + $\frac{h}{2}s$ & 1 & $1 + h s$ & 1 & 1 \\
\multicolumn{1}{|c}{} & \multicolumn{1}{|c|}{tc} & 1 + $\frac{h}{2}s$ & $1 + h s$ & $1 + h s$ & $1 + s$ & $1 + h s$ & $1 + s$ \\
\hline
\end{tabular}
\end{center}
\begin{flushleft}{\small{
Fitnesses of different male-female gametic combinations when there is a beneficial partially recessive interaction between an autosomal allele and an X-linked allele (males are the heterogametic sex). T/t - trans-acting autosomal gene; C/c - cis-acting X-linked locus; 0 - indicates a male gamete carrying a Y chromosome; $s$ - selection coefficient; $h$ - dominance coefficient.}}
\end{flushleft}
\label{fitnesses}
\end{table}

\end{document}